\documentclass[11pt,a4paper]{article}

\usepackage{mymacro}

\newcommand{\hBPS}[1][2]{\ensuremath{\CO_{#1}}}
\newcommand{\qBPS}[1][02]{\ensuremath{\CO_{#1}}}
\newcommand{\qBPSb}[1][02]{\ensuremath{\COb_{#1}}}

\newcommand{\stry}[1]{\ensuremath{\mathsf{y}_{#1}}}
\newcommand{\strSSb}[1]{\ensuremath{\mathsf{S}_{#1}}}
\newcommand{\strTrY}[1]{\ensuremath{\mathsf{Y}_{#1}}}
\newcommand{\strSYSb}[3]{\ensuremath{\mathsf{J}^{#1#3}_{#2}}}
\newcommand{\strSYS}[3]{\ensuremath{\mathsf{K}^{#1#3}_{#2}}}
\newcommand{\strSbYSb}[3]{\ensuremath{\overbarUp{\mathsf{K}}^{#1#3}_{#2}}}

\newcommand{\THHHH}[1]{\ensuremath{\BBT_{#1}}}                                                     % <O2 O2 O2 O2>
\newcommand{\TQHHH}[1]{{\hypersetup{linkcolor=.}\hyperlink{eqref:TQHHHDef}{\ensuremath{\BBT_{#1}^{\mathrm{\llsp I}}}}}}      % <(2,0,2) O2 O2 O2>
\newcommand{\TQQHH}[1]{{\hypersetup{linkcolor=.}\hyperlink{eqref:TQQHHDef}{\ensuremath{\BBT_{#1}^{\mathrm{\llsp II}}}}}}     % <(2,0,2) (2,0,2) O2 O2>
\newcommand{\THQQH}[1]{{\hypersetup{linkcolor=.}\hyperlink{eqref:THQQHDef}{\ensuremath{\BBT_{#1}^{\mathrm{III}}}}}}          % <O2 (2,0,2) (2,0,2) O2>
\newcommand{\TQQQQ}[1]{{\hypersetup{linkcolor=.}\hyperlink{eqref:TQQQQDef}{\ensuremath{\BBT_{#1}^{\mathrm{IV}}}}}}           % <(2,0,2) x 4 >
\newcommand{\TQHHHtilde}[1]{{\hypersetup{linkcolor=.}\hyperlink{eqref:TQHHHtildeDef}{\ensuremath{\tilde{\BBT}_{#1}^{\mathrm{\llsp I}}}}}}      % <(2,0,2) O2 O2 O2> (alternative basis)

\newcommand{\kinPref}[1]{{\hypersetup{linkcolor=.}\hyperlink{eqref:kinPref}{\ensuremath{\CK_{#1}}}}}
\newcommand{\kinPrefz}[1]{{\hypersetup{linkcolor=.}\hyperlink{eqref:kinPrefz}{\ensuremath{\mathfrak{K}_{#1}}}}}
\newcommand{\kinPrefzb}[1]{{\hypersetup{linkcolor=.}\hyperlink{eqref:kinPrefz}{\ensuremath{\overbar{\mathfrak{K}}_{#1}}}}}

\newcommand{\Dt}{\Delta_{34}}
\newcommand{\Do}{\Delta_{12}}
 \newcommand{\shortminus}{\scalebox{0.75}[1.0]{\( - \)}}

\title{\textbf{Rebooting quarter-BPS operators in $\boldsymbol{\mathcal{N}=4}$ Super Yang-Mills}}

\author{Agnese Bissi,}
\author{Giulia Fardelli}
\author{and Andrea Manenti}

\affiliation{Department of Physics and Astronomy, Uppsala University, Box 516, SE-751 20 Uppsala, Sweden}

\makeatletter
\renewcommand{\@email}[1]{#1}
\makeatother

\emailAdd{\{\href{mailto:agnese.bissi@physics.uu.se}{\tt agnese.bissi},
            \href{mailto:giulia.fardelli@physics.uu.se}{\tt giulia.fardelli},
            \href{mailto:andrea.manenti@physics.uu.se}{\tt andrea.manenti}\}\texttt{\textcolor{blue}{@physics.uu.se}}
}

\abstract{We start a systematic study of quarter-BPS operators in four-dimensional ${\CN=4}$ Super Yang-Mills with gauge group $\SU(N)$ making use of recently developed tools in conformal field theory. We adapt the technology of embedding space tensor structures in four dimensions to the problem of computing R-symmetry tensor structures, and we use the underlying chiral algebra to obtain the superconformal Ward identities. This allows us to fix the protected part of the four-point correlators, up to few ambiguities. As applications, we use the Lorentzian inversion formula to study 
the leading order OPE data in the large~$N$ supergravity limit and we make contact with the OPE limit of the five-point function of half-BPS operators.}

\preprint{UUITP-55/21}

\begin{document}

\maketitle

\section{Introduction}

In the last years, there has been a tremendous progress in studying four-point correlators of half-BPS operators in four dimensional Superconformal Field Theories (SCFTs) with conformal bootstrap techniques,
in particular in the context of $\mathcal{N}=4$ supersymmetry, with the seminal papers~\cite{Beem:2013qxa, Beem:2016wfs}.\footnote{ Several results have been obtained also in the context of $\mathcal{N}=2$ supersymmetry, started in~\cite{Beem:2014zpa}.} 
The half-BPS superconformal primaries $\mathcal{O}_p$ are scalar operators, of protected dimension $\Delta=p$, transforming under the $(0,p,0)$ representation of the $\mathrm{SU}(4)$ R-symmetry group. They are single trace operators made by the six scalar fields of $\mathcal{N}=4$ super Yang-Mills. 
The reason for these extensive studies is mainly twofold. These operators preserve the highest amount of supersymmetry,  their conformal dimension as well as three-point correlators involving all such operators are protected, meaning that they are fixed by symmetries and do not contain dynamical information. 
Moreover, higher-point correlators are strongly constrained by supersymmetry, both in their space-time and R-symmetry dependence. Remarkably, the existence of a chiral algebra, essentially linked to the $\mathcal{N}=2$ superconformal invariance, has been shown to emerge when studying the meromorphic part of four-point correlators, thus this piece of information can be completely fixed by the free-field values for Super Yang-Mills (SYM) theories.
These properties have been crucial in applying conformal bootstrap techniques to these correlators, numerically and analytically, see the recent reviews~\cite{Poland:2018epd, Bissi:2021spj}. 
From a different perspective, the study of these correlators is linked to amplitudes in the dual superstring theory description.  In particular, the  half-BPS operator with the lowest conformal dimension, $\Delta=2$, is the scalar component of the stress-energy supermultiplet, which is dual to the super-graviton multiplet. The entire tower of Kaluza-Klein modes of the graviton are instead  dual to the $\Delta=p$ with $p>2$ half-BPS operators. Thus the study of these correlators using conformal bootstrap techniques, gives an operative method to explore super-graviton amplitudes on $\mathrm{AdS}$ background, which have been inaccessible for several years, due to the increasing complications intrinsic to perturbative methods in this setup~\cite{Aharony:2016dwx,Alday:2017xua,Aprile:2017bgs}. 

Driven by the conceptual advancements and the plethora of results obtained by studying this class of correlators, mostly in providing information on dynamical, coupling dependent quantities, in this paper we would like to revive the study of quarter-BPS operators in the context of $\mathcal{N}=4$ Super Yang-Mills theories in four dimensions, with $\mathrm{SU}(N)$ gauge group. These operators have protected conformal dimension, they transform in the $(q,p,q)$ representation of the $\mathrm{SU}(4)$ R-symmetry group and are a combination of a double trace and a single trace term, the latter being sub-leading in the large $N$ expansion. These operators appear in all the context mentioned above, in particular they are present in the operator product expansion of half-BPS operators.

In this paper, we start the study of  four-point correlators involving at least one quarter-BPS operator. We focus in particular on operators transforming under the representation $(2,0,2)$, since they are the first non-trivial ones and they appear in the operator product expansion of $\mathcal{O}_2 \times \mathcal{O}_2$, making it possible to consider mixed correlator with such operators.
We exploit the power of the symmetries, using together the superconformal Ward identities and the underlying chiral algebra, to constrain, at least partially, the protected structure of such correlators. We computed free theory results and, by requiring the absence of higher\nobreakdash-spin currents and other mild assumptions, we managed to partially resolve the ambiguities. In addition, using the inversion formula, we compute the sub-leading large-$N$ correction to the CFT data, in the supergravity regime. We make contact with the recently computed five-point function of half-BPS operators of protected dimension two~\cite{Goncalves:2019znr}, by taking the OPE limit and projecting in the R-symmetry structure that we are interested in, and we found perfect agreement with all the checks that we made.

The paper is structured as follows. In section~\ref{sec:generalities} we introduce the quarter-BPS operators, first from a representation theoretic point of view and then we specialize to the kinematics of their correlators. In section~\ref{sec:chialg} we briefly review the chiral algebra construction in order to subsequently apply it to our correlators of interest. Then in section~\ref{sec:cases} we consider mixed correlators involving the $(2,0,2)$ quarter-BPS operator $\qBPS$, while in section~\ref{sec:Results} we present large-$N$ results for the averaged anomalous dimensions of unprotected operators appearing in their OPE, after a review of the method presented in section~\ref{sec:Lorentz}. In section~\ref{sec:fivepoint} we prove that the correlator with one $\qBPS$ and three $\hBPS$ is protected at large $N$. Finally, section~\ref{sec:outlook} contains some discussions and future directions. Several technical details are left to six appendices.

\subsection{General idea}

In this paper we focus on quarter-BPS operators $\qBPS[pq]$ transforming in the $(q,p,q)$ representation of the $\SU(4)$ R-symmetry and which satisfy a shortening condition of the $\CB$ type --- also known as $B\overbar{B}$ type. 

When considering operators transforming in these representations, it is important to deal efficiently with all the R-symmetry indices that start appearing. To this aim, we introduce auxiliary null vectors, satisfying precise properties, contracting $\SO(6)$ and $\SU(4)$ (anti)fundamental indices
\eqn{
\qBPS[pq](S,\Sb,y) \equiv (\qBPS[pq])^{m_1\cdots m_q}_{n_1\cdots n_q\,,\,M_1\cdots M_p} \, \Sb_{m_1}\cdots\Sb_{m_q}\,S^{n_1}\cdots S^{n_q}\,y^{M_1}\cdots y^{M_p}\,.
}[]
This allows us to define all the tensor structures, now functions of products or combinations of these vectors. This is necessary in order to construct correlators of these operators as well as to analyze the different representations exchanged in the OPE. 

When expanded in $\mathcal{N}=2$ supermultiplets, the operators $\qBPS[pq]$ contain half-BPS operators, which, according to the chiral algebra construction of~\cite{Beem:2013sza}, are of the Schur type. It is essentially this fact that allows us to identify a protected subsector in their correlation functions. More precisely, we are able to derive superconformal Ward identities for the four-point functions under analysis by imposing that the correlator should be meromorphic when computed in a special, space-time dependent, configuration of the $\SU(4)$ polarizations, with the operator positions restricted to a plane. This configuration simultaneously selects the specific $\CN=2$ half-BPS operator inside $\qBPS[pq]$ and performs the chiral algebra twist
\eqn{
\langle \qBPS[p_1q_1]\qBPS[p_2q_2]\qBPS[p_3q_3]\qBPS[p_4q_4]\rangle \quad \xrightarrow[\substack{
\quad\, y_i\cdot y_j \,=\, \frac12\etat_i\etat_j\, (\zb_i - \zb_j)\,,\quad\\
S_i\cdot\Sb_j \,=\, \zb_i - \zb_j + \etat_i\etat_j\quad
}]{} \quad  \BBT(\etat_1,\ldots,\etat_4)\,\gothf(z) + \cdots
}[]
In the above schematic equation the $\etat_i$ are some newly introduced $\SU(2)$ polarizations that contract the flavor indices of the Schur operator inside $\qBPS[pq]$, $\BBT(\etat_1,\ldots,\etat_4)$ is a tensor structure\footnote{In principle, there could be more than one structure, but in our cases there will always be only one.} that scales as $\lambda^{p_i}$ when $\etat_i\to\lambda\etat_i$ and $z_i,\zb_i$ are the positions on the chiral algebra plane of the four operators. The ellipsis denotes terms that scale differently in the $\etat_i$ polarizations. The function $\gothf(z)$ is meromorphic in $z$ and is coupling independent, so it can be computed from the free theory.

The basic idea behind these identities is that they provide a way to separate a correlator in a protected part, encoding the information deriving from the chiral algebra, from a set of unprotected and dynamical functions $\mathcal{H}_m$. Unfortunately this splitting is not unique: there exists always an ambiguity, namely a function that vanishes under the chiral algebra map, but it is not responsible for the exchange of unprotected multiplets. In this paper we fix this ambiguity as much as possible by imposing simple consistency requirements on the conformal block expansion of the correlator.

\subsection{Summary of results}

In order to illustrate the construction outlined above,  we first apply it to the well-studied example of the correlator of four $\hBPS$ operators, reproducing the well-known results~\cite{Beem:2013qxa,Beem:2016wfs}.

Then we focus on four-point functions containing respectively one, two or four $\qBPS$ operators.  Given the presence of Schur-type operators in the $\mathcal{N}=2$ decomposition of $\qBPS$, one could think of using $\mathcal{N}=2$ superconformal blocks~\cite{Dolan:2001tt,Beem:2014zpa} to study these correlators. Unfortunately, the $\mathcal{N}=2$ half-BPS operator in $\qBPS$ comes together with long multiplets,  for  whose correlators an expansion in superconformal blocks is not known yet.

In analysing the various four-point functions, we start by fixing the protected part as much as possible,  then we extract corrections at large $N$ to the OPE data of operators in the OPE $\hBPS \times \qBPS$ and $\qBPS \times \qBPS$ by means of the Lorentzian inversion formula. We cannot yet compute the anomalous dimensions of the single eigenstates due to the inevitable mixing of degenerate operators that will take place even at tree level. For this reason we only quote the results for the averaged quantities
\eqn{
\langle a^{(0)}\,\gamma^{(1)}\rangle = \sum_I a^{(0)}_{\CO_I}\,\gamma^{(1)}_{\CO_I}\,,\qquad
\langle a^{(1)} \rangle = \sum_I a^{(1)}_{\CO_I}\,,
}[]
where $\gamma^{(1)}$ is the first correction to the conformal dimension, $a^{(i)}$ is the $i$-th order OPE coefficient squared and the sum is over all operators with same bare dimension.

By taking the OPE limit of the supergravity five-point function of all $\hBPS$ in~\cite{Goncalves:2019znr}, we prove that $\langle \qBPS \hBPS \hBPS \hBPS \rangle$ is protected. This result was not known  in the literature as far as we are aware.\footnote{Although~\cite{Goncalves:2019znr} argued for some OPE coefficient in their block expansion to be protected.} In doing this analysis, we develop a machinery to deal with five-point tensor structures and to project them to four-point ones that can be easily generalized to higher-points. 

\section{Generalities of quarter-BPS operators}\label{sec:generalities}

\subsection{Superconformal representation theory}

Studying correlators of quarter-BPS scalar primaries involves various technical challenges, mainly due to the rapid growth of R-symmetry tensor structures. In this section, however, we would like to introduce the operators of interest from a purely representation theoretic point of view, keeping the technical details to a minimum. The quarter-BPS operator \qBPS[pq], for $q>0$, is defined to be the superconformal primary of the following multiplet
\eqn{
B\overbar{B}[0\llsp;0]^{(q,p,q)}_{2q+p}\,,\qquad \text{or} \qquad \mathcal{B}_{[q,p,q](0,0)}\,,
}[]
where in the left we use the notation of~\cite{Cordova:2016emh} and in the right that of~\cite{Dolan:2002zh}. We will use the former for the rest of the paper. When $q>0$ the operator satisfies a quarter-BPS shortening condition because it is annihilated by four supercharges, namely
\eqn{
Q^{(1,0,0)}_+\,,\qquad \Qb^{(0,0,1)}_{\dot+}\,,\qquad
Q^{(1,0,0)}_-\,,\qquad \Qb^{(0,0,1)}_{\dot-}\,,
}[]
where the superscript denotes the $\SU(4)$ representation and the subscript the $\SU(2)$ spin.\footnote{If instead $q=0$ the superconformal primary is annihilated by two additional supercharges, namely $
Q^{(-1,1,0)}_+$ and $\Qb^{(0,1,-1)}_{\dot+}$,
consequently, the shortening condition becomes half-BPS and indeed it yields the familiar multiplet $\hBPS[p]$. }

Naively, the simplest such multiplet would be $\qBPS[01]$, which transforms in the $(1,0,1)$ of $\SU(4)$ and has dimension two. However this operator only appears in free theories because it has a higher spin conserved current at level six
\eqn{
Q^3 \Qb{}^3\lsp \qBPS[01] \sim  [3\lsp;\lnsp3]^{(0,0,0)}_{5}\,.
}[]
Moreover, given its conformal dimension, it must be built out of two fundamental fields $\varphi^M_I(x)$ with the gauge index contracted and the $\SU(4)$ index antisymmetrized.\footnote{Refer to table~\ref{tab:indices} in appendix~\ref{app:notation} for the naming of the indices.} This contraction, without any additional indices, is vanishing. The next simplest operators are $\qBPS[11]$ and $\qBPS[21]$, which have dimensions $3$ and $4$ respectively.  In this case, it can be shown that the trace over $\SU(N)$ vanishes and therefore no such operators can be constructed~\cite{DHoker:2003csh}.\footnote{This is proven in detail in appendix~\ref{app:low_ops} for the interested reader.}
Thus, the first non trivial operator that one can consider is $\qBPS$, which has dimension four and transforms in the $(2,0,2)$ representation of $\SU(4)$.

The superprimary of each of these multiplets can be built out of the six scalars $\varphi^M_I(x)$. As it is well known, the representations of the type $(0,p,0)$ are obtained by taking a symmetric traceless combination of the $\SO(6)$ indices $M_i$, as follows
\eqn{
\hBPS[p] = \tr\lsp(T^{I_1}\cdots T^{I_p})\, \varphi^{(M_1}_{I_1}\cdots \varphi^{M_p)}_{I_p} -  \mbox{traces}\,.
}[]
While the representations $(q,0,q)$ are built by antisymmetrizing the indices in pairs. More precisely, we can make use of the six-dimensional rotation matrices $(\Sigma^{MN})_m^{\phantom{m}n}$, $(\Sigmab{}^{MN})^m_{\phantom{m}n}$, defined in appendix~\ref{app:notation}. These operators normally are a linear combination of single and double traces~\cite{Ryzhov:2001bp,DHoker:2003csh}. The coefficients of the linear combination are fixed by imposing that the two-point function is unit-normalized and the operator is a short superconformal primary. If, for the moment, we leave aside the $\SU(N)$ part, we can write
\eqn{
\qBPS[0q]= \tr\lsp(T^{I_1}\cdots )\,\tr\lsp(\cdots T^{I_{2q}})\, \varphi^{M_1}_{I_1}\cdots \varphi^{M_{2q}}_{I_{2q}}\, (\Sigma_{M_1M_2})_{m_1}^{\phantom{m_1}{n_1}} \cdots (\Sigma_{M_{2q-1}M_{2q}})_{m_q}^{\phantom{m_q}{n_q}}\,.
}[]
The detailed expression will be given only for the operator under study, $\qBPS$. General operators $\qBPS[pq]$ with both labels nonzero are obtained with a combination of the two index contractions presented above.

In section~\ref{sec:chialg} we will make use of the chiral algebra of~\cite{Beem:2013sza} for finding the superconformal Ward identities satisfied by the four-point functions of $\qBPS[pq]$. It is therefore beneficial to learn how to expand these multiplets into $\CN=2$ submultiplets, since the chiral algebra is an $\CN=2$ construction. In particular, we want to look for operators which satisfy two equalities between the conformal dimension $\Delta$, the spin $j,\jb$, the $\rmU(1)_R$ R-charge $r$ and the $\SU(2)_R$ R-charge\footnote{Note that our convention is to denote representations by their Dynkin labels, so spin-$\frac12$ would be $R=1$.} $R$
\eqn{
r=j-\jb\,,\qquad 2\Delta = j+\jb+2R\,.
}[eq:schurcondition]
These are termed Schur operators. The quantum numbers $r$ and $R$ arise from the familiar breaking of the $\SU(4)_R$ R-symmetry group into
\eqn{
\SU(4)_R \;\longrightarrow\; \SU(2)_R \times \rmU(1)_R \times \SU(2)_F\,.
}[eq:breaking]
The equalities~\eqref{eq:schurcondition} define a superconformal primary only if $r=j=0$. In this case the operator belongs to an $\CN=2$ half-BPS multiplet. At level zero of $\qBPS[pq]$ we find precisely these multiplets, with a multiplicity of $p+1$
\eqn{
B\bar B[0\llsp;\lnsp0]^{(q,p,q)}_{2q+p} \quad \supset  \quad (p+1)\,B\bar B[0\llsp;\lnsp0]^{(2q+p;\llsp0)}_{2q+p}\,.
}[eq:submultiplet]
The multiplicity implies that the Schur operator extracted from $\qBPS[pq]$ is also transforming in the charge-$p$ representation of the flavor group $\SU(2)_F$ which appears in~\eqref{eq:breaking}. So, to summarize, the quarter-BPS operator $\qBPS[pq]$ contains, at level zero, an $\CN=2$ half-BPS primary operator which has $\SU(2)_R$ R-charge equal to $2q+p$ and flavor charge equal to~$p$. This is the operator that we will exploit in order to derive the superconformal Ward identities.

\subsection{Protected three-point functions}
In this section, we collect known facts about three-point functions involving quarter-BPS operators.
In \cite{DHoker:2001jzy}, the authors analyze three-point functions including quarter-BPS operators in a weak coupling expansions and they find that certain classes of correlators are protected at order $g^2$ for any value of $N$ based on SU(4), SU($N$) arguments and explicit space-time computations.  These are
\begin{enumerate}
\item $\langle \hBPS[p] \hBPS[q] \mathcal{O}\rangle $, where $\mathcal{O}$ is a $\frac{1}{2}$-,  $\frac{1}{4}$- or $\frac{1}{8}$-BPS operator.
\item $\langle \qBPS[pq]\qBPSb[rs] \hBPS[2]\rangle$ either vanish because $(q,p,q) \not \in (s,r,s) \otimes (0,2,0)$ or can be related to
\begin{subequations}\label{eq:3qqh}
\begin{gather}
\langle \qBPS[pq] \qBPSb[(p-2)q] \hBPS[2] \rangle\, , \quad
\langle \qBPS[pq] \qBPSb[(p+2)(q-2)] \hBPS[2] \rangle\, ,\quad
\langle \qBPS[pq] \qBPSb[p(q-2)] \hBPS[2] \rangle\, , \label{eq:3qqh123}\\
\langle \qBPS[pq] \qBPSb[(p+2)(q-1)] \hBPS[2] \rangle\, ,\quad \label{eq:3qqh4}
\langle \qBPS[pq] \qBPSb[p q] \hBPS[2] \rangle\, .
\end{gather}
\end{subequations}
The three-point functions \eqref{eq:3qqh123} turn out to be protected since they are extremal, namely that the dimension of one operator is equal to the sum of the remaining ones. Then, the first one in \eqref{eq:3qqh4} can be proved to  vanish based on a SU$(N)$ and SU(4) reasoning.  Finally the last three-point function can be shown to be protected through more complicated arguments. 
\item Generic $\langle \qBPS[pq]\qBPSb[rs] \hBPS[k]\rangle$  are protected at order $g^2$ if one of the following conditions is satisfied
\begin{subequations}
\begin{align}
\begin{rcases}
2s+r=2q+p+k\\
2q+p= 2s+r+k\\
2q+p+2s+r=k\\
\end{rcases}& \qquad \text{ extremality condition\,,}\\
\begin{rcases}
2s+r \leq k+p\\
2q+p \leq k+r \\
\end{rcases}&\qquad  \text{ ``three flavours'' condition}\,,\\
k \leq p+r &\qquad  \text{ ``two flavours'' condition}\,.
\end{align}
\end{subequations}
Specializing to operators with scaling dimension $\Delta \leq 7$, the only three-point function not included in the previous cases is $\langle \qBPS[13] \qBPS[13] \hBPS[4]\rangle$, with $\qBPS[13]$ as defined in \cite{Ryzhov:2001bp}. This is proved to be protected as well. 
\item $\langle \qBPS[pq] \qBPS[rs] \qBPS[lk] \rangle$ with $2k+l \leq 2q+p \leq 2s+r$ are guaranteed to not receive $g^2$ corrections if they satisfy either the extremality condition
\eqn{
2s+r=2q+p+2k+l
}[]
or all of the ``three flavours'' conditions 
\eqna{
2s+r & \leq 2k+l+p \, , \\
2s+r & \leq 2q+p+l \, , \\
2q+p & \leq 2k+l+s \, .
}[]
Notice that these conditions are for instance satisfied by $\langle \qBPS \qBPS \qBPS \rangle$. Then among quarter-BPS operators with dimension less than seven, the only  three-point functions left are
%\eqn{
\begin{gather}
\langle \qBPS[02] \qBPS[02] \qBPS[22] \rangle \, , \quad
\langle \qBPS[02] \qBPS[12] \qBPS[32] \rangle \, , \quad
\langle \qBPS[02] \qBPS[12] \qBPS[13] \rangle \, , \\
\langle \qBPS[02] \qBPS[13] \qBPS[32] \rangle \, , \quad
\langle \qBPS[02] \qBPS[13] \qBPS[14] \rangle \, , 
\end{gather}
%}[]
which are shown to be protected.  
\end{enumerate}
Similar analyses can be found in \cite{Heslop:2003xu,Heslop:2001gp}. 

A parallel and complementary study of three-point functions can be done resorting to the underlying chiral algebra.\footnote{We thank Xinan Zhou for suggesting this approach to us.} The idea is that if, in a given three-point function,  it appears only one structure that survives to the chiral algebra map, this directly implies that the correlator is protected in the full $\mathcal{N}=4$ theory.  We have performed this analysis for the mixed three-point functions of $\hBPS$ and $\qBPS$ and for $\langle \qBPS \qBPS \qBPS \rangle$.  Proceeding in this way, we confirmed the protected nature of  $\langle \qBPS \hBPS \hBPS \rangle$.  Unfortunately, for the remaining two three-point functions the chiral algebra argument is not that constraining.  We find that in $\langle \qBPS \qBPS \hBPS \rangle$, the chiral map kills the only SU(4) structure not giving us any new insight. The case of all $\qBPS$ operators is  even different.  Here, it turns out that a particular linear combination of the appearing tensor structures  is indeed protected, but this is not enough to conclude that the entire function does not get renormalized as the weak-coupling analysis is suggesting.  It seems that the conditions imposed by the chiral algebra are weaker than the ones outlined above.  One of the reason can hinge on the fact that, differently from the $\hBPS[p]$ case, when decomposed in $\mathcal{N}=2$ supermultiplets, $\qBPS$ does not contain only a Schur operator but other non protected ones.  The presence of this additional operators can possibly justify why the chiral algebra is insufficient or less powerful in studying correlators containing this quarter-BPS operators.

\subsection{R-symmetry structures}

\subsubsection{Polarizations for \texorpdfstring{$\boldsymbol{\CN=4}$}{N=4}}

Due to the high proliferation of indices and the complicated symmetrizations and subtractions that one needs to do in order to construct quarter-BPS operators, it is convenient to introduce an index-free formalism for $\SU(4)$ structures.

The same idea was used in~\cite{Nirschl:2004pa} to greatly simplify the computation of tensor structures, and has also been applied in the context of spinning operators~\cite{Costa:2011mg, Elkhidir:2014woa, Cuomo:2017wme}. In fact, the embedding formalism in four dimensions differs from our setup simply by a signature: $\SU(2,2)$ versus $\SU(4)$.\footnote{Another obvious difference with the spinning structure formalism is that the six-dimensional vectors there are positions, so they can appear in the denominator. Here they are polarizations so they cannot.}

We contract all $\SO(6)$ fundamental indices with a six-dimensional complex vector $y^M$ and all $\SU(4)$ (anti)fundamental indices with a four-vector $S^m$ ($\Sb_m$). One can trade an $\SO(6)$ index with an antisymmetrized pair of $\SU(4)$ indices using the Dirac matrices defined in appendix~\ref{app:notation}, thus we can also define
\eqn{
\rmy_{mn} \equiv y_M \Sigma^M_{mn}\,,\qquad
\bar{\rmy}^{mn} \equiv y_M \Sigmab{}^{M\lsp mn}\,.
}
These polarizations must satisfy some constraints following from the irreducibility of the representation to which they are attached. The list of constraints reads
\eqn{
y\cdot y = 0\,,\qquad S\cdot\Sb=0\,,\qquad \rmy\lsp S=0\,,\qquad \bar\rmy\lsp\Sb=0\,.
}[eq:constraints]
An operator in the $(q,p,q)$ will be a field with homogeneity $p$ in $y$ and $q$ in $S,\Sb$
\eqn{
\qBPS[pq](\lambda S,\bar \lambda\Sb,\mu y) = (\lambda \bar\lambda)^q \mu^p \,\qBPS[pq](S,\Sb,y)\,.
}[eq:opScaling]
In order to recover the tensor form of this operator we must differentiate with respect to the polarizations. However, we need to be careful because the polarizations are constrained and so their derivatives are not free. This problem can be solved by defining differential operators which are interior to the constraints~\eqref{eq:constraints}. With the aid of those operators we can recover
\eqn{
(\CO_{pq})^{m_1\cdots m_q}_{n_1\cdots n_q\,,\,M_1\cdots M_p} = \partial_{n_1}\cdots \partial_{n_q}\,\bar\partial^{m_1}\cdots \bar\partial^{m_q}\, \CD_{M_1}\cdots \CD_{M_p}\CO_{pq}(S,\Sb,y)\,.
}[]
In~\eqref{eq:diffop} we show an explicit definition of $\CD_M$, $\partial_m$ and $\bar\partial^m$, which previously appeared in~\cite{Bargmann:1977gy, Nirschl:2004pa, Costa:2011mg, Cuomo:2017wme}. This definition solves only the first two constraints and therefore can be used only when either $p$ or $q$ is zero, which is enough for our cases.\footnote{In principle there are also unconstrained parametrizations of $y$ and $S, \Sb$~\cite{Simmons-Duffin:2012juh}. Since this is not strictly necessary for our goals we have chosen not to pursue this direction.}

The various polarizations can be contracted into index-free expressions which then combine to give the $n$-point tensor structures. A complete basis of such building blocks is\footnote{For six or more points, one also has to consider the Levi-Civita tensor $\epsilon_{MNPQRS}\, y_i^M y_j^N y_k^P y_l^Q y_m^R y_n^S$.}
\eqna{
\stry{ij} &= y_i\cdot y_j\,,\qquad&
\strSSb{ij} &=S_i\cdot\Sb_j\,,\\
\strTrY{i_1i_2i_3\cdots} &= \tr\lsp(\rmy_{i_1}\bar\rmy_{i_2}\rmy_{i_3}\cdots)\,,\qquad&
\strSYSb{i}{j_1\cdots j_{2p}}{k} &= S_i\lsp \rmy_{j_1}\lnsp\cdots\lsp\bar{\rmy}_{j_{2p}}\Sb_k\,,\\
\strSYS{i}{j_1\cdots j_{2p+1}}{k} &= S_i\lsp \rmy_{j_1}\lnsp\cdots\lsp\rmy_{j_{2p+1}}S_k\,,\qquad&
\strSbYSb{i}{j_1\cdots j_{2p+1}}{k} &= \Sb_i\lsp \bar\rmy_{j_1}\lnsp\cdots\lsp\bar{\rmy}_{j_{2p+1}}\Sb_k\,,\\
\CE_{ijkl} &= \epsilon_{mnpq}\lsp S_i^m S_j^n S_k^p S_l^q \,,\qquad&
\overbar{\CE}_{ijkl} &= \epsilon^{mnpq}\lsp \Sb_{im} \Sb_{jn} \Sb_{kp} \Sb_{lq}\,, \\
}[eq:buildingBlocks]
With this notation, the most general two-point function of an operator $\CO_\Delta^{(q,p,\qb)}$ of dimension $\Delta$ transforming in the $(q,p,\qb)$ of $\SU(4)$ and its conjugate can be written as follows:
\eqn{
\langle \CO_\Delta^{(q,p,\qb)}(x_1) \lsp \CO_\Delta^{(\qb,p,q)}(x_2)\rangle = \frac{(\stry{12})^p(\strSSb{12})^q(\strSSb{21})^{\qb}}{(x_{12}^2)^\Delta}\,.
}[eq:twopf]

\subsubsection{Polarizations for \texorpdfstring{$\boldsymbol{\CN=2}$}{N=2}}
We also need a formalism for dealing with R-symmetry and flavor indices of $\CN=2$ tensor structures. This is again analogous to four dimensions, but in position space. Indeed, we have $\mathfrak{su}(2)\oplus\mathfrak{su}(2)$ which differs from $\mathfrak{so}(3,1)$ by the signature.

We contract all R-symmetry indices with a complex two-vector $\eta^a$ and all flavor indices with another two-vector $\etat{}^{a'}$. Indices are raised and lowered with the Levi-Civita tensors $\epsilon_{ab}$ and $\epsilon_{a'b'}$. The only possible building blocks are
\eqn{
\eta_{ij} = \eta_i^a \eta_{j\lsp a}\,,\qquad
\etat_{ij} = \etat_{i\lsp a'} \etat_j^{a'}\,.
}[]
Clearly, due to the antisymmetry of $\epsilon$, the above expressions are nonzero for $i\neq j$. For this reason, we do not need to put further constraints on these vectors and there are no issues in taking derivatives. The two-point functions are also easy to write down. Given an $\CN=2$ primary $O_{R,F}$ with R-charge $R$, flavor charge $F$ and dimension $\Delta$ one has
\eqn{
\langle O_{R,F}(x_1) \lsp \overbar{O}_{R,F}(x_2)\rangle = \frac{(\eta_{12})^R\,(\etat_{12})^F}{(x_{12}^2)^\Delta}\,.
}[]

\subsubsection{Tensor structures as Casimir eigenvectors}\label{sec:casimir}

Let us consider a four-point function of four quarter-BPS operators
\eqn{
G(p_1,q_1;\lsp\ldots;\lsp p_4,q_4) = \langle \qBPS[p_1q_1](x_1,\bfS_1)\lsp \qBPS[p_2q_2](x_2,\bfS_2) \lsp \qBPS[p_3q_3](x_3,\bfS_3) \lsp \qBPS[p_4q_4](x_4,\bfS_4)\rangle\,,
}[]
where $\bfS_i$ collectively denotes $S_i,\Sb_i$ and $y_i$. This correlator can be expanded into a certain number $N_\mathrm{str}$ of functions. Each of these functions is multiplied by a combination of the polarizations defined previously, which we call $\BBT_k$, where each $k$ is associated to a given representation exchanged. The allowed representations are contained in the following tensor product
\eqn{
\CR \in (q_1,p_1,q_1) \otimes (q_2,p_2,q_2) \; \cap \; (q_3,p_3,q_3) \otimes (q_4,p_4,q_4)\,.
}[]
We can therefore write the correlator as
\eqn{
G(p_1,q_1;\lsp\ldots;\lsp p_4,q_4) = \kinPref{2q_1+p_1,\,\ldots,\,2q_4+p_4}\lsp \sum_{k=1}^{N_\mathrm{str}} \BBT_k\, g_k(z,\zb)\,.
}[]
Above we defined the four-point prefactor $\kinPref{}$ and the cross-ratios $z,\zb$, as \hypertarget{eqref:kinPref}{follows}
\eqna{
&\kinPref{\Delta_1\Delta_2\Delta_3\Delta_4} = \frac{\left(\frac{x_{24}^2}{x_{14}^2}\right)^{\frac12\Delta_{12}}\left(\frac{x_{14}^2}{x_{13}^2}\right)^{\frac12\Delta_{34}}}{(x_{12}^2)^{\frac12(\Delta_1+\Delta_2)}(x_{34}^2)^{\frac12(\Delta_3+\Delta_4)}}\,,\qquad \Delta_i = 2q_i+p_i\,,\quad\Delta_{ij} = \Delta_i-\Delta_j\,,\\
&\frac{x_{12}^2\lsp x_{34}^2}{x_{13}^2\lsp x_{24}^2} = u = z\zb\,,\qquad
\frac{x_{14}^2\lsp x_{23}^2}{x_{13}^2\lsp x_{24}^2} = v = (1-z)(1-\zb)\,.
}[eq:KzzbDef]

One way to find the tensor structures $\BBT_k$ is to first write down a basis using all possible products of the monomials in~\eqref{eq:buildingBlocks}, while making sure that they satisfy the scaling~\eqref{eq:opScaling} for all four operators, and then to rotate this basis to one that diagonalizes the $\SU(4)$ Casimir operators. The eigenvalues associated to each tensor structure will tell us the representation to which they correspond. In our cases of interest, the quadratic Casimir is not enough because different representations can have the same eigenvalue. It is therefore necessary to consider one higher order Casimir as well and for simplicity we use the quartic one instead of the cubic.\footnote{The reason is that the cubic Casimir contains an $\epsilon$ tensor which makes the index contractions more involved.}

Any operator in the $\SU(4)$ universal enveloping algebra can be represented in terms of differential operators acting on the polarizations $S_i,\Sb_i$ and $y_i$. Concretely, the three Casimirs of $\SU(4)$ can be represented as~\cite{Karateev:2017yoq}
\eqna{
\CC_2(\partial_{\bfS}) &= \frac12\lsp L_{MN}\lsp L^{NM}\,,\\
\CC_3(\partial_{\bfS}) &= \frac1{24i}\lsp \epsilon^{MNPQRS} \lsp L_{MN}\lsp L_{PQ}\lsp L_{RS}\,,\\
\CC_4(\partial_{\bfS}) &= \frac12\lsp L_{MN}\lsp L^{NP}\lsp L_{PQ}\lsp L^{QM}\,,
}[]
where $L_{MN}$ are the generators of $\SU(4)$. Since we want to act on the first two points, their expression is
\eqna{
L_{MN} &= L_{1,\,MN} + L_{2,\,MN}\,,\\
L_{i,\,MN} &= -\left(y_{iM}\frac{\partial}{\partial y^N_i} - y_{iN}\frac{\partial}{\partial y^M_i}\right) - S^m_i\lsp \Sigma_{MN}{}_m^{\phantom{m}n}\frac{\partial}{\partial S^n_i}- \Sb_{im}\lsp \overbarUp\Sigma_{MN}{}^m_{\phantom{m}n}\frac{\partial}{\partial \Sb_{in}}\,.
}[]
If $\BBT_k$ exchanges the representation $(q,p,\qb)$ then it must satisfy the following eigenvalue equations
\eqn{
\CC_r(\partial_{\bfS_1},\partial_{\bfS_2}) \,\BBT_k(\bfS_1,\ldots,\bfS_4) = C_r\,\BBT_k(\bfS_1,\ldots,\bfS_4)\,,\qquad r = 2,4\,,
}[]
with the eigenvalues given by~\cite{Perelomov2:1966}
\eqna{
C_2 &= p \left(\qb+q+4\right)+\frac{1}{4} \left(3 \qb^2+2 (q+6) \qb+3 q (q+4)\right)+p^2\,,\\
C_4 &= \frac{1}{16} \big( \left(\qb+2 p+q\right)^2 \left(\qb+2 p+q+8\right)^2+24 \left(\qb+2 p+q\right) \left(\qb+2 p+q+8\right)\\&\;\quad\quad\;\;\,-2\left(q-\qb\right)^2 \left(\qb+q+2\right)^2+4 \left(\qb \left(\qb+2\right)+q (q+2)\right)^2\big)\,.\\
}[]
For this work, we implemented the Casimir differential operators in a \textsl{Mathematica} notebook which can be made available upon request.

\subsection{Relation with older superspace formulations}

Quarter-BPS operators in $\CN=4$ Super Yang-Mills have been studied in detail since the early 2000s~\cite{Ryzhov:2001bp, DHoker:2003csh}. In this subsection we will show how to connect our formalism to the approach adopted in the past which made use of superspace.

There are various families of superspaces for $\nobreak{\CN=4}$. Any given superspace formulation is designed to make the shortening conditions appear ``simple.'' This means that imposing that a certain operator is short simply amounts to declaring that it depends on only a subset of the superspace coordinates. The most familiar example is a chiral multiplet in a four-dimensional $\CN=1$ theory. Without a formulation of chiral superspace one would have to introduce a field $\CO(x,\theta,\thetab)$ and require $\Db_\alpha\CO(x,\theta,\thetab)=0$. But if we define $y=x+i\theta\sigma\thetab$ then it suffices to take an arbitrary (unconstrained) field $\CO(y,\theta)$ and it is going to be automatically chiral. Generalizing this idea to extended supersymmetry is the main challenge that has been undertaken in the early nineties and that we wish to briefly review now.

The family of superspaces that we wish to review was introduced in~\cite{Hartwell:1994rp} and it takes the name of $(\CN,\mathrm{p},\mathrm{q})$ superspace. The construction roughly goes as follows: we start with complexified super Minkowski space $\BBC^{4|4\CN}$ on which we require that the $x^\mu$ coordinates are real and $\thetab_{m\alphad}=(\theta^m_\alpha)^*$. Let us call this space $M_\CN$. Next we enlarge this space by a ``flag manifold'' $\BBF_{\mathrm{p},\mathrm{q}}$ defined as the coset space
\eqn{
\BBF_{\mathrm{p},\mathrm{q}} \equiv \frac{\SU(\CN)}{\mathrm{S}(\rmU(\mathrm{p})\times \rmU(\CN-\mathrm{p}-\mathrm{q}) \times \rmU(\mathrm{q}))}\,.
}[cosetdef]
This coset is parametrized by a special unitary matrix $u_R^{\phantom{R}m}$ such that the index $m$ is acted upon by the full $\SU(\CN)$ group and the index $R$ by the isotropy group --- i.e. the group at the denominator of \cosetdef. All in all, this means that the coordinates of our superspace are $(x^\mu,\theta^m_\alpha,\thetab_{m\alphad},u_R^{\phantom{R}m})$.

Superfields carry a representation of the so-called Levi subalgebra\footnote{The origin of this comes from interpreting the full superspace as a coset. This in turn follows from interpreting $M_\CN$ as a coset of $\mathrm{SU}(2,2|\CN)$.}
\eqn{
\mathfrak{l} = \mathfrak{sl}(2|\mathrm{p}) \oplus \mathfrak{sl}(2|\mathrm{q}) \oplus \mathfrak{sl}(\CN-\mathrm{p}-\mathrm{q})  \oplus \BBC^2\,.
}[levisubalgebra]
The bosonic parts of the first two summands and the last are the spin labels $j,\jb$. The two complex dimensions are $\Delta$ and the $r$-charge --- which for $\CN=4$ disappears and $\BBC^2$ is replaced by $\BBC$ --- and the remaining parts are R-symmetry quantum numbers. Let us call this representation $\CR$ with carrier space $V_\CR$. Then superfields are defined as sections of a $V_\CR$ bundle over $\BBM_\CN \equiv M_\CN\times\BBF_{\mathrm{p},\mathrm{q}}$. Equivalently, they are functions from $\BBM_{\CN}$ to $V_\CR$ that satisfy an equivariance property
\eqna{
f:\BBM_{\CN} &\to V_\CR\,,\\
f(g x) &= \CR(g)f(x)\qquad \forall\; g\in \exp(\mathfrak{l})\,.
}[]
In order to impose shortening conditions it is necessary to require that certain covariant derivatives annihilate the superfield. The whole point of this construction is that, for the fields of interest, requiring $Df =0$ amounts to an holomorphicity constraint. So we do not need to deal with constrained superfields but we simply have superfields that depend on a subset of variables. There are two notions of holomorphicity: G\nobreakdash-analiticity and H\nobreakdash-analiticity. A superfield which is both G-analytic and H-analytic is called CR\nobreakdash-analytic or simply analytic. A field is G\nobreakdash-analytic if it satisfies\footnote{See e.g.~\cite{Hartwell:1994rp} for the explicit definitions of the derivatives.}
\eqn{
u_r^{\phantom{r}m}D_{\alpha m} \,f = \Db_\alphad^{m} (u^\dagger)_m^{\phantom{m}r'}\,f =0\,,
}[]
where $D_{\alpha i}$ and $\Db_\alphad^i$ are the covariant derivatives on $M_\CN$, the index $r$ is on $\SU(\mathrm{p})$ and the index $r'$ is on $\SU(\mathrm{q})$. Next one defines left-invariant vector fields of $\SU(\CN)$, $D_{R}^{\phantom{R}S}$, and splits the indices $R = (r,r'',r')$ with $r,r'$ as before and $r''$ being on $\SU(\CN-\mathrm{p}-\mathrm{q})$. With this definition a field is H-analytic if it satisfies
\eqn{
D_r^{\phantom{r}s'}\,f = D_r^{\phantom{r}s''}\,f = D_{r''}^{\phantom{r''}s'}\,f=0\,.
}[]
The operator of $B_1\overbar{B}_1$ type, namely half-BPS $(0,p,0)$ or quarter-BPS $(q,p,q)$, are realized as CR\nobreakdash-analytic superfields.

Let us fix $\CN=4$ and take $\mathrm{p}$ and $\mathrm{q}$ as in~\cite{DHoker:2003csh}, namely $\mathrm{p}=\mathrm{q}=1$. By looking at~\levisubalgebra we see that, other than the usual conformal quantum numbers, we have an extra $\mathfrak{su}(2)$ index (recall that for $\CN=4$ the last factor is just $\BBC$). So fields are functions of the superspace variables of the form
\eqn{
V_{r_1\cdots\lsp r_n}(x,\theta^m_\alpha,\thetab_{m\alphad},u_R^{\phantom{R}m})\,.
}[]
The vector multiplet, in particular, has a single index and is CR\nobreakdash-analytic
\eqn{
W_r(x,\theta^{2,3,4}_\alpha,\thetab_{\alpha\,1,2,3},u)\,,\qquad \mbox{with}\qquad (D_1^{\phantom1s},\,D_1^{\phantom14},\,D_r^{\phantom{r}4})W_r = 0\,.
}[]
Observe that this superspace is not optimal for considering the vector multiplet as we still have 12 supercharges instead of the expected 8. Indeed we still need to impose the H\nobreakdash-analyticity on $W$. Actually the H\nobreakdash-analyticity follows automatically here as it will be easy to check from the next paragraph. One could argue that there are more convenient ways to study this operator, such as $(\CN,2,2)$ superspace~\cite{Hartwell:1994rp} or  the $\SU(\CN)/\rmU(1)^3$ coset superspace~\cite{Andrianopoli:1999vr}. We will however stick with $\mathrm{p}=\mathrm{q}=1$ in order to keep pursuing the comparison with~\cite{DHoker:2003csh}.

The explicit form of the bottom component of the analytic superfield can be given in terms of the more familiar $\varphi^M$ scalars
\eqn{
W_r(x,u) \equiv u_1^{\phantom1m}u_{r+1}^{\phantom{r+1}n} \,\Sigma_{M\lsp mn}\,\varphi^M\,,\qquad r=1,2\,.
}[Wrdef]
As we can see, it is not so different from the principle of contracting indices with polarizations. In this case only part of the $\SU(4)$ indices are ``transferred'' to the $u$ dependence. Consistently with the rest of our formalism, let us introduce a new $\SU(2)$ polarization $\chi^r$ and contract the $r$ index of $W_r$ with it
\eqn{
W(x,u,\chi) \equiv W_r(x,u)\,\chi^r\,.
}[]
In order to translate from these variables to the polarizations we are used to, we can consider the two simplest cases: the $(0,2,0)$ and the $(2,0,2)$. For the former we have a totally symmetric trace
\eqna{
\Tr\lsp(W\,W) &= \chi^r \lsp u_1^{\phantom1m}u_{r+1}^{\phantom{r+1}n} \,\Sigma_{M\lsp mn}\; \chi^s\lsp u_1^{\phantom1p}u_{s+1}^{\phantom{s+1}q} \,\Sigma_{N\lsp pq}\;\Tr\lsp(\varphi^M\varphi^N) \\&= y_M\lsp y_N\,\Tr\lsp(\varphi^M\varphi^N)\,.
}[]
It follows simply
\eqn{
y^M = \chi^r \lsp u_1^{\phantom1m}u_{r+1}^{\phantom{r+1}n} \,\Sigma^M_{mn}\,.
}[ymap]
The case $(2,0,2)$ instead requires a small computation
\eqna{
\Tr\lsp(W_r\,W_s)\Tr\lsp(\overbar{W}^r\,\overbar{W}^s) &= u_1^{\phantom1m}u_r^{\phantom{r}n} \, \Sigma_{M\,mn}\;  u^4_{\phantom1p}u^r_{\phantom{r}q} \, \overbarUp{\Sigma}_P^{pq} \\&\quad\times
u_1^{\phantom1m'}u_r^{\phantom{r}n'} \, \Sigma_{N\,m'n'}\;  u^4_{\phantom1p'}u^r_{\phantom{r}q'} \, \overbarUp{\Sigma}_Q^{p'q'} \\&\quad\times
\Tr\lsp(\varphi^M\varphi^N) \Tr\lsp(\varphi^P\varphi^Q)
\\&= \eqref{Osingtrace}\,.
}[]
The matrices $u_i^{\phantom{i}j}$ and $u^i_{\phantom{i}j}$ are one the inverse of the other, so they yield a $\delta$ of the external indices which in turn contract a pair $\Sigma,\overbarUp{\Sigma}$ together. Then we use the definition~$\eqref{SigmaMNdef}$ to exactly match the right hand side, provided we identify
\eqn{
S^m = 4\lsp u_1^{\phantom1m}\,,\qquad
\Sb_m = 4\lsp u^4_{\phantom1m}\,.
}[SSbmap]
We should check that \ymap and \SSbmap are compatible with the constraints~\eqref{constraints}. This can be done very simply
\fourseqn{
y\cdot y &\propto \chi^r\chi^{s} \lsp u_1^{\phantom1m}u_{r+1}^{\phantom{r+1}n}u_1^{\phantom1p}u_{s+1}^{\phantom{s+1}q}\,\epsilon_{mnpq} \propto \epsilon_{mnpq} u_1^{\phantom1m}u_1^{\phantom1p} = 0\,,
}[]{
S\cdot \Sb &\propto u_1^{\phantom1m}u^4_{\phantom1m} = \delta^1_4 = 0\,.
}[]{
S\rmy &\propto u_1^{\phantom1m} \,\chi^r \lsp u_1^{\phantom1n}u_{r+1}^{\phantom{r+1}p} \epsilon_{mnpq} \propto \epsilon_{mnpq} u_1^{\phantom1m}u_1^{\phantom1n} = 0\,,
}[]{
\Sb\bar\rmy &\propto u^4_{\phantom1m} \,\chi^{r} \lsp u_1^{\phantom1m}u_{r+1}^{\phantom{r+1}n} \propto \delta^4_1 = 0\,.
}[][]
So, to summarize, for our purposes going to $(\CN,1,1)$ superspace is equivalent to choosing a specific parametrization of the polarizations $S,\Sb,y$ given by~\ymap and~\SSbmap.

\section{Chiral algebra and Ward identities}\label{sec:chialg}

\subsection{Chiral algebra review}

Every four-dimensional $\CN=2$ superconformal field theory admits a protected subsector which is described by a vertex operator algebra or chiral algebra. This was discovered in~\cite{Beem:2013sza} and from that seminal paper there were many developments aimed at finding similar constructions in other dimensions~\cite{Beem:2014kka, Chester:2014mea, Beem:2016cbd} and also at better understanding the structure of the chiral algebra and what it can teach us about the original superconformal field theory~\cite{Beem:2014rza, Beem:2017ooy}. While this story has now taken on a somewhat formal route, its source of inspiration came from a very concrete observation stemming from the results of~\cite{Nirschl:2004pa}. The seminal work~\cite{Nirschl:2004pa} provided a solution to the superconformal Ward identities for four-point functions of $\CN=2,4$ superconformal field theories. This solution is expressed in terms of a meromorphic function of a single cross-ratio $\gothf(z)$, which can be obtained by evaluating the four-point function with the R\nobreakdash-symmetry polarizations in a specific $\zb$\nobreakdash-dependent configuration. The function $\gothf(z)$ was later interpreted in~\cite{Beem:2013sza} as the four-point function in the two-dimensional chiral algebra.

We plan to apply the same idea to our correlators of interest. However, let us first quickly introduce the chiral algebra construction so that all subsequent steps in deriving the Ward identities will be clear.

The first step is to choose a two-dimensional plane $\BBR^2\cong\BBC$ in $\BBR^4$ along with a nilpotent supercharge $\funnyQ$. The supercharge is chosen in such a way that holomorphic transformations in the two-dimensional plane $\BBC$ are $\funnyQ$-closed and the anti-holomorphic transformations are $\funnyQ$-exact. A concrete choice that satisfies these requirements is
\eqn{
\funnyQ = Q^1_- + \overbar{S}^{2\dot-}\,,
}[]
with the plane $\BBC$ being $x^1=x^2=0$, $z = x^3 + i x^4$ and $\zb=x^3-ix^4$. Then, once an operator $\CO$ is restricted to such plane, its $\zb$ dependence can be dropped by passing to the $\funnyQ$-cohomology. The operators surviving this cohomology are precisely the Schur operators mentioned around~\eqref{eq:schurcondition}, located at the origin.

The key aspect of this construction is that, in order to keep the operator holomorphic when leaving the origin of the $\BBC$ plane, one has to move the position in lockstep with the R\nobreakdash-symmetry indices. This operation is known as twisted translation. In our notation it simply means that the polarization $\eta$ must become $\zb$-dependent in the following way
\eqn{
\chi\big[\CO_{\mathrm{Schur}}\big](z) \equiv \eta_{a_1}(\zb)\cdots \eta_{a_R}(\zb)\lsp \CO_{\mathrm{Schur}}^{a_1\cdots a_R}(z,\zb)\,,\qquad \eta(\zb) \equiv (1,\zb)\,.
}[]
The equality is understood to hold inside correlation functions, in the sense that the $\zb$ dependence on the right hand side will drop out. The resulting correlator is necessarily coupling independent because the chiral algebra is rigid under marginal deformations, thus it can be computed from the free theory.\footnote{In our examples we compute the protected function $\mathfrak{f}$ by means of Wick contractions, made possible by the Lagrangian formulation of our theory.  An  alternative approach can be found in~\cite{Rastelli:2017ymc, Behan:2021pzk}. }
In particular, we can apply the map $\chi$ on four-point functions to obtain a meromorphic and protected function of a single cross ratio
\eqna{
&\chi\big[\langle \CO_{\mathrm{Schur}}(x_1,\eta_1)\CO_{\mathrm{Schur}}(x_2,\eta_2)\CO_{\mathrm{Schur}}(x_3,\eta_3)\CO_{\mathrm{Schur}}(x_4,\eta_4) \rangle\big]\\
%&\langle \CO_{\mathrm{Schur}}(x_1,\eta_1)\CO_{\mathrm{Schur}}(x_2,\eta_2)\CO_{\mathrm{Schur}}(x_3,\eta_3)\CO_{\mathrm{Schur}}(x_4,\eta_4) \rangle\Big|_{x_i^\mu\to\left(\substack{0\\0\\z_i+\zb_i\\i(\zb_i-z_i)}\right),\,\eta_i\to\left(\substack{1\\\zb_i}\right)}=\\
&\equiv\langle \CO_{\mathrm{Schur}}(x_1,\eta_1)\CO_{\mathrm{Schur}}(x_2,\eta_2)\CO_{\mathrm{Schur}}(x_3,\eta_3)\CO_{\mathrm{Schur}}(x_4,\eta_4) \rangle\Big|_{\substack{x_k^\mu\to\bigish(0,\lsp 0,\lsp \frac{z_k+\zb_k}2,\lsp \frac{z_k-\zb_k}{2i}\bigish)\\\,\eta_k\lsp\to(1,\lsp \zb_k)\hfill}}\\
&=\kinPrefz{R_1R_2R_3R_4}\,\gothf(z)\,.
}[]
The prefactor $\kinPrefz{}$ and the cross ratio $z$ are defined \hypertarget{eqref:kinPrefz}{as}
\eqn{
\kinPrefz{h_1h_2h_3h_4} = \frac{\Bigish(\frac{z_{14}}{z_{13}}\Bigish)^{\frac12(h_3-h_4)}\Bigish(\frac{z_{24}}{z_{14}}\Bigish)^{\frac12(h_1-h_2)}}{z_{12}^{\frac12(h_1+h_2)}\,z_{34}^{\frac12(h_3+h_4)}} \,,\qquad
z  = \frac{z_{12}\lsp z_{34}}{z_{13}\lsp z_{24}} = 1- \frac{z_{14}\lsp z_{23}}{z_{13}\lsp z_{24}}\,,
}[eq:Kzdef]
where $z_{ij}=z_i-z_j$.

This construction holds for all $\CN=2$ four dimensional superconformal field theory. Therefore, in particular, it holds for $\CN=4$ super Yang-Mills. The chiral algebra of $\CN=4$ super Yang-Mills contains a Virasoro subalgebra with central charge $c_{2d} = -12 \lsp c$ where $c$ is the four\nobreakdash-dimensional central charge
\eqn{
c = \frac{N^2-1}{4}\,.
}[]
The Schur operators we are interested in are the superconformal primaries of the multiplets $B\overbar{B}[0\lsp;\lnsp0]^{(R;\llsp0)}_{R}$, as we previously mentioned. Furthermore, these primaries can transform in a representation $F$ of the flavor group $\SU(2)$. We denote them as $O_{R,F}$. Under the chiral map $\chi$, the operators $O_{R,F}$ are mapped to single traces of free symplectic bosons, each being in a flavor doublet
\eqn{
\chi\big[O_{R,F}\big] = \tr\lsp(q^{a'_1}\cdots q^{a'_R})\, \Pi_{a_1\cdots a_R}^{b_1\cdots b_F} \, \etat_{b'_1}\cdots\etat_{b'_F}\,,\qquad q_{a'}^I(z)\,\lsp q_{b'}^J(0) \sim \frac{\epsilon_{a'b'}\,\delta^{IJ}}{z}\,,
}[twodWick]
where $\Pi_{a_1\cdots a_R}^{b_1\cdots b_F}$ is a tensor responsible for contracting $R-F$ indices so that the end result has $R$ fields and transforms in the charge\nobreakdash-$F$ flavor representation.\footnote{Note that when $O_{R,F}$ is taken as an $\CN=2$ sub-multiplet of an $\CN=4$ multiplet $\hBPS[p]$, one always has $R=F$. Thus, the tensor $\Pi_{a_1\cdots a_R}^{b_1\cdots b_F}$ is just a product of Kronecker $\delta$'s. In the $\qBPS[0q]$ instead $F=0$ so the tensor is a combination of Levi-Civita tensors that contracts all the indices.} Thanks to this map, computing $\gothf(z)$ simply requires taking all Wick contractions among the $q$'s by using the singular OPE shown above. In order to obtain the Ward identities one has to consider different variants of the chiral algebra construction considered so far. One is obtained by exchanging the roles of $z$ and $\zb$ and one is obtained by exchanging the roles of the flavor and the R-symmetry $\SU(2)$. Let us denote these four variants as
\eqn{
\chi_{z,\etat} \equiv \chi\,,\qquad
\chi_{\zb,\etat}\,,\qquad
\chi_{z,\eta} \,,\qquad
\chi_{\zb,\eta}\,.
}[eq:variants]
The Ward identities then read
\eqna{
\chi_{z,\etat}\big[\langle \CO_1\cdots\CO_4\rangle\big] &= \kinPrefz{}\,\widetilde{\CP}_4\,\gothf(z,\etat)\,,\qquad&
\chi_{\zb,\etat}\big[\langle \CO_1\cdots\CO_4\rangle\big] &= \kinPrefzb{}\,\widetilde{\CP}_4\,\gothf(\zb,\etat)\,,\\
\chi_{z,\eta}\big[\langle \CO_1\cdots\CO_4\rangle\big] &= \kinPrefz{}\,\CP_4\,\gothf(z,\eta)\,,\qquad&
\chi_{\zb,\eta}\big[\langle \CO_1\cdots\CO_4\rangle\big] &= \kinPrefzb{}\,\CP_4\,\gothf(\zb,\eta)\,,
}[]
where the function $\gothf(z,\eta)$ is the same for all equations, $\kinPrefzb{}$ is the same as $\kinPrefz{}$ in~\eqref{eq:Kzdef} with $z_i\to\zb_i$ and, finally, $\CP_4$ and $\widetilde{\CP}_4$ are kinematic prefactors defined as follows, for non\nobreakdash-increasing $F_i$~\cite{Aprile:2017xsp}
\eqna{
\CP_4 &= \left\lbrace
\begin{aligned}
&\eta_{12}^{\frac12(F_1+F_2-F_3+F_4)} \eta_{13}^{\frac12(F_1-F_2+F_3-F_4)} \eta_{23}^{\frac12(-F_1+F_2+F_3-F_4)} \eta_{34}^{F_4} &\mbox{if}\quad F_2+F_3\geq F_1+F_4\\
&\eta_{12}^{F_2}\lsp \eta_{13}^{\frac12(F_1-F_2+F_3-F_4)} \eta_{14}^{\frac12(F_1-F_2-F_3+F_4)} \eta_{34}^{\frac12(-F_1+F_2+F_3+F_4)}&\mbox{if}\quad F_2+F_3\leq F_1+F_4
\end{aligned}\right.\,,\\
\widetilde{\CP}_4 &= \CP_4\big|_{\eta_i\to\etat_i}\,.
}[]
Often two out of four of these identities are redundant.

\subsection{Action on tensor structures}

Now that we understand the construction of the chiral algebra in $\CN=2$ we can work out the concrete action of the map $\chi$ directly in the $\CN=4$ tensor structures. The breaking~\eqref{eq:breaking} can be seen explicitly at the level of the polarizations $y^M$, $S^m$, $\Sb_m$. Let us start from the six\nobreakdash-dimensional vector: we can study the breaking in its matrix form $\rmy_{mn}$. The decomposition amounts to subdivide this matrix in $2\times2$ blocks. The diagonal blocks correspond to $\rmU(1)_r$ polarizations, which we do not need, while the off-diagonal ones correspond to the two $\SU(2)$s. More precisely, we can write
\eqn{
\setlength{\arrayrulewidth}{.25pt}
\setstretch{1.2}
\rmy_{mn} = \left(
\begin{array}{c:c}
0 & \eta_a \otimes \etat^{b'} \\
\hdashline
-\etat^{a'} \otimes \eta_{b}  & 0 \\
\end{array}
\right)\,.
\setstretch{1}
}[]
Another way to state the same equation is to take the first four components of $y^M$ and impose
\eqn{
y_A\, \sigma^A_{aa'} = \eta_a \etat_{a'}\,.
}[]
One can easily check that this identification respects the constraint $y\cdot y = 0$.\footnote{Use $\sigma_{aa'}\cdot \sigma_{bb'} \propto \epsilon_{ab}\epsilon_{a'b'}$ together with $\eta_a\eta^a = \eta_a\eta_b \epsilon^{ab} = 0$.\label{foot:etasq}} Similarly, we can write mappings between the $\SU(4)$ fundamental polarizations and the $\eta,\etat$. This is done by splitting the components in two halves
\eqna{
S^m = \begin{cases}
\eta^a & m = 1,2 \\
\etat_{a'} & m = 3,4
\end{cases}\,,\qquad
\Sb_m = \begin{cases}
\eta_a & m = 1,2 \\
\etat^{a'} & m = 3,4
\end{cases}\,.
}
As before, this is compatible with $S\cdot \Sb = 0$. Based on the decomposition~\eqref{eq:submultiplet} we must single out the charge-$p$ flavor component. This is equivalent to selecting terms proportional to $\lambda^p_i$ under the rescaling
\eqn{
\etat_i \to \lambda_i \lsp \etat_i\,.
}[]
In particular, if the operator under study transforms in the $(q,0,q)$, then one has to simply set $\etat$ to zero for that operator.
To give a few examples, the simplest building blocks are decomposed as follows
\eqn{
\stry{ij} = \frac12\lsp \eta_{ij}\,\etat_{ij}\,,\qquad
\strSSb{ij} = \eta_{ij} + \etat_{ij}\,.
}[]
Assuming that there are only $\hBPS[p]$ and $\qBPS[0q]$ operators, meaning that we can throw away the $\etat$ inside $S^m$ and $\Sb_m$, the other tensor structures can be decomposed as
\twoseqn{
\strSYS{i}{j_1\cdots j_{2n+1}}k &= O(\etat_i,\etat_k) = \strSbYSb{i}{j_1\cdots j_{2n+1}}k\,,
}[]{
\strSYSb{i}{j_1\cdots j_{2n}}k &= (-1)^n\, \eta_i\eta_{j_1}\;\etat_{j_1}\etat_{j_2}\;\eta_{j_2}\eta_{j_3} \cdots\lsp \etat_{j_{2n-1}}\lnsp\etat_{j_{2n}}\;\eta_{j_{2n}}\lnsp\eta_k + O(\etat_i,\etat_k)\,.
}[][eq:morePolarizations]
When all operators have a nonzero $p$ Dynkin label, it is possible to form cross ratios in the $y^M$ vectors
\eqn{
\sigma = \frac{\stry{13}\lsp\stry{24}}{\stry{12}\lsp\stry{34}}=\alpha\alphab\,,\qquad
\tau = \frac{\stry{14}\lsp\stry{23}}{\stry{12}\lsp\stry{34}}=(1-\alpha)(1-\alphab)\,.
}[]
The decomposition of these cross ratios is remarkably simple as it suffices to send $\alpha$ and $\alphab$ to the following ratios\footnote{The equalities of the form $\eta_{12}\lsp\eta_{34} + \eta_{14}\lsp\eta_{23} = \eta_{13}\lsp\eta_{24}$ follow from $\epsilon_{[ab}\lsp\epsilon_{c]d}=0$.}
\eqn{
\alpha = \nu \equiv \frac{\eta_{13}\lsp\eta_{24}}{\eta_{12}\lsp\eta_{34}} = 1 + \frac{\eta_{14}\lsp\eta_{23}}{\eta_{12}\lsp\eta_{34}}\,,\qquad
\alphab = \tilde\nu \equiv \frac{\etat_{13}\lsp\etat_{24}}{\etat_{12}\lsp\etat_{34}} = 1 + \frac{\etat_{14}\lsp\etat_{23}}{\etat_{12}\lsp\etat_{34}}\,.
}[]
When the map $\chi$ is applied to the above cross ratios we get the familiar result of~\cite{Nirschl:2004pa}, namely
\eqn{
\chi[\nu] = \frac{1}{\zb}\,.
}[]
This follows trivially from $\chi[\eta_{ij}] = \zb_{ij}$.

When considering half-BPS operators one has a further Ward identity $\mathfrak{f}(z,\tilde\nu=1/z)=k$ with $k$ a constant. This is equivalent to the topological twist of~\cite{Drukker:2009sf}. With correlators of quarter-BPS operators this property is no longer true --- in our specific cases trivially so because $\mathfrak{f}$ does not even depend on $\tilde\nu$.

\subsection{Ambiguity and Ward identitites}\label{sec:ambiguity}
With the methods discussed so far we are able to obtain the Ward Identities. Note that we have not proved that this represents a complete set,  although for half-BPS it happens to be the case by inspection.
Let us start with any four-point function of quarter-BPS or half-BPS operators $\CG$
\eqn{
\langle \CO_1\CO_2\CO_3\CO_4\rangle = \kinPref{} \; \CG(z,\zb;\bfS_1,\,\ldots,\,\bfS_4)\,,
}[eq:4pt_generic]
with $\kinPref{}$ being the kinematic prefactor defined in~\eqref{eq:KzzbDef}. We can expand $\CG$ in the R\nobreakdash-symmetry tensor structures obtaining $N_\mathrm{str}$ functions of the cross ratios
\eqn{
\CG(z,\zb,\bfS_1,\,\ldots,\,\bfS_4) =  \sum_{k=1}^{N_\mathrm{str}} g_k(z,\zb)\;\BBT_k(\bfS_1,\,\ldots,\,\bfS_4)\,.
}[]
The map $\chi$ acting on the correlator will produce a function of $z$ and the flavor polarizations~$\etat$
\eqn{
\chi\big[\langle \CO_1\CO_2\CO_3\CO_4\rangle\big] = \kinPrefz{} \;\mathfrak{f}(z,\etat)\,,
}[]
with the kinematic prefactor $\mathfrak{K}_4$ defined in~\eqref{eq:Kzdef}. Similarly, the three other variants~\eqref{eq:variants} will produce their own Ward identity. This allows us to separate the functions $g_k(z,\zb)$ into a contribution from $\mathfrak{f}(z,\etat)$ and an unprotected piece
\eqn{
g_k(z,\zb) = w_k(z,\zb) + \sum_{m=1}^{N_\mathrm{u}} \CH_m(z,\zb)\;v_k^{(m)}(z,\zb)\,,
}[eq:decompositionwv]
In the above decomposition, the vector $\vec{w}$ contains the contribution of the chiral algebra result while the vectors ${\vec{v}^{\,(m)} = (v_k^{(m)})}$ span the kernel of the map $\chi$ in the space of tensor structures. This means that only the functions $\CH_m$ will contain unprotected contributions and $N_\mathrm{u}$ denotes both the dimension of the kernel of $\chi$ and the number of such unprotected functions. By definition, the above vectors must satisfy these conditions\footnote{Here it is understood that also the equations for the other variants hold. Here for brevity we indicate only the equations for $\chi_{z,\etat}$.}
\twoseqn{
\chi \Biggl[\lsp\sum _{k=1}^{N_\mathrm{str}} \mathbb{T}_k w_k\Biggr]&= \mathfrak{f}\left(z,\etat\right)\,,
}[]{
\chi \Biggl[\lsp\sum _{k=1}^{N_\mathrm{str}} \mathbb{T}_k v_k^{(m)}\Biggr] &=0\,,\qquad \mbox{for}\; m=1,\,\ldots,\; \dim  (\ker  \chi )\equiv N_\mathrm{u}\,.
}[][eq:definitionWandV]
Clearly this decomposition immediately leads to an arbitrariness given by
\eqn{
w_k(z,\zb) \;\sim \;w_k(z,\zb) + \sum_{m=1}^{N_\mathrm{u}} \CA_m(z,\zb)\;v_k^{(m)}(z,\zb)\,.
}[eq:ambiguityOnw]
Our goal is to fix the functions $\CA_m(z,\zb)$ as much as possible. For convenience let us introduce this terminology: the vector $\vec{w}$ is called an ``uplift'' of the chiral algebra and the functions $\CA_m(z,\zb)$ are called ``ambiguities.'' There are a few criteria that one can use to partially fix the functional form of $\CA_m(z,\zb)$.
\begin{enumerate}
\item When there are degeneracies in the tensor structures, meaning that there are more $\BBT_k$ associated to a given $\CR$, it could be that unitarity and Bose symmetry force some of these structures to be zero.
\item The operator of dimension two transforming in the $(0,2,0)$ must be the superconformal primary $\hBPS$ and therefore it must appear in $\vec{w}(z,\zb)$ with the same coefficient as it appears in the free theory.\footnote{This is because it belongs to the same multiplet as the stress tensor, whose OPE coefficient and conformal dimensions are protected. One cannot make the same reasoning for the stress tensor or the R-current: see remark before equation~\eqref{eq:lambdasqTJ}.}
\item The identity, if present, must contribute with OPE coefficient $1$ as a normalization condition.
\item The disconnected $O(N^0)$ part has to match the free theory computation.
\item The correlator cannot exchange operators of twist two with spin higher than two due to the Maldacena-Zhiboedov theorem later extended to four dimension by Alba and Diab~\cite{Maldacena:2011jn, Alba:2013yda}.
\end{enumerate}
In particular the last condition is very strong because the solutions to the Ward identities $\vec{w}(z,\zb)$ will typically exchange operators of twist two that must disappear. Removing such contributions requires adding towers of operators in other R\nobreakdash-symmetry structures in order to keep the full ambiguity within the kernel of $\chi$.

Let us end with a few additional comments about these criteria: first of all, these requirements solve any possible issue related to multiplet recombination involving the Konishi operator. However for twist greater than two, these are no longer enough to achieve such a recombination.

A second important remark regards whether or not these criteria fully incorporate all the conditions dictated by unitarity of superconformal representations.  As a general rule, the answer is not: since we are expanding in usual conformal blocks, all the exchanged primary operators manifestly satisfy the unitarity constraints just for the conformal group.  Exploiting the full power of superconformal symmetry would require using superconformal blocks, which are known only for correlators of all half-BPS operators.  Superblocks organize the expansion in term of superprimaries, this allows really to tell if a given operator is below unitarity, in the full superconformal sense,  and hence must be canceled. Quite surprisingly, it turns out that for $\langle \hBPS \hBPS \hBPS \hBPS \rangle$ criteria 1--5 are equivalent to the one imposed by the full superconformal symmetry since they completely fix the ambiguity.

\subsection{A familiar example revisited} \label{subsec:HalfBPSAmbiguity}

In order to get acquainted with this slightly more general point of view for imposing Ward identities, let us revisit a familiar example: the four-point function of the half-BPS operator $\hBPS$.
In this case the chiral algebra result $\mathfrak{f}$ is given by
\eqna{
\gothf(z,\tilde\nu)&=\hat{\gothf}(z,\tilde\nu) + \hat{\gothf}\Big(\frac{z}{z-1},1-\tilde\nu\Big)\,,\qquad \hat{\gothf}(z,\tilde\nu) \equiv \frac1{32}+\frac{\tilde{\nu}^2\lsp z^2}{16}+\frac{
\tilde{\nu}(2-\tilde{\nu})\lsp z}{4\lsp(N^2-1)}\,.
}[]
There are six tensor structures given by the first six polynomials in equation (B.14) of~\cite{Nirschl:2004pa}, to give a few:
\eqn{
\THHHH1 = 1\,,\quad \ldots\, ,\quad \THHHH6 = \sigma^2+\tau^2 +4\llsp\sigma\tau - \frac{4}{5}(\sigma+\tau) + \frac{1}{10}\,.
}[tensO2]
If we try to uplift the function $\mathfrak{f}$ to the full correlator we find that there is just one ambiguity degree of freedom, consistent with the fact that we expect only one $\CH(z,\zb)$ function. The ambiguity $\CA$ in this case is obvious: it is just a redefinition of $\CH \to \CH - \CA$.

In the well known derivation of the partial wave decomposition~\cite{Dolan:2004iy}, the function $\CA(z,\zb)$ appears as a consequence of the fact that the chiral algebra result needs to be properly uplifted in order to get the protected contribution to the four-point function. In other words: the correlator always admits a splitting as a constant, a one-variable degree of freedom and a two-variables degree of freedom. The chiral algebra twist fixes the constant and the one-variable function. However, the protected spectrum contributes to the two-variable function as well. The precise way in which this happens can be derived from the knowledge of the superconformal blocks of short multiplets. Since we will not have this knowledge for the cases of quarter-BPS operators, our goal for the rest of this section is to obtain the same result as Dolan and Osborn without using the form of the superconformal blocks. We will indeed see that criteria 1--5 of section~\ref{sec:ambiguity} are sufficient for this.

Following the definitions above, the vectors $\vec w$ and $\vec{v}^{\lsp(1)}$ can be chosen as
\eqna{
\vec{w} &= \big\{ *, \ldots, *, 0\big\}\,,\\
\vec{v}^{\lsp(1)} &= \big\{\tfrac{3 z^2 \zb^2-12 z^2 \zb+10 z^2-12 z \zb^2+64 z \zb-60 z+10 \zb^2-60 \zb+60}{60 z^2 \zb^2},
\\&\quad\hspace{.93em} \tfrac{(z-2) (\zb-2) (z \zb-z-\zb)}{4 z^2 \zb^2},
\tfrac{6 z^2 \zb^2-15 z^2 \zb+10 z^2-15 z \zb^2+10 z \zb+10 \zb^2}{60 z^2 \zb^2},
\\&\quad\hspace{.93em} \tfrac{2 z \zb-3 z-3 \zb+6}{6 z \zb},\tfrac{z \zb-z-\zb}{2 z \zb}, \tfrac16\big\}\,,
}[eq:naiveUplift]
where in place of the $*$'s there are some functions of $\mathfrak{f}(z,\tilde\nu)$, $\mathfrak{f}(z,\nu)$, $\mathfrak{f}(\zb,\tilde\nu)$ and $\mathfrak{f}(\zb,\nu)$ that we will not specify here for brevity. For concreteness, one can check that the four-point function comes out as expected, namely
\eqna{
\sum_{k=1}^6\THHHH{k}\Big(w_k(z,\zb) + \CH(z,\zb)\, v_k^{(1)}\Big) &= \frac{(\alpha z -1)(\bar\alpha \zb -1) \,\mathfrak{f}(z,\bar\alpha) + (\alpha\leftrightarrow \bar\alpha\;\mathrm{or}\; z\leftrightarrow \zb)}{(z-\zb)(\bar\alpha-\alpha)} \\&\quad\;+ \frac{(\alpha z - 1)(\alpha \zb - 1)(\bar\alpha z - 1)(\bar\alpha \zb - 1)}{z^2 \zb^2}\CH(z,\zb)\,.
}[]
The fact that this choice satisfies~\eqref{eq:definitionWandV} is readily verified by noticing that here $\chi$ is simply the replacement $\alpha\to1/\zb,\,\bar\alpha\to\tilde\nu$.\footnote{We obviously need to match also all other chiral algebra limits, i.e. $\bar\alpha \to 1/z,\alpha\to1/z$ and $\bar\alpha\to1/\zb$. Here for simplicity of notation we indicated only one of them.}

Now, the vector $\vec w$ defined as in~\eqref{eq:naiveUplift} contains higher spin twist twos and twist zeros in all nonzero five entries. We can cancel them by adding various higher twist contributions in the ambiguity $\CA(z,\zb)$, which will contribute to the correlator as in~\eqref{eq:ambiguityOnw} with $N_\mathrm{u}=1$. These will appear in the other entries with lower twist thanks to the recursion relations written in appendix~\ref{app:blocksRecRel}. For example, a single block $g_{6,2}$ in $\CA(z,\zb)$ will contribute to the representation $(1,2,1)$ --- structure $\THHHH5$ --- as follows
\eqn{
\CA(z,\zb) = g_{6,2}(z,\zb)\quad\Longrightarrow\quad \langle \hBPS\hBPS\hBPS\hBPS\rangle\big|_{(1,2,1)} = \tfrac32\lsp g_{5,1} + 6 \lsp g_{5,3} + \tfrac18\lsp g_{7,1} + \tfrac8{21}\lsp g_{7,3}\,.
}[eq:recrelExample]
The other entries are more involved but they can be obtained with repeated applications of the relations in appendix~\ref{app:blocksRecRel}. Notice that in the above example a twist-four contribution was able to generate a twist-two contribution in a different representation, as advertised. In order to do this more systematically we can take an ansatz for $\CA(z,\zb)$ made of an infinite sum of conformal blocks\footnote{The convention that we use for conformal blocks is given in appendix~\ref{app:notationBlocks}.} of twist four and six with arbitrary coefficients
\eqn{
\CA(z,\zb) = \sum_{\tau=4,6}\sum_{\ell=0}^\infty a_{\tau+\ell,\ell}\,g_{\tau+\ell,\ell}(z,\zb)\,.
}[]
Then we truncate this sum up to a maximal spin $\ell_{\mathrm{max}}$ and impose that the twist two and zero up to $\ell_\mathrm{max}$ cancel. Remarkably, the solution is unique for each $\ell_\mathrm{max}$. This will not be true for quarter-BPS operators or for higher half-BPS operators $\hBPS[p]$. We can then easily spot a pattern and extend this solution to the non-truncated equation. The coefficients that enter in $\CA(z,\zb)$ are the ones found in (3.11) of~\cite{Dolan:2004iy}, modulo some factors due to the conventions. For convenience we report them here
\twoseqn{
a_{\ell+4,\ell} &= \frac{2^\ell((\ell+1)!)^2\,\big((\ell+1)(\ell+2)+\frac1c\big)}{3(2\ell+2)!}\,,
}[]{
a_{\ell+6,\ell} &= \frac{2^\ell((\ell+2)!)^2\,\big((\ell+1)(\ell+4)-\frac3c\big)}{6(2\ell+4)!}\,.
}[][]
After resumming these coefficients with their blocks, which can be done following the steps described in appendix~\ref{app:resumming}, we reproduce exactly the known result
\eqn{
\CA(z,\zb) = \frac{1}{6}\,z^2 \zb^2\,\big(\lsp \mbox{equation (2.31) of \cite{Beem:2016wfs}}\,\big)\,.
}[]
The overall factor with respect to~\cite{Beem:2016wfs} is due to a different choice of normalization and conventions. By explicit inspection we see that
\eqn{
\sum_{k=1}^6\BBT_k\Big(w_k(z,\zb) + \CA(z,\zb)\, v_k^{(1)}\Big)
}[w&AO2]
is free of twist zero and twist two operators. We can therefore take $\CH(z,\zb)$ as having only unprotected contributions with anomalous dimensions. Note that in this computation we did not use at all the form of the superconformal blocks. In fact, it was not even needed. In the subsequent cases having the superconformal blocks would instead be very beneficial but, unfortunately, the blocks for external quarter-BPS operators are not known yet. We therefore have no choice but using this method, which, as we will see, is still able to produce some useful results.

\section{Cases of interest}\label{sec:cases}
In this section we introduce the quarter-BPS operator transforming in the (2,0,2) SU(4) representation, which will be at the center of our investigation. We proceed by discussing in details various four-point functions including this operator. We report the results for correlator with one, two or four $\qBPS$'s. The case $\langle \qBPS \qBPS \qBPS \hBPS\rangle$ is not very illuminating: there are no constraints coming from the ambiguity resolution discussed in subsection~\ref{sec:ambiguity} and it does not provide new information about unknown OPE data.  Thus, although we analyzed it,  we decided not to include this example in the main discussion.

\subsection[The \texorpdfstring{$(2,0,2)$}{(2,0,2)} quarter-BPS operator]{The $\boldsymbol{(2,0,2)}$ quarter-BPS operator}\label{sec:qBPSdef}

Let us define in detail the quarter-BPS operator transforming in the $(2,0,2)$ in terms of free fields. It turns out that we need to take a specific linear combination if we want to end up with a quarter-BPS operator in the interacting theory as well. As before, we can write both a single trace and a double trace version
\twoseqn{
\qBPS^{(\mathrm{st},1)}(x,\bfS)=\tr\mleft(\varphi^{M_1}\varphi^{M_2}\varphi^{M_3}\varphi^{M_4}\mright)\,S\cdot\Sigma_{M_1M_2}\lnsp\cdot\Sb \;S\cdot\Sigma_{M_3M_4}\lnsp\cdot\Sb\,,
}[Osingtrace]{
\qBPS^{(\mathrm{st},2)}(x,\bfS)=\tr\mleft(\varphi^{M_1}\varphi^{M_2}\varphi^{M_3}\varphi^{M_4}\mright)\,S\cdot\Sigma_{M_1M_3}\lnsp\cdot\Sb \;S\cdot\Sigma_{M_2M_4}\lnsp\cdot\Sb\,.
}[Osingtrace2][]
In the single trace case there are, a priori, two possibilities. Whereas for double trace we find
\eqn{
\qBPS^{(\mathrm{dt})}(x,\bfS)=\tr\mleft(\varphi^{M_1}\varphi^{M_2}\mright)\,\tr\mleft(\varphi^{M_3}\varphi^{M_4}\mright)\,S\cdot\Sigma_{M_1M_3}\lnsp\cdot\Sb \;S\cdot\Sigma_{M_2M_4}\lnsp\cdot\Sb\,.
}[]
If we compute the two-point function of $\CO^{(\mathrm{st},2)}_{02}$ we notice that it vanishes so the operator must be identically zero, as explicitly shown in appendix \ref{app:low_ops}. Then~\cite{Ryzhov:2001bp,DHoker:2003csh} teach us that the correct linear combination that remains protected is\footnote{Note that this and the following expressions in the present section are exact in $N$.}
\eqn{
\qBPS^{(\mathrm{dt})}(x,\bfS)+\frac{2}{N}\,\qBPS^{(\mathrm{st},1)}(x,\bfS)\,.
}[]
This is also consistent with the fact that at large $N$ the single trace operators disappear. The orthogonal combination to this one, namely $\qBPS^{(\mathrm{st},1)}$, is actually not a superconformal primary. It is the $Q^2\Qb{}^2$ superdescendant of the free Konishi operator as we will show in subsection~\ref{sec:check5point}. As it is well known, this operator will be lifted by quantum corrections in the interacting theory.

We want to define our operator such that its two-point function is given by~\eqref{eq:twopf}, with $p=0$ and $q=\qb=2$. A simple computation shows that the correct normalization is the following
\eqn{
\qBPS(x,\bfS)=\frac{\sqrt2}{\sqrt3\sqrt{(N^2-4)(N^2-1)}}\,\left(\qBPS^{(\mathrm{dt})}(x,\bfS)+\frac{2}{N}\,\qBPS^{(\mathrm{st},1)}(x,\bfS)\right)\,.
}[eq:qBPSfields]
For completeness, let us also give the normalized $\hBPS$ operator
\eqn{
\hBPS(x,y) = \frac{\sqrt2}{\sqrt{N^2-1}} \,\tr\lsp(\varphi\cdot y)^2\,.
}[eq:hBPSfields]

\subsection[Correlator \texorpdfstring{$\langle\qBPS\hBPS\hBPS\hBPS\rangle$}{<4222>}]{Correlator $\boldsymbol{\langle\qBPS\hBPS\hBPS\hBPS\rangle}$}\label{sec:QHHH}
\subsubsection{Tensor structures}
As a warm up, let us consider a four-point function with a single insertion of $\qBPS$. The correlator $\langle\qBPS\hBPS\hBPS\hBPS\rangle$ exchanges three different representations.
The label assignment we chose is summarized in table~\ref{tab:irrepsQHHH} and the associated tensor structures are given \hypertarget{eqref:TQHHHDef}{below}
\eqna{
\TQHHH1 &= \frac{1}{2} \strSYSb{1}{34}{1}\lsp (\strSYSb{1}{32}{1}\lsp \stry{24}-\strSYSb{1}{24}{1}\lsp \stry{23})\,,\\
\TQHHH2 &= \frac{1}{4} (2 \strSYSb{1}{32}{1}\lsp \strSYSb{1}{34}{1}\lsp \stry{24}+\strSYSb{1}{24}{1}\lsp (2 \strSYSb{1}{34}{1}\lsp \stry{23}-\strSYSb{1}{32}{1}\lsp \stry{34}))\,,\\
\TQHHH3 &= \frac{1}{4} \strSYSb{1}{24}{1}\lsp \strSYSb{1}{32}{1}\lsp \stry{34}\lsp\,.
}[eq:TQHHHDef]

\begin{table}[h]
\centering
\begin{tabular}{c?ccc}
\Hhline
$k$ & 1 & 2 & 3 \\[-.4ex]
\hline
$(q,p,\qb)$ & $(121)$ & $(202)$ & $(020)$ \\
\Hhline
\end{tabular}
\caption{Dictionary between the tensor structure $\protect\TQHHH{k}$ and the representation exchanged in the $(12)(34)$ OPE of $\langle\qBPS\hBPS\hBPS\hBPS\rangle$.}\label{tab:irrepsQHHH}
\end{table}

\subsubsection{Free theory}

With the definitions in~\eqref{eq:qBPSfields} and~\eqref{eq:hBPSfields} we can compute the free theory four-point function by taking Wick contraction and performing the traces over the various indices. The result can be written as follows
\eqn{
\langle \qBPS(x_1,\bfS_1)\hBPS(x_2,y_2)\hBPS(x_3,y_3)\hBPS(x_4,y_4)\rangle = \kinPref{4222}\,\sum_{k=1}^3 g_k(z,\zb)\, \TQHHH{k}\,,
}[eq:QHHH4pf]
We will not give the explicit expression of the functions $g_k(z,\zb)$ here. However, in table~\ref{tab:freeExchangedQHHH} we provide a summary of the operators exchanged in each tensor structure. Let us emphasize that the operators that we show are conformal primaries, not superconformal primaries. To know the expansion in the latter one would have to use the superconformal blocks which are not yet known. Note however that, unlike the following cases, no operators of twist higher than four are exchanged. The reason for this is that the entire correlation function is actually protected. Thus there are no long operators at threshold that are expected to gain anomalous dimensions when the coupling is turned on. A detailed proof of the protected nature of this correlator is given in section~\ref{sec:fivepoint} and it follows from the study of the OPE limit of the $\hBPS$ five-point function.

\begin{table}[h]
\centering
\begin{tabular}{c?cccc}
\Hhline
$k$ & 1 & 2 & \multicolumn{2}{c}{3} \\
\hline
$\tau$ & 4   & 4    & 2 & 4    \\
$\ell$ & odd & even & $\ell=0$ & even \\
\Hhline
\end{tabular}
\caption{Conformal primaries exchanged in the various structures $g_k(z,\zb)$ of the correlator $\langle\qBPS\hBPS\hBPS\hBPS\rangle$ in the free theory. We define $\tau=\Delta-\ell$.}\label{tab:freeExchangedQHHH}
\end{table}

The chiral algebra contribution to this four-point function can be obtained by applying the chiral algebra map defined in section~\ref{sec:chialg}. The result is quite simple and it reads\footnote{As a consistency check, here and in all the following examples,  we have computed $\mathfrak{f}(z)$ by applying the chiral algebra map to the  free-theory result in $4d$ as well as by direct computation of the $2d$ correlator by means of~\eqref{twodWick}.}
\eqn{
\frac{\chi\big[\langle\qBPS\hBPS\hBPS\hBPS\rangle\big]}{\kinPrefz{4222}} \equiv \gothf_{\qBPS\hBPS\hBPS\hBPS}(z) = - \frac{\sqrt{N^2-4}}{2\sqrt3(N^2-1)}\left(z^2 + \frac{z}{z-1}\right)\,.
}[eq:gothfQHHH]
The contribution of $\gothf$ to the full correlator can be written as follows --- we omit the subscript for brevity
\eqn{
\frac{\langle\qBPS\hBPS\hBPS\hBPS\rangle}{\kinPref{4222}}\big|_{\gothf} =  \TQHHH1\, \frac{2z\zb\,\bigl(z\lsp \gothf(\zb)-\zb\lsp\gothf(z)\bigr)}{z-\zb}  + \TQHHH3\,\frac{4\,\bigl(z^2(\zb-2)\lsp\gothf(\zb) - \zb^2(z-2)\lsp\gothf(z)\bigr)}{z-\zb}\,.
}[]
In a way, this is the same as making a choice of the vector $\vec{w}$ introduced in subsection~\ref{sec:ambiguity}.
It can be checked that the action of $\chi$ and all of its variants --- see~\eqref{eq:variants} --- acting on the expression above yield $\gothf$.

\subsubsection{Ward identity and ambiguity}

Since the entire correlator is protected, the analysis that follows here is inessential. We will however show the steps in preparation to the subsequent cases where they become substantially more computationally involved. The Ward identities follow from requiring that the chiral algebra map~$\chi$ applied on the interacting correlator gives the same result as~\eqref{eq:gothfQHHH}. This forces a generic unprotected contribution to $g_k(z,\zb)$, let us call it $\delta\lnsp g_k(z,\zb)$,  to satisfy
\eqn{
\delta\lnsp g_1(z,\zb) = 0\,,\qquad \delta\lnsp g_3(z,\zb) = - \delta\lnsp g_2(z,\zb)\,.
}[]
Following the notation introduced in subsection~\ref{sec:ambiguity}, this fact can be encoded by defining the vectors $\vec{v}^{\lsp(m)}$ and $\vec{w}$ as follows:
\eqna{
\vec{v}^{\,(1)}\! &= \{0,1,-1\}\,,\\
\vec{w} &= \frac{\sqrt{N^2-4}}{\sqrt3\lsp(N^2-1)}\bigg\{
\frac{z^2 \zb^2 (z \zb-z-\zb)}{(z-1) (\zb-1)},\,0,\,\frac{2 z \zb \left(z^2 \zb^2-z^2 \zb-z \zb^2+2 z+2 \zb-2\right)}{(z-1) (\zb-1)}
\bigg\}\,.
}[]
There is only one ambiguity $\CA_1(z,\zb)$. In this case, since the correlator is protected, we know that 
\eqn{
\CA_1 = \frac{\sqrt{N^2-4}}{\sqrt3\lsp(N^2-1)}\frac{z^2\zb^2(z\zb-z-\zb+2)}{(z-1)(\zb-1)}\,.
}[]
and $\CH_1=0$. However, the criteria described in subsection~\ref{sec:ambiguity} would not have allowed us to conclude this because there are no twist-two operators that need to be cancelled.

By expanding $\vec{w}$ in conformal blocks we see that in the $(1,2,1)$ we exchange only twist four while in the $(0,2,0)$ we exchange $\hBPS$ and other operators of twist four. In detail we have
\eqna{
w_1(z,\zb) &= \frac{\sqrt{N^2-4}}{\sqrt3(N^2-1)}\sum_{\ell=1,\,\mathrm{odd}}^\infty \frac{2^\ell \,\ell!\lsp(\ell+2)!}{(2\ell+1)!}\,g_{\ell+4,\ell}^{2,0}\,,\\
w_3(z,\zb) &= \frac{\sqrt{N^2-4}}{\sqrt3\lsp(N^2-1)}\bigg(-4\lsp g_{2,0}^{2,0} +  \sum_{\ell=0,\,\mathrm{even}}^\infty \frac{2^{\ell+1} \,\ell!\lsp(\ell+2)!}{(2\ell+1)!}\,g_{\ell+4,\ell}^{2,0}\bigg)\,.
}[]

\subsection[Correlator \texorpdfstring{$\langle\qBPS\qBPS\hBPS\hBPS\rangle$}{<4422>}]{Correlator $\boldsymbol{\langle\qBPS\qBPS\hBPS\hBPS\rangle}$}\label{sec:QQHH}
\subsubsection{Tensor structures}

Now we repeat the same analysis for all the other four-point functions. Let us proceed with the correlator $\langle\qBPS\qBPS\hBPS\hBPS\rangle$. It has ten different structures exchanging six distinct representations.
The label assignment we chose is summarized in table~\ref{tab:irrepsQQHH} and the associated tensor structures are given \hypertarget{eqref:TQQHHDef}{below}
\def\tonewline{\\&\;\quad}
\eqnal{
\TQQHH1 &= \frac{1}{30} \bigl(30\lsp\strSbYSb{1}{3}{2}\lsp \strSSb{12}\lsp \strSYS{1}{4}{2}\lsp \strSYSb{2}{34}{1}+30\lsp\strSbYSb{1}{3}{2}\lsp \strSSb{21}\lsp \strSYS{1}{4}{2}\lsp \strSYSb{1}{34}{2}-56\lsp\strSbYSb{1}{3}{2}\lsp \strSSb{21}\lsp \strSSb{12}\lsp \strSYS{1}{4}{2}\lsp \stry{34}+30\lsp(\strSbYSb{1}{3}{2})^2\lsp(\strSYS{1}{4}{2})^2
\tonewline
-8\lsp\strSSb{21}\lsp (\strSSb{12})^2\lsp\strSYSb{2}{34}{1}\lsp \stry{34}-8\lsp(\strSSb{21})^2\lsp\strSSb{12}\lsp \strSYSb{1}{34}{2}\lsp \stry{34}+5\lsp(\strSSb{12})^2\lsp(\strSYSb{2}{34}{1})^2+10\lsp\strSSb{21}\lsp \strSSb{12}\lsp \strSYSb{1}{34}{1}\lsp \strSYSb{2}{34}{2}
\tonewline
+5\lsp(\strSSb{21})^2\lsp(\strSYSb{1}{34}{2})^2-2\lsp(\strSSb{21})^2\lsp(\strSSb{12})^2\lsp(\stry{34})^2\bigr)\,,\\
\TQQHH2 &= \frac{1}{4} \bigl(2\lsp\strSbYSb{1}{3}{2}\lsp \strSSb{12}\lsp \strSYS{1}{4}{2}\lsp \strSYSb{2}{34}{1}+2\lsp\strSbYSb{1}{3}{2}\lsp \strSSb{21}\lsp \strSYS{1}{4}{2}\lsp \strSYSb{1}{34}{2}-\strSSb{21}\lsp (\strSSb{12})^2\lsp\strSYSb{2}{34}{1}\lsp \stry{34}-(\strSSb{21})^2\lsp\strSSb{12}\lsp \strSYSb{1}{34}{2}\lsp \stry{34}
\tonewline
+(\strSSb{12})^2\lsp(\strSYSb{2}{34}{1})^2+2\lsp\strSSb{21}\lsp \strSSb{12}\lsp \strSYSb{1}{34}{1}\lsp \strSYSb{2}{34}{2}+(\strSSb{21})^2\lsp(\strSYSb{1}{34}{2})^2-2\lsp(\strSSb{21})^2\lsp(\strSSb{12})^2\lsp(\stry{34})^2\bigr)\,,\\
\TQQHH3 &= -\frac{1}{8} \lsp(\strSSb{21}\lsp \strSYSb{1}{34}{2}-\strSSb{12}\lsp \strSYSb{2}{34}{1}) (4\lsp\strSbYSb{1}{3}{2}\lsp \strSYS{1}{4}{2}+2\lsp\strSSb{21}\lsp \strSYSb{1}{34}{2}+2\lsp\strSSb{12}\lsp \strSYSb{2}{34}{1}-5\lsp\strSSb{12}\lsp \strSSb{21}\lsp \stry{34})\,,\\
\TQQHH4 &= -\frac{1}{60}\lsp \strSSb{12}\lsp \strSSb{21}\lsp \bigl(-10\lsp\strSbYSb{1}{3}{2}\lsp \strSYS{1}{4}{2}\lsp \stry{34}-5\lsp\strSSb{21}\lsp \strSYSb{1}{34}{2}\lsp \stry{34}-5\lsp\strSSb{12}\lsp \strSYSb{2}{34}{1}\lsp \stry{34}+16\lsp\strSSb{12}\lsp \strSSb{21}\lsp (\stry{34})^2
\tonewline
-10\lsp\strSYSb{1}{34}{1}\lsp \strSYSb{2}{34}{2}\lsp\bigr)\,,\\
\TQQHH5 &= \frac{1}{30}\lsp (\strSSb{12})^2 \bigl(-10\lsp\strSSb{21}\lsp \strSYSb{2}{34}{1}\lsp \stry{34}+4\lsp(\strSSb{21})^2\lsp(\stry{34})^2+5\lsp(\strSYSb{2}{34}{1})^2\bigr)\,,\\
\TQQHH6 &= \frac{1}{30}\lsp (\strSSb{21})^2 \bigl(-10\lsp\strSSb{12}\lsp \strSYSb{1}{34}{2}\lsp \stry{34}+4\lsp(\strSSb{12})^2\lsp(\stry{34})^2+5\lsp(\strSYSb{1}{34}{2})^2\bigr)\,,\\
\TQQHH7 &= -\frac{1}{60}\lsp \strSSb{12}\lsp \strSSb{21}\lsp \stry{34}\lsp (-6\lsp\strSbYSb{1}{3}{2}\lsp \strSYS{1}{4}{2}-3\lsp\strSSb{21}\lsp \strSYSb{1}{34}{2}-3\lsp\strSSb{12}\lsp \strSYSb{2}{34}{1}+8\lsp\strSSb{12}\lsp \strSSb{21}\lsp \stry{34})\,,\\
\TQQHH8 &= -\frac{1}{16}\lsp \strSSb{12}\lsp \strSSb{21}\lsp \stry{34}\lsp (-\strSSb{21}\lsp \strSYSb{1}{34}{2}-\strSSb{12}\lsp \strSYSb{2}{34}{1}+2\lsp\strSSb{12}\lsp \strSSb{21}\lsp \stry{34})\,,\\
\TQQHH9 &= \frac{1}{16}\lsp \strSSb{12}\lsp \strSSb{21}\lsp \stry{34}\lsp (\strSSb{12}\lsp \strSYSb{2}{34}{1}-\strSSb{21}\lsp \strSYSb{1}{34}{2})\,,\\
\TQQHH{10} &= (\strSSb{12})^2\lsp(\strSSb{21})^2\lsp(\stry{34})^2\,.
}[eq:TQQHHDef]

\begin{table}[h]
\centering
\begin{tabular}{c?cccccccccc}
\Hhline
$k$ & 1 & 2 & 3 & 4 & 5 & 6 & 7 & 8 & 9 & 10 \\[-.4ex]
\hline
$(q,p,\qb)$ & $(040)$ & \multicolumn{2}{c}{$(121)$} & \multicolumn{3}{c}{$(202)$} & $(020)$ & \multicolumn{2}{c}{$(101)$} & $(000)$ \\
\Hhline
\end{tabular}
\caption{Dictionary between the tensor structure $\protect\TQQHH{k}$ and the representation exchanged in the $(12)(34)$ OPE of $\langle\qBPS\qBPS\hBPS\hBPS\rangle$.}\label{tab:irrepsQQHH}
\end{table}

\subsubsection{Free theory}

The free theory result can be written as follows
\eqn{
\langle \qBPS(x_1,\bfS_1)\qBPS(x_2,\bfS_2)\hBPS(x_3,y_3)\hBPS(x_4,y_4)\rangle = \kinPref{4422}\,\sum_{k=1}^{10} g_k(z,\zb)\, \TQQHH{k}\,,
}[eq:QQHH4pf]
Again, we will are not providing the explicit expression of the functions $g_k(z,\zb)$. In table~\ref{tab:freeExchangedQQHH} we show the summary of the operators exchanged in each tensor structure.

\begin{table}[h]
\centering
\begin{tabular}{c?ccccccccc}
\Hhline
$k$ & 1 & 3 & 4 & 5 & 6 & 7 & 9 & \multicolumn{2}{c}{10} \\
\hline
$\tau=2h$ & $h\geq2$ & $h\geq2$ & $h\geq2$ & $h\geq2$& $h\geq2$ &   $h\geq1$ &   $h\geq1$ &$h=0$&$h\geq 1$\\
$\ell$      & even      & odd      & even      & even      & even          & even & odd & $\ell=0$ & even\\\Hhline
\end{tabular}
\caption{Conformal primaries exchanged in the various structures $g_k(z,\zb)$ of the correlator $\langle\qBPS\qBPS\hBPS\hBPS\rangle$ in the free theory. Structures $2$ and $8$ are zero.}\label{tab:freeExchangedQQHH}
\end{table}

The result of acting with the chiral algebra map~$\chi$ reads
\eqn{
\frac{\chi\big[\langle\qBPS\qBPS\hBPS\hBPS\rangle\big]}{\kinPrefz{4422}} \equiv \gothf_{\qBPS\qBPS\hBPS\hBPS}(z) = \frac3{12} + \frac{N^2-4}{6\lsp(N^2-1)}\left(z^2+\frac{z^2}{(z-1)^2}\right)\,.
}[eq:gothfQQHH]

\subsubsection{Ward identity and ambiguity}

The Ward identities are specified by an uplift vector $\vec{w}$ and eight ambiguity vectors $\vec{v}^{\,(m)}$.
The nonzero entries of the vector $\vec{w}$ read
\eqna{
w_9(z,\zb) &= -\frac{N^2-4}{N^2-1}\left(\hat{w}_9(z,\zb) + \hat{w}_9\Biggish(\frac{z}{z-1},\frac{\zb}{\zb-1}\Biggish)\right) \,,\\
w_{10}(z,\zb) &= 1 + \frac{N^2-4}{N^2-1}\left(\hat{w}_{10}(z,\zb)- \hat{w}_{10}\Biggish(\frac{z}{z-1},\frac{\zb}{\zb-1}\Biggish)\right) \,,
}[]
where we have defined
\eqna{
\hat{w}_9(z,\zb) &= \frac83\big(z^2\zb+z\zb^2\big)\,,\\
\hat{w}_{10}(z,\zb) &= \frac13\big(z^2\zb+z\zb^2-2z\zb\big)\,.
}[]
The expressions for the ambiguity vectors instead are given by
\eqn{
v^{(m)}_k (z,\zb) = \delta^m_k + \hat{v}^{(m)}_k (z,\zb)\,,\qquad m=1,\lsp\ldots,\lsp8\,,
}[]
with the nonzero entries of $\hat{v}^{(m)}_k (z,\zb)$ being
\eqna{
-2\hat{v}^{(1)}_{10} =  \hat{v}^{(7)}_{10}  &= \frac1{30} \,,\\
\hat{v}^{(3)}_9 &= 2\,,\\
-\hat{v}^{(4)}_9 = \hat{v}^{(5)}_9 = \hat{v}^{(6)}_9 &= \frac{8\lsp (z+\zb - z\zb)}{3\lsp z\zb}\,,\\
-\frac1{30} - \hat{v}^{(4)}_{10} = \hat{v}^{(5)}_{10} = \hat{v}^{(6)}_{10} &=\frac{3 z \zb-5 z-5 \zb+10}{15\lsp z \zb}\,.
}[]
As before, we can expand the uplift vector in conformal blocks to see what are the contributions coming from the chiral algebra. We obtain the following result
\eqna{
w_9(z,\zb) &= \frac{N^2-4}{3(N^2-1)}\sum_{\ell=1,\,\mathrm{odd}}^\infty \frac{2^{\ell+3}\lsp \ell!\lsp (\ell+1)!}{(2\ell-1)!} g_{\ell+2,\ell}\,,\\
w_{10}(z,\zb) &= g_{0,0} - \frac{N^2-4}{3(N^2-1)} \sum_{\ell=0,\,\mathrm{even}}^\infty \frac{2^\ell\lsp \ell\lsp(\ell^2+\ell+2)\lsp\big((\ell-1)!\big)^2}{(2\ell-1)!} g_{\ell+2,\ell}\,.
}[]
The above expressions show the presence of a tower of higher spin conserved currents in both protected structures. As we remarked in section~\ref{sec:ambiguity}, these should disappear in the interacting theory. We therefore have to tweak the ambiguity such that all the twist-two contributions vanish with the exception of spin zero, one and two, which all belong to the $\hBPS$ multiplet. Furthermore we also have to make sure that the OPE coefficient of $g_{2,0}$ in the $(0,2,0)$ is not modified. As before, in order to achieve this in practice we make an ansatz with a finite number of blocks for the ambiguities $\CA_m$, we impose that the twist-two contributions vanish up to a certain maximal spin $\ell_{\mathrm{max}}$, and then we extrapolate our results for infinite $\ell_{\mathrm{max}}$.

The conjugation properties of the structures $\TQQHH{k}$ allow us to set to zero $\CA_8$ and also to set $\CA_5 = \CA_6$. Furthermore, since $\CA_4$ and $\CA_5$ exchange the same representation, we can assume that they will be proportional to each other. After these remarks we can fix the twist-four contributions of $\CA_4$, $\CA_5$ and $\CA_6$ up to a single constant $\kappa$ and the twist-two contribution of $\CA_7$. On the other hand, we cannot say anything about $\CA_1$, $\CA_2$ and $\CA_3$. Presumably the knowledge of the superconformal blocks would allow us to fix their twist-four sector as well, but at the moment we do not have access to this information. All in all, the ambiguities that could be fixed read
\eqna{
\CA_5 = \CA_6 = \frac1\kappa \CA_4 &=-\frac{2\lsp(N^2-3)}{(\kappa-2)(N^2-1)}g_{4,0} - \frac{N^2-4}{N^2-1} \sum_{\ell=2,\,\mathrm{even}}^\infty \frac{2^\ell\lsp(\ell+1)!(\ell+2)!}{(2\ell+1)!(\kappa-2)}\,g_{\ell+4,\ell}\,,\\
\CA_7 &= - \frac{40}{N^2-1} \, g_{2,0}\,.
}[]
The sum over spins can be performed with the methods described in appendix~\ref{app:resumming}. The result is
\eqna{
\CA_5(z,\zb) = \frac{N^2-4}{(N^2-1)(\kappa-2)}\bigg[&a(z,\zb)\lsp\log(1-\zb) + a(\zb,z)\lsp\log(1-z)\\& -24 \log(1-z)\log(1-\zb)
\bigg]-\frac{2\lsp(N^2-3)}{(\kappa-2)(N^2-1)}g_{4,0}\,,\\
}[]
with 
\eqn{
a(z,\zb) = \frac{z \zb}{z-\zb}\left(\hat{a}(z)+\hat{a}\Biggish(\frac{z}{z-1}\Biggish)\right)\,,\qquad \hat{a}(z) = 12 + z^2\,.
}[]

Notice that, while we chose $\CA_7$ such that the OPE coefficient of $\tr(y\cdot\varphi)^2$ matches the free theory value, we did not do that  for the other twist-two operators, namely the R-current $J_\mu$ and the stress-tensor $T_{\mu\nu}$. Indeed their free theory values extracted naively are incorrect as they are contaminated by superdescentants of the free Konishi operator $\tr\varphi^2=A_2\bar{A}_2[0;0]^{(0,0,0)}$ which are lifted in the interacting theory. More precisely, at levels $2l=2$ and $2l=4$ the Konishi operator has a ``fake'' R-current and stress tensor descendant, respectively: $Q^l\Qb{}^{\llsp l}\,\tr\varphi^2$. The true values differ from the free theory ones by a simple multiplicative factor
\twoseqn{
\lambda_{\qBPS\qBPS J_{\mu}}\lambda_{\hBPS\hBPS J_{\mu}} = \frac13\,\lambda_{\qBPS\qBPS J_{\mu}}^{\mathrm{free}}\lambda_{\hBPS\hBPS J_{\mu}}^{\mathrm{free}} &= -\frac{32}{3(N^2-1)}\,,
}[]{
\lambda_{\qBPS\qBPS T_{\mu\nu}}\lambda_{\hBPS\hBPS T_{\mu\nu}} = \frac15\,\lambda_{\qBPS\qBPS T_{\mu\nu}}^{\mathrm{free}}\lambda_{\hBPS\hBPS T_{\mu\nu}}^{\mathrm{free}} &= \frac{16}{45(N^2-1)}\,,
}[][eq:lambdasqTJ]
where the OPE coefficient of the current is taken from the structure \TQQHH{9}, the other structure in the $(1,0,1)$ being zero.
Getting the correct OPE coefficient for the twist-two contributions will be crucial for the results of section~\ref{sec:Results} because they give the only singular contribution to the crossed correlator, which is consequently the only one picked up the $\mathrm{dDisc}$.

\subsection[Correlator \texorpdfstring{$\langle\hBPS\qBPS\qBPS\hBPS\rangle$}{<2442>}]{Correlator $\boldsymbol{\langle\hBPS\qBPS\qBPS\hBPS\rangle}$}\label{sec:HQQH}
\subsubsection{Tensor structures}

The correlator $\langle\hBPS\qBPS\qBPS\hBPS\rangle$ is a permutation of the previous one, therefore it also has ten different structures.
The label assignment we chose is summarized in table~\ref{tab:irrepsHQQH} and the associated tensor structures are given below.
Unfortunately, the expressions look quite involved so we will write down only a few \hypertarget{eqref:THQQHDef}{structures}.\footnote{The remaining ones will be provided in an ancillary file attached to the arXiv version of the paper.}
\begin{small}
\eqnal{
\THQQH1 &= \frac1{2800}\Bigl(-45\lsp\strSbYSb{2}{1}{3}\lsp \strSSb{23}\lsp \strSYS{2}{4}{3}\lsp \strSYSb{3}{14}{2}+63\lsp\strSbYSb{2}{1}{3}\lsp \strSSb{32}\lsp \strSYS{2}{4}{3}\lsp \strSYSb{2}{14}{3}-736\lsp\strSbYSb{2}{1}{3}\lsp \strSSb{32}\lsp \strSSb{23}\lsp \strSYS{2}{4}{3}\lsp \stry{14}+9\lsp(\strSbYSb{2}{1}{3})^2\lsp(\strSYS{2}{4}{3})^2
\tonewline
+230\lsp\strSSb{32}\lsp (\strSSb{23})^2\lsp\strSYSb{3}{14}{2}\lsp \stry{14}-1120\lsp(\strSSb{32})^2\lsp\strSSb{23}\lsp \strSYSb{2}{14}{3}\lsp \stry{14}+16\lsp(\strSSb{23})^2\lsp(\strSYSb{3}{14}{2})^2-133\lsp\strSSb{32}\lsp \strSSb{23}\lsp \strSYSb{2}{14}{2}\lsp \strSYSb{3}{14}{3}
\tonewline
+70\lsp(\strSSb{32})^2\lsp(\strSYSb{2}{14}{3})^2+2276\lsp(\strSSb{32})^2\lsp(\strSSb{23})^2\lsp(\stry{14})^2\Bigr)\,,\\
\THQQH6 &= \frac{1}{144} \bigl(-\strSbYSb{2}{1}{3}\lsp \strSSb{23}\lsp \strSYS{2}{4}{3}\lsp \strSYSb{3}{14}{2}+3\lsp\strSbYSb{2}{1}{3}\lsp \strSSb{32}\lsp \strSYS{2}{4}{3}\lsp \strSYSb{2}{14}{3}-16\lsp\strSbYSb{2}{1}{3}\lsp \strSSb{32}\lsp \strSSb{23}\lsp \strSYS{2}{4}{3}\lsp \stry{14}+(\strSbYSb{2}{1}{3})^2\lsp(\strSYS{2}{4}{3})^2
\tonewline
-10\lsp\strSSb{32}\lsp (\strSSb{23})^2\lsp\strSYSb{3}{14}{2}\lsp \stry{14}-2\lsp(\strSSb{23})^2\lsp(\strSYSb{3}{14}{2})^2+15\lsp\strSSb{32}\lsp \strSSb{23}\lsp \strSYSb{2}{14}{2}\lsp \strSYSb{3}{14}{3}+28\lsp(\strSSb{32})^2\lsp(\strSSb{23})^2\lsp(\stry{14})^2\bigr)\,,\\
\THQQH7 &= \frac{1}{400} \bigl(2\lsp\strSbYSb{2}{1}{3}\lsp \strSSb{32}\lsp \strSYS{2}{4}{3}\lsp \strSYSb{2}{14}{3}-4\lsp\strSbYSb{2}{1}{3}\lsp \strSSb{23}\lsp \strSSb{32}\lsp \strSYS{2}{4}{3}\lsp \stry{14}+(\strSbYSb{2}{1}{3})^2\lsp(\strSYS{2}{4}{3})^2-(\strSSb{23})^2\lsp(\strSYSb{3}{14}{2})^2
\tonewline
-2\lsp\strSSb{23}\lsp \strSSb{32}\lsp \strSYSb{2}{14}{2}\lsp \strSYSb{3}{14}{3}+4\lsp(\strSSb{23})^2\lsp(\strSSb{32})^2\lsp(\stry{14})^2\bigr)\,,\\
%
%\THQQH8 &= \frac{1}{112} \bigl(-\strSbYSb{2}{1}{3}\lsp \strSSb{23}\lsp \strSYS{2}{4}{3}\lsp \strSYSb{3}{14}{2}-\strSbYSb{2}{1}{3}\lsp \strSSb{32}\lsp \strSYS{2}{4}{3}\lsp \strSYSb{2}{14}{3}-(\strSbYSb{2}{1}{3})^2\lsp(\strSYS{2}{4}{3})^2-2\lsp\strSSb{32}\lsp (\strSSb{23})^2\lsp\strSYSb{3}{14}{2}\lsp \stry{14}-\strSSb{32}\lsp \strSSb{23}\lsp \strSYSb{2}{14}{2}\lsp \strSYSb{3}{14}{3}
%\tonewline
%+4\lsp(\strSSb{32})^2\lsp(\strSSb{23})^2\lsp(\stry{14})^2\bigr)\,,\\
%%
%\THQQH9 &= \frac{1}{56} \bigl(\strSbYSb{2}{1}{3}\lsp \strSSb{23}\lsp \strSYS{2}{4}{3}\lsp \strSYSb{3}{14}{2}+\strSbYSb{2}{1}{3}\lsp \strSSb{32}\lsp \strSYS{2}{4}{3}\lsp \strSYSb{2}{14}{3}+3\lsp(\strSbYSb{2}{1}{3})^2\lsp(\strSYS{2}{4}{3})^2+10\lsp\strSSb{32}\lsp (\strSSb{23})^2\lsp\strSYSb{3}{14}{2}\lsp \stry{14}
%\tonewline
%-2\lsp(\strSSb{23})^2\lsp(\strSYSb{3}{14}{2})^2+\strSSb{32}\lsp \strSSb{23}\lsp \strSYSb{2}{14}{2}\lsp \strSYSb{3}{14}{3}-12\lsp(\strSSb{32})^2\lsp(\strSSb{23})^2\lsp(\stry{14})^2\bigr)\,,\\
%
\THQQH{10} &= \frac1{1008}\big(\strSbYSb{2}{1}{3}\lsp \strSYS{2}{4}{3}+\strSSb{23}\lsp \strSYSb{3}{14}{2}-2\lsp\strSSb{23}\lsp \strSSb{32}\lsp \stry{14}\big)^2\,.
}[eq:THQQHDef]
\end{small}

\begin{table}[h]
\centering
\begin{tabular}{c?cccccccccc}
\Hhline
$k$ & 1 & 2 & 3 & 4 & 5 & 6 & 7 & 8 & 9 & 10 \\[-.4ex]
\hline
$(q,p,\qb)$ & $(222)$ & \multicolumn{2}{c}{$(311)$} &  \multicolumn{2}{c}{$(400)$} & $(121)$ & $(202)$ & \multicolumn{2}{c}{$(210)$}  & $(020)$ \\
\Hhline
\end{tabular}
\caption{Dictionary between the tensor structure $\protect\THQQH{k}$ and the representation exchanged in the $(12)(34)$ OPE of $\langle\hBPS\qBPS\qBPS\hBPS\rangle$.  The representations $(311),\, (400),\, (210)$ are meant in combination with their complex conjugate.}\label{tab:irrepsHQQH}
\end{table}

\subsubsection{Free theory}

The free theory result can be written as follows
\eqn{
\langle \hBPS(x_1,y_1)\qBPS(x_2,\bfS_2)\qBPS(x_3,\bfS_3)\hBPS(x_4,y_4)\rangle = \kinPref{2442}\,\sum_{k=1}^{10} g_k(z,\zb)\, \THQQH{k}\,,
}[eq:HQQH4pf]
This is just like the previous correlator but in a different ordering. Since our convention is to always consider the OPE $(12)(34)$, here we are doing the conformal block expansion in a different channel. Of course, crossing will relate the following results with those of the previous subsection, but in a nontrivial way that will be explored in subsection~\ref{sec:invMixed}. In table~\ref{tab:freeExchangedHQQH} we show the summary of the operators exchanged in each tensor structure.

\begin{table}[h]
\centering
\begin{tabular}{c?cccccccc}
\Hhline
structure   & 1&2&4&6&7&8&\multicolumn{2}{c}{10}\\
\hline
$\tau=2n+2$ & $n\geq2$ & $n\geq2$ & $n\geq2$ & $n\geq1$ & $n\geq1$ & $n\geq1$ & $n=0$ & $n>0$\\
$\ell$      & all      & all      & all      & all      & all      & all      & 0 & all\\
excluded $(\Delta,\ell)$ &&&& $(4,0)$ & $(5,1)$ & $(4,0)$ && \\
\Hhline
\end{tabular}
\caption{Conformal primaries exchanged in the various structures $g_k(z,\zb)$ of the correlator $\langle\hBPS\qBPS\qBPS\hBPS\rangle$ in the free theory. Structures $3$, $5$ and $9$ are zero.}\label{tab:freeExchangedHQQH}
\end{table}

The result of acting with the chiral algebra map~$\chi$ reads
\eqn{
\frac{\chi\big[\langle\hBPS\qBPS\qBPS\hBPS\rangle\big]}{\kinPrefz{2442}} \equiv \gothf_{\hBPS\qBPS\qBPS\hBPS}(z) = \frac1{24}\bigg(\hat{f}(z) + z^2\lsp \hat{f}\Bigggish(\frac1{z}\Bigggish)\bigg)\,,
}[eq:gothfHQQH]
with
\eqn{
\hat{f}(z) = \frac{N^2-4}{N^2-1}\frac{4z}{(z-1)^2} + \frac{3z^3}{(z-1)^4}\,.
}[]

\subsubsection{Ward identity and ambiguity}

The Ward identities are specified by an uplift vector $\vec{w}$ and eight ambiguity vectors $\vec{v}^{\,(m)}$. The nonzero entries of the vector $\vec{w}$ read
\eqna{
w_6(z,\zb) &= \frac{z^2\zb^2}{(z-1)^2(\zb-1)^2}\left(\frac{3\,(2z\zb-z-\zb)}{2\,(z-1)^2(\zb-1)^2} - \frac{N^2-4}{N^2-1} (z+\zb)\right)\,,\\
w_{10}(z,\zb) &= \frac{7\lsp z\zb}{(z-1)^2(\zb-1)^2}\bigg( \frac{3\lsp z\zb\,(10 z \zb+7 z+7 \zb-24)}{2\,(z-1)^2(\zb-1)^2}\\&\hspace{9.05em} + \frac{N^2-4}{N^2-1} (7 z^2 \zb+7 z \zb^2-24 z \zb+24)\bigg)\,.
}[]
The ambiguity vectors instead can be expressed as follows
\eqna{
v^{(m)}_k (z,\zb) &= \delta^m_k + \hat{v}^{(m)}_k (z,\zb)\,,&\qquad m &= 1,\,\ldots,\,5\,,\\
v^{(m)}_k (z,\zb) &= \delta^m_{k+1} + \hat{v}^{(m)}_k (z,\zb)\,,&\qquad m &= 6,\,7,\,8\,,\\
}[]
with the nonzero entries of $\hat{v}^{(m)}_k (z,\zb)$ being
\eqnal{
\hat{v}_6^{(1)} &= -\frac{3 (4 z \zb-5 z-5 \zb)}{10 z \zb}	 \,, &\qquad \hat{v}_{10}^{(1)} &= \frac{3 (506 z \zb-1225 z-1225 \zb+4200)}{50 z \zb} \,,\\
\hat{v}_6^{(2)} &= -\frac{2}{5}	\,,&\qquad \hat{v}_{10}^{(2)} &= -\frac{214}{25} \,,\\
\hat{v}_6^{(3)} &= -\frac{3}{5}	\,,&\qquad \hat{v}_{10}^{(3)} &= -\frac{321}{25} \,,\\
&&\hat{v}_{10}^{(4)} &= -14	\,,\\
&&\hat{v}_{10}^{(6)} &= -\frac{63}{25}	\,,\\
&&\hat{v}_{10}^{(7)} &= -9	\,,\\
&&\hat{v}_{10}^{(8)} &= 54	\,.
}[]
From the vector $\vec{w}$ we can obtain the conformal blocks expansion of the protected sector. Here is the result
\eqna{
w_6(z,\zb) &= \sum_{\ell=0}^\infty\,\left(
\frac{N^2-4}{N^2-1} c_{6} + d_{6}
\right) g_{\ell+4,\ell}^{-2,2}\,,\\
w_{10}(z,\zb) &= \frac{168(N^2-4)}{N^2-1}\, g_{2,0}^{-2,2} + \sum_{\ell=0}^\infty\,\left(
\frac{N^2-4}{N^2-1} c_{10} + d_{10}
\right) g_{\ell+4,\ell}^{-2,2}\,.
}[]
with
\eqna{
c_6    &=  \frac{ 2^{\ell-1} \,\ell\, \ell!\lsp (\ell+3)!}{(\ell+1) (2 \ell+1)!}\,,\\
d_6    &=  \frac{(-1)^{\ell+1} 2^{\ell-3} \,\ell\lsp (\ell+2)\, \ell!\lsp (\ell+3)!}{(2 \ell+1)!}\,,\\
c_{10} &= -\frac{7 \cdot 2^{\ell-1} (\ell+2) \big(7\lsp\ell\lsp(\ell+3)+24\big) (\ell!)^2}{(2 \ell+1)!}\,,\\
d_{10} &= \frac{7 \cdot 2^{\ell-3} (-1)^{\ell+1}(\ell+1) (\ell+2)^2 \big(17\lsp\ell\lsp(\ell+3)+72\big) (\ell!)^2}{(2 \ell+1)!}\,.
}[]

The only twist-two operator exchanged is precisely $\hBPS$ since the representation exchanged in the structure $\THQQH{10}$ is the $(0,2,0)$. One can check that the coefficient is the same as in the free theory so we do not have to modify anything. The other contributions are of twist four and therefore we cannot say anything about the ambiguities $\CA_m$.

\subsection[Correlator \texorpdfstring{$\langle\qBPS\qBPS\qBPS\qBPS\rangle$}{<4444>}]{Correlator $\boldsymbol{\langle\qBPS\qBPS\qBPS\qBPS\rangle}$}\label{sec:QQQQ}
\subsubsection{Tensor structures}

The correlator $\langle\qBPS\qBPS\qBPS\qBPS\rangle$ has 42 different structures exchanging 19 distinct representations.
The label assignment we chose is summarized in table~\ref{tab:irrepsQQQQ} and the associated tensor structures are given below. Here we introduce a shorthand notation
\eqn{
\BBS^{i_1j_1k_1l_1}_{i_2j_2k_2l_2} \equiv \prod_{n=1,2}\strSSb{1i_n}\strSSb{2j_n}\strSSb{3k_n}\strSSb{4l_n}\,.
}[]
For example $\BBS^{2143}_{2143}=(\strSSb{12})^2(\strSSb{21})^2(\strSSb{34})^2(\strSSb{43})^2$.
Unfortunately, the expressions look quite involved so we will write down only a few \hypertarget{eqref:TQQQQDef}{structures.}\footnote{The remaining ones will be provided in an ancillary file attached to the arXiv version of the paper.}
\begin{small}
\eqnal{
\TQQQQ1 &= \frac1{22680}\Bigl(224\lsp\BBS^{2112}_{3443}+224\lsp\BBS^{2112}_{4343}-21\lsp\BBS^{2113}_{2443}+112\lsp\BBS^{2113}_{4423}+224\lsp\BBS^{2121}_{3443}+224\lsp\BBS^{2121}_{4343}-21\lsp\BBS^{2123}_{4143}
\tonewline
-21\lsp\BBS^{2141}_{2343}+112\lsp\BBS^{2141}_{3342}-21\lsp\BBS^{2142}_{3143}+3\lsp\BBS^{2143}_{2143}+112\lsp\BBS^{2311}_{2443}-1008\lsp\BBS^{2311}_{3442}-504\lsp\BBS^{2311}_{4342}-1008\lsp\BBS^{2311}_{4423}
\tonewline
-504\lsp\BBS^{2312}_{4413}-504\lsp\BBS^{2321}_{3441}-252\lsp\BBS^{2321}_{4341}+14\lsp\BBS^{2341}_{2341}-504\lsp\BBS^{2411}_{3423}-252\lsp\BBS^{2412}_{3413}+14\lsp\BBS^{2413}_{2413}-252\lsp\BBS^{3112}_{3442}
\tonewline
-504\lsp\BBS^{3112}_{4342}-1008\lsp\BBS^{3112}_{4423}-504\lsp\BBS^{3121}_{3442}-1008\lsp\BBS^{3121}_{4342}-504\lsp\BBS^{3121}_{4423}+112\lsp\BBS^{3122}_{4143}+14\lsp\BBS^{3142}_{3142}\tonewline
+10080\lsp\BBS^{3311}_{4422}+2520\lsp\BBS^{3312}_{4412}+2520\lsp\BBS^{3321}_{4421}+2520\lsp\BBS^{3411}_{3422}+630\lsp\BBS^{3412}_{3412}+630\lsp\BBS^{3421}_{3421}-504\lsp\BBS^{4112}_{4323}
\tonewline
-252\lsp\BBS^{4121}_{4323}+14\lsp\BBS^{4123}_{4123}+2520\lsp\BBS^{4311}_{4322}+630\lsp\BBS^{4312}_{4312}+630\lsp\BBS^{4321}_{4321}\Bigr)\,,\\
\TQQQQ{37} &= \frac1{5040}\Bigl(6\lsp\BBS^{2112}_{3443}-6\lsp\BBS^{2112}_{4343}-3\lsp\BBS^{2113}_{2443}-6\lsp\BBS^{2121}_{3443}+6\lsp\BBS^{2121}_{4343}-3\lsp\BBS^{2123}_{4143}-3\lsp\BBS^{2141}_{2343}-3\lsp\BBS^{2142}_{3143}+2\lsp\BBS^{2143}_{2143}\Bigr)\,,\\
%\TQQQQ{38} &= \frac{1}{672} \lsp\bigl(4\lsp\BBS^{2123}_{4143}-\BBS^{2143}_{2143}\bigr)\,,\\
%\TQQQQ{39} &= \frac{1}{336} \lsp\bigl(\BBS^{2113}_{2443}-\BBS^{2142}_{3143}\bigr)\,,\\
\TQQQQ{40} &= \frac1{1344}\Bigl(\BBS^{2113}_{2443}+\BBS^{2123}_{4143}+\BBS^{2141}_{2343}+\BBS^{2142}_{3143}-\BBS^{2143}_{2143}\Bigr)\,,\\
\TQQQQ{41} &= \frac1{1344}\Bigl(\BBS^{2113}_{2443}-\BBS^{2123}_{4143}-\BBS^{2141}_{2343}+\BBS^{2142}_{3143}\Bigr)\,,\\
\TQQQQ{42} &= \BBS^{2143}_{2143}\lsp\,.
}[eq:TQQQQDef]
\end{small}

\begin{table}[h]
\centering
\begin{tabular}{c?cccccccccc}
\Hhline
$i$ & 1&2&3&4&5&6&7&8&9--12 &13\\[-.4ex]
\hline
$(q,p,\qb)$ & $(404)$ &  \multicolumn{2}{c}{$(412)$} &  \multicolumn{2}{c}{$(420)$} & $(222)$ &  \multicolumn{2}{c}{$(230)$} & $(303)$ & $(040)$ \\
\Hhline
\multicolumn{11}{c}{}\\[-1em]
\Hhline
$i$ &14--21  & 22--25& 26--34 & 35 & 36 & 37& 38--41&42 \\[-.4ex]
\hline
$(q,p,\qb)$ &  $(311)$ & $(121)$ & $(202)$ &  \multicolumn{2}{c}{$(210)$} & $(020)$ & $(101)$ & $(000)$ \\
\Hhline
\end{tabular}
\caption{Dictionary between the tensor structure $\protect\TQQQQ{k}$ and the representation exchanged in the $(12)(34)$ OPE of $\langle\qBPS\qBPS\qBPS\qBPS\rangle$. The representations $(412),\,(420), \, (230),\, (311),\,(210)$ are meant in combination with their complex conjugate.}\label{tab:irrepsQQQQ}
\end{table}

\subsubsection{Free theory}

The free theory result can be written as follows
\eqn{
\langle \qBPS(x_1,\bfS_1)\qBPS(x_2,\bfS_2)\qBPS(x_3,\bfS_3)\qBPS(x_4,\bfS_4)\rangle = \kinPref{4444}\,\sum_{k=1}^{42} g_k(z,\zb)\, \TQQQQ{k}\,,
}[eq:QQQQ4pf]
Computing the free theory value of this correlator is somewhat challenging because each operator is a sum of two terms each being a product of four fields~\eqref{eq:qBPSfields}. The number of Wick contractions grows factorially and furthermore one has to take the traces over $\SU(N)$ and $\SO(6)$ indices. In practice we computed the correlators with $n$ $\qBPS^{(\mathrm{dt})}$ and $4-n$ $\qBPS^{(\mathrm{st},1)}$, in some ordering, and then found the full correlator by permuting the points in the results.

In table~\ref{tab:freeExchangedQQQQ} we show the summary of the operators exchanged in each tensor structure.

\begin{table}[H]

\begin{tabular}{c?cccccccccc}
\Hhline
structure   & 1&2&4&6&7&11 & 12 & 13 & 18 \\
\hline
$\tau=2n$ & $n\geq4$ & $n\geq4$ & $n\geq4$ & $n\geq3$ & $n\geq3$ & $n\geq4$ & $n\geq3$& $n\geq2$& $n\geq3$\\
$\ell$      & even     & odd     & even     & even      & odd      & even      & odd &even & even\\
\Hhline
\multicolumn{11}{c}{}\\[-1em]
\Hhline
structure   &20&24&25 &30 & 32 &33 &34 &35 & 37\\
\hline
$\tau=2n$ & $n\geq3$  & $n\geq3$& $n\geq2$ & $n\geq2$ & $n\geq2$ & $n\geq2$ & $n\geq3$ & $n\geq2$ &  $n\geq1$ \\
$\ell$    & odd & even & odd      & even      & even      & even      & odd      & odd      & even \\
\Hhline
\multicolumn{11}{c}{}\\[-1em]
\Hhline
structure  & \multicolumn{2}{c}{40} &41& \multicolumn{2}{c}{42}\\
\hline
$\tau=2n$ & $n=2$ &$n\geq3$ &  $n\geq1$ &   $n\geq1$  &  $n=0$ \\
$\ell$ & $\ell \geq 2$, even &even&odd  $\ell$    &  even & $\ell=0$\\
\Hhline
\end{tabular}
\caption{Conformal primaries exchanged in the various structures $g_k(z, \zb)$ of the correlator $\langle\CO_{02}\CO_{02}\CO_{02}\CO_{02}\rangle$ in the free theory. Structures 3, 5, 8--10, 14--17, 19, 21--23, 26--29, 31, 36 and 38--39 are zero.\label{tab:freeExchangedQQQQ}}
\end{table}
The result of acting with the chiral algebra map~$\chi$ reads
\eqn{
\frac{\chi\big[\langle\qBPS\qBPS\qBPS\qBPS\rangle\big]}{\kinPrefz{4444}} \equiv \gothf_{\qBPS\qBPS\qBPS\qBPS}(z) = 1 + \hat{f}(z) + \hat{f}\Biggish(\frac{z}{z-1}\Biggish)\,,
}[eq:gothfQQQQ]
with
\eqn{
\hat{f}(z) = z^4  + \frac{8 \lsp (N^4-7 N^2+13)}{3\lsp (N^2-4)(N^2-1)}\lsp\bigl(z+z^2\bigr)\,.
}[]
The rational function of $N$ above will appear in the next subsection as well, so let us define a shorthand for it
\eqn{
R_N \equiv \frac{N^4-7 N^2+13}{(N^2-4)(N^2-1)}\,.
}[eq:rNdef]

\subsubsection{Ward identity and ambiguity} \label{subsec:WI4444}

The Ward identities are specified by an uplift vector $\vec{w}$ and 40 ambiguity vectors $\vec{v}^{\,(m)}$. The nonzero entries of the vector $\vec{w}$ read
\eqna{
w_{41}(z,\zb) &= 336\left(\hat{w}_{41}(z,\zb) - \hat{w}_{41}\Biggish(\frac{z}{z-1},\frac{\zb}{\zb-1}\Biggish) \right)\,,\\
w_{42}(z,\zb) &= 1+\hat{w}_{42}(z,\zb) - \hat{w}_{42}\Biggish(\frac{z}{z-1},\frac{\zb}{\zb-1}\Biggish) \,,
}[]
where we have introduced two functions which, calling $\hat{z} \equiv z+\zb$ and using $R_N$ from~\eqref{eq:rNdef}, read
\eqna{
\hat{w}_{41}(z,\zb) &= z \zb\hat{z}\lsp\bigl(z^2+\zb^2\bigr) + \frac83\lsp R_N\lsp z \zb\lsp (\hat{z}+1)\,,\\
\hat{w}_{42}(z,\zb) &= \frac12\lsp  z \zb \,\bigl( \hat{z}^3 -2\lsp \hat{z}^2 -2 \lsp z \zb\lsp(\hat{z}-1) \bigr) + \frac43\lsp R_N\lsp z \zb\lsp (\hat{z}-1)\,.
}[]
The ambiguity vectors instead can be expressed as follows
\eqn{
v^{(m)}_k (z,\zb) = \delta^m_k + \hat{v}^{(m)}_k (z,\zb)\,.
}[eq:vhatdef4444]
The expressions of the nonzero entries of $\hat{v}^{(m)}_k (z,\zb)$ are quite lengthy so we report them in appendix~\ref{app:supplement}. From the vector $\vec{w}$ we can obtain the conformal blocks expansion of the protected sector. Here is the result\footnote{$(\ell-1)_n = (\ell-1)\ell(\ell+1)\cdots (\ell+n-2)$ is the Pochhammer symbol.}
\eqna{
w_{41}(z,\zb) &= \sum_{\ell=1,\,\mathrm{odd}}^\infty\frac73\lsp\frac{2^{\ell+2}\lsp\ell!\lsp(\ell-1)!}{(2\ell-1)!}\left((\ell-2)_6 + 96\lsp R_N\lsp\big(\ell^2+\ell-1\big)\right)g_{\ell+2,\ell}\,,\\
w_{42}(z,\zb) &= g_{0,0} - \sum_{\ell=0,\,\mathrm{even}}^\infty\frac{2^\ell\lsp(\ell!)^2}{3\lsp(2\ell)!}\left(\frac{(\ell-1)_4\lsp(\ell^2+\ell+12)\lsp}{12} + 8\lsp R_N\lsp\big(\ell^2+\ell+1\big)\right)g_{\ell+2,\ell} \,,
}[]
with $R_N$ defined in~\eqref{eq:rNdef}.

Here, as in the $\langle\qBPS\qBPS\hBPS\hBPS\rangle$ case, we have to resolve an ambiguity. Indeed there is a tower of twist-two operators coming from the protected function $\gothf$. In order to resolve this ambiguity we impose that the twist-two contributions completely vanish and also that the OPE coefficient of $g_{2,0}$ in the $(0,2,0)$ is the same as the free theory one. Furthermore, we can make an assumption similar to the one made before. Namely, since structures $26$--$34$ are associated to the same representation, we can assume their contributions to the ambiguity to be proportional to each other. Actually, structures $28$ and $31$ can be set to zero altogether since their tensor structures $\TQQQQ{28}$ and $\TQQQQ{31}$ are purely imaginary under complex conjugation. Furthermore, if we assume also structure $34$ to be proportional to the others, the resolution of the ambiguity requires it to be set to zero, so we will directly omit $\CA_{34}$ from the ansatz below.\footnote{One can see from table~\ref{tab:freeExchangedQQQQ} that this structure is different as it exchanges operators from twist six onward.} This does not fix all the structures and leaves us with a large ambiguity. We will however see in subsection~\ref{sec:invSingle} that some OPE data can still be fixed unambiguously.

After the above remarks, we find that a minimal ansatz for the ambiguity reads as follows
\eqna{
\CA_{33}(z,\zb) &= a^{(33)}_{4,0}\,g_{4,0} + \sum_{\ell=2,\,\mathrm{even}}^\infty a^{(33)}_{\ell+4,\ell}\,g_{\ell+4,\ell}\,,\\
\CA_m(z,\zb) &= \lambda_m\,a^{(33)}_{4,0}\,g_{4,0} + \kappa_m\sum_{\ell=2,\,\mathrm{even}}^\infty \,a^{(33)}_{\ell+4,\ell}\,g_{\ell+4,\ell}\,,\qquad m = 26,\,27,\,29,\,30,\,32,\\
\CA_{37}(z,\zb) &= \frac43\frac{7!}{N^2-1}\,g_{2,0}\,.
}[eq:ambigAnsatzQQQQ]
for some constants $\kappa_m$ and $\lambda_m$ --- we have defined two independent sets of constants for the spin zero and spin $\ell\geq2$ part since the former does not follow the same pattern of the latter. The coefficients $a^{(m)}_{\Delta,\ell}$ that follow from imposing the points discussed in subsection~\ref{sec:ambiguity} are given by
\eqna{
a^{(33)}_{4,0} &= - \frac{143\cdot(5!)^2\lsp(N^2-3)^2}{\CD(\{\lambda_m\})\lsp (N^2-4)(N^2-1)}\,,\\
a^{(33)}_{4+\ell,\ell} &=-\frac{21450}{\CD(\{\kappa_m\})}\frac{2^\ell\lsp\ell! (\ell+1)!}{(2\ell+1)!}\left(
	3\cdot2^5 \big(\ell^2+3 \ell+1\big) \lsp R_N + (\ell-1)_6
\right)\,,\qquad \ell\geq2\,,
}[]
with
\eqn{
\CD(\{\kappa_m\}) = 1662 -19305 \lsp \kappa_{26}+77220\lsp  \kappa_{27}+77220 \lsp \kappa_{29}-64922\lsp \kappa_{30}-3198 \lsp \kappa_{32}\,.
}[]
As before, this expression can be resummed using the methods of appendix~\ref{app:resumming}, leading to 
\eqna{
\CA_{33}(z,\zb) &= a^{(33)}_{4,0}\,g_{4,0} + \frac{1}{\CD(\{\kappa_m\})} \Big[\hspace{-.7em}&&\frac{z \zb}{z-\zb}\Bigish(a(z,\zb) + R_N \, b(z,\zb) - a(\zb,z) - R_N \, b(\zb,z)\Bigish) \\&&&-1716\cdot(5!)^2\lsp R_N \log(1-z)\log(1-\zb)\Big]\,,
}[]
with
\eqna{
a(z,\zb) &= -21450\cdot(3!)^2\left( \hat{a}(\zb) + \hat{a}\Biggish(\frac{\zb}{\zb-1}\Biggish)\right)\log(1-z)\,,\\
b(z,\zb) &= -143\cdot(5!)^2\left(12 + \hat{b}(\zb) + \hat{b}\Biggish(\frac{\zb}{\zb-1}\Biggish)\right)\log(1-z)\,,\\
\hat{a}(z) &= \, z^4\,,\\
\hat{b}(z) &= \big(z+z^2\big)\,.
}[]
The requirements discussed in subsection~\ref{sec:ambiguity} also impose that $\CA_{40}$ exchanges no twist-two operators at all. The other components remain in principle arbitrary, including possibly contributions to $\CA_{26},\ldots,\CA_{34}$ that have twist higher than four and thus not proportional to $\CA_{33}$ found above.

Just like the discussion before equation~\eqref{eq:lambdasqTJ}, also here the OPE coefficients of the stress tensor and R-current get a simple multiplicative factor with respect to their free theory naive value
\twoseqn{
(\lambda_{\qBPS\qBPS J_{\mu}})^2 = \frac13\,(\lambda_{\qBPS\qBPS J_{\mu}}^{\mathrm{free}})^2 &= \frac{7\cdot2^8}{N^2-1}\,,
}[]{
(\lambda_{\qBPS\qBPS T_{\mu\nu}})^2 = \frac15\,(\lambda_{\qBPS\qBPS T_{\mu\nu}}^{\mathrm{free}})^2 &= \frac{32}{45(N^2-1)}\,,
}[][]
where the OPE coefficient of the current is taken from the structure \TQQQQ{41}, while the other structures in the $(1,0,1)$ do not have any twist-two contribution. As before, this is a consequence of carefully studying the constraints stemming from the ambiguity resolution. In this case the twist-two operators are not the only ones giving a singular contribution in the crossed channel, and therefore the anomalous dimensions may depend on other unknown terms. It is still true, however, that the large spin asymptotics is fixed and it depends on these factors of $1/3$ and $1/5$.

\section{Lorentzian inversion formula}\label{sec:Lorentz}

\subsection{Review of the Lorentzian inversion formula}
Let us consider again a generic four-point function of quarter-BPS and/or half-BPS operators as in \eqref{eq:4pt_generic}
\eqna{
\CG(z,\zb,\bfS_1,\,\ldots,\,\bfS_4) &=  \sum_{k=1}^{N_\mathrm{str}} g_k(z,\zb)\;\BBT_k(\bfS_1,\,\ldots,\,\bfS_4)\, \\
g_k(z, \zb)&=\sum_{\Delta, \ell}a_{k,\Delta, \ell}\lsp g_{\Delta, \ell}^{\Delta_{12}\Delta_{34}}(z, \zb)\, .
}[]
The Lorentzian inversion formula \cite{Caron-Huot:2017vep} allows us to extract the $s$-channel OPE data directly from the double discontinuity of the correlator in terms of the sum of two  functions, analytic
in  the spin, depending on the two other channels contributions
\twoseqn{
c_k (\Delta, \ell)&=c^t_k(\Delta, \ell)+(-1)^\ell c^u_k(\Delta, \ell)\, ,
}[cTot]{
c^t_k(\Delta, \ell)&=
\begin{aligned}[t]
&\frac{\tilde{\kappa}^{(\Delta_{12},\Delta_{34})}_{\Delta+\ell} (-2)^\ell}{2}\int_0^1 \frac{dz}{z^2}\frac{d\zb}{\zb^2} \left[(1-z)(1-\zb)\right]^{\frac{\Dt-\Do}{2}} 
\times \\ & \times 
\kappa_{4-2h}^{\Do, \Dt}(z)\lsp\kappa_{2h+2\ell}^{\Do, \Dt} (\zb)\,\text{dDisc}\mleft[\frac{\zb-z}{z\zb}g_k(z, \zb)\mright] \, ,
\end{aligned}
}[cT][]
where we have introduced $h=\frac{\Delta-\ell}{2}$, $\tilde{\kappa}^{(r,s)}_{\beta}=\frac{\Gamma\mleft(\frac{\beta+r}{2}\mright)\,\Gamma\mleft(\frac{\beta-r}{2}\mright)\,\Gamma\mleft(\frac{\beta+s}{2}\mright)\,\Gamma\mleft(\frac{\beta-s}{2}\mright)}{2\pi^2 \Gamma(\beta-1)\Gamma(\beta)}$ and
\eqn{
\kappa_\beta^{(r,s)}(z) = z^{\beta/2}\lsp {}_2F_1\mleft(\frac{\beta-r}{2},\frac{\beta+s}{2};\beta;z\mright)\,.
}[]
The function $c^u_k(\Delta, \ell)$ can be easily obtained by its $t$-channel counterpart by replacing $\Delta_1 \leftrightarrow \Delta_2$ and $g_k(z, \zb)$ with 
\eqn{
\left((1-z)(1-\zb)\right)^{-\frac{\Delta_{34}}2} (\mathbb{M}_{1\leftrightarrow 2}^\text{T})_{k k^\prime}\left[g_{k^\prime}\mleft(\frac{z}{z-1},\frac{\zb}{\zb-1}\mright)\right]_{\Delta_1 \leftrightarrow \Delta_2}\, ,
}[uCrossing]
where $\mathbb{M}_{i\leftrightarrow j}$ is the change of basis matrix between the original tensor structures and the one with the indices $i$ and $j$ exchanged
\eqn{
\mathbb{T}^{i\leftrightarrow j}_{k^ \prime}=\left(\mathbb{M}_{i\leftrightarrow j}\right)_{k^\prime k}\mathbb{T}_k \, .
}[eq:CrossingMatrices]
The function $c_k(\Delta, \ell)$ encodes the OPE data: it is constructed in such a way that for fixed integer spins it develops poles in correspondence of the dimensions of the exchanged operators, whose residues represent the OPE coefficients according to 
\eqn{
 c_k (\Delta, \ell)\underset{\Delta \to \Delta_{\mathrm{ex}}}{\sim}\frac{a_{k, \Delta, \ell}}{\Delta_{\mathrm{ex}}-\Delta}\, .
}[eq:Cfun]
Notice that the reconstruction of the OPE data through the inversion formula can be in general trusted only for spin $\ell \geq 2$ \cite{Caron-Huot:2017vep}. However, in a large $N$ expansion, higher orders can suffer of worse ambiguities thus invalidating the results for  a larger though finite number of low spins~\cite{Alday:2017vkk}.
The double discontinuity appearing in \eqref{cT} is defined as the difference between the Euclidean correlators and its two possible analytic continuations around $\zb=1$, namely
\begin{align}\begin{aligned}
\text{dDisc}\left[g_k(z, \zb)\right]&=\cos\left(\pi \alpha \right)g_k(z, \zb)-\frac{1}{2}e^{i\pi \alpha }g_k^\circlearrowleft(z, \zb)-\frac{1}{2}e^{-i\pi\alpha}g_k^\circlearrowright(z, \zb)\, ,\\ \alpha&=\frac{\Dt-\Do}{2} \, . \end{aligned}
\end{align}

Similarly to what was done in \cite{Alday:2017vkk, Caron-Huot:2018kta},  in order to reconstruct the $s$-channel OPE we would like to employ only the information coming from the protected contributions in the cross-channels and in our case encoded in $w_k$ of~\eqref{eq:decompositionwv} together with the associated ambiguity~\eqref{eq:ambiguityOnw}.  To do so,  we will not consider directly  the dDisc of the correlator as it is in \eqref{cT}, but we will use crossing symmetry to re-express $g_k(z, \zb)$ as
\eqn{
g_k(z, \zb)=\frac{(z \zb)^{\frac{\Delta_1+\Delta_2}{2}}}{\left((1-z)(1-\zb)\right)^{\frac{\Delta_2+\Delta_3}{2}}}(\mathbb{M}^T_{1 \leftrightarrow 3})_{k k^\prime} \left[g_{k^\prime} (1-z, 1-\zb) \right]_{\Delta_1 \leftrightarrow \Delta_3}\, .
}[]
The same transformation has to be performed in $c^u_k$.  At this stage, it is important to mention that  differently from the all half-BPS case,  we might not be able  to construct  the full tree-level correlator in this way. This is due to the fact that the only contribution we can completely fix is the one coming  from  twist-two operators and the identity, when present. Nonetheless we will try to draw some interesting conclusions about the anomalous dimensions and OPE coefficients of non-protected operators appearing in the OPE involving  our quarter-BPS operators.

Let us now briefly review how the inversion formula is applied to our cases of interest. First of all, the only terms with non vanishing dDisc that we are going to encounter are 
\begin{align}
\text{dDisc}\left[\left(\frac{1-\zb}{\zb}\right)^\lambda \right]=\text{dDisc}[e^{\lambda \log\left(\frac{1-\zb}{\zb}\right)}]=\left(\frac{1-\zb}{\zb}\right)^\lambda 2 \sin(\pi \lambda)\sin(\pi(\lambda+\alpha))\, .
\end{align}
Also $\log(1-\zb)$ appears in our expressions, but it is straightforward to see that its dDisc vanishes.  All the relevant integrals appearing in $c^t_k(\Delta, \ell)$ then are going to be of the form
\eqna{
&\int_0^1 \frac{d\zb}{\zb^2} (1-\zb)^{\frac{\Dt-\Do}{2}}\kappa_{2h+2\ell}^{\Do, \Dt} (\zb) \lsp \zb^{-\frac{\Dt}{2}} \text{dDisc}\left[\left(\frac{1-\zb}{\zb}\right)^\lambda\right] \times \\ & \qquad\qquad \qquad \int_0^1 \frac{d z}{z^2} (1-z)^{\frac{\Dt-\Do}{2}}\kappa_{4-2h}^{\Do, \Dt} (z) \lsp f(\boldsymbol{x}(\tilde{\lambda}), \boldsymbol{z} (\tilde{\lambda}), \log z)\,,
}[genericInt]
for some generic function $f$  depending on the variables $\boldsymbol{x}(\tilde{\lambda})=z^{-\frac{\Dt}{2}}\big(\frac{z}{1-z}\big)^{\tilde{\lambda}}$,  $  \boldsymbol{z}(\tilde{\lambda})=z^{-\frac{\Dt}{2}}z^{\tilde{\lambda}}$ and possible $\log z$.   Explicit computations of these integrals can be found in appendix \ref{app:InversionIntegrals}.  We will further consider  a large  central charge $c$ expansion for our correlators.  At $c\to \infty$ the dimensions and OPE coefficients of non-protected operators acquire corrections with the respect to their bare values of the form\footnote{Notice that we have slightly changed notation, from $a^{(k)}_{\Delta, \ell}$ to $a_{k, \Delta, \ell}$.}
\twoseqn{
\Delta_k &=\Delta^{(0)}_k+\frac{\gamma^{(1)}_{k, \Delta,\ell}}{c}+\mathcal{O}(c^{-2})\,,
}[]{
a_{k, \Delta, \ell}&=a_{k, \Delta, \ell}^{(0)}+\frac{a_{k, \Delta, \ell}^{(1)}}{c}+\mathcal{O}(c^{-2})\, ,
}[]
where we have defined the anomalous dimension $\gamma^{(1)}_{k, \Delta,\ell}$.  Accordingly, \eqref{eq:Cfun} gets expanded as
\eqn{
c_k(\Delta, \ell)\sim -\frac{1}{2} \bigg \langle \frac{a^{(0)}_{k, \Delta, \ell}}{h-\frac{\Delta^{(0)}-\ell}{2}}\bigg \rangle-\frac{1}{c}\Bigg(\frac{1}{4}  \bigg \langle \frac{a^{(0)}_{k, \Delta, \ell}\gamma^{(1)}_{k, \Delta, \ell}}{\big(h-\frac{\Delta^{(0)}-\ell}{2}\big)^{\mbox{\lnsp\tiny2}}}\bigg \rangle+\frac{1}{2} \bigg \langle \frac{a^{(1)}_{k, \Delta, \ell}}{h-\frac{\Delta^{(0)}-\ell}{2}}\bigg \rangle \Bigg)+\mathcal{O}(c^{-2})\, ,
}[largeCInversion]
where the brackets stand for averages over all the possible degenerate operators with the same twist.  From this formula it is clear that at order $c^0$ we expect  simple poles for the $h$ of the exchanged operators, whose OPE coefficients  should recover  the free theory results computed at $c\to \infty$. At the next order, double poles arise in correspondence of those operators developing an anomalous dimension. Notice that the information one can extract for this type of poles are not exactly the anomalous dimensions, but rather the products $\langle a^{(0)}_{k, \Delta, \ell}\gamma^{(1)}_{k, \Delta,\ell}\rangle$. Determining the $\gamma^{(1)}_{k, \Delta,\ell}$'s by themselves would indeed require solving a mixing problem at order $c^{0}$ able to distinguish all the possible degeneracies.  Finally the simple poles at order $c^{-1}$ take into account the corrections to the OPE coefficients.

\subsection{Half-BPS case}
In the same spirit of subsection~\ref{subsec:HalfBPSAmbiguity}, let us first review the well-known example of four half\nobreakdash-BPS $\mathcal{O}_2$ operators and let us apply the inversion formula to this correlator without relying on its expansion in superconformal blocks.  

The way to proceed is the following: we start by considering the large $c$ expansion of \eqref{w&AO2} and then we plug this in \eqref{cT} after passing to the crossed channel as explained before. The $u$-contribution can be  obtained similarly.  At order $c^0$ we should be able to recover the free theory coefficients and indeed we find that the exchanged operators have twist four and higher and the corresponding  $a^{(0)}_{k, \Delta, \ell}$ are exactly the expected ones. After this preliminary check, we can try to extract corrections at order $c^{-1}$. The anomalous dimensions arise from those terms in the correlator proportional to $\log z$ once passed to the $t$-channel. These are in fact the only contributions in \eqref{genericInt} that can develop double poles.  Let us denote $\Gamma^{k}_{h, \ell}=\langle a^{(0)}_{k, \Delta, \ell} \gamma^{(1)}_{k, \Delta, \ell}\rangle$, with $\Delta=2h+\ell$, for each tensor structure $k$ in \eqref{tensO2} and let us define 
\eqn{
\Xi_{t}^{(\Do, \Dt)}= \frac{ \Gamma \mleft( t-\frac{\Do}{2}\mright)\Gamma \mleft( t+\frac{\Dt}{2}\mright)}{4\Gamma \mleft(2t -1\mright)}\, ,  \qquad \qquad \Xi_{t}\equiv \Xi_t^{(0,0)} \, .
}[]
Then our results read
\eqnal{
\Gamma^{1}_{h, \ell} &= -\frac{4}{5} \lsp \Xi_{h+\ell} \Xi_{h-1}2^\ell(1+(\shortminus 1)^{\ell }) \left(h^4-6 h^3+27 h^2-54 h+\frac{136}{3}\right)  & \text{for }h\geq 2 \,, \ell\geq 4\, ,\\
\Gamma^{2}_{h, \ell} &= -4\lsp  \Xi_{h+\ell} \Xi_{h-1}2^\ell(1-(\shortminus  1)^{\ell }) \left(h^4-6 h^3+23 h^2-42 h+32\right)  & \text{for }h\geq 2 \,, \ell\geq 3 \, ,\\
\Gamma^{3}_{h, \ell} &= -\frac{8}{5}\lsp  \Xi_{h,+\ell} \Xi_{h-1}2^\ell(1+(\shortminus 1)^{\ell }) \left(h^4-6 h^3+21 h^2-36 h+\frac{80}{3}\right)  & \text{for }h\geq 2 \,,\ell\geq 2 \, ,\\
\Gamma^{4}_{h, \ell} &= -\frac{16}{3}\lsp  \Xi_{h+\ell} \Xi_{h-1}2^\ell(1+(\shortminus 1)^{\ell })(h-2) (h-1) \left(h^2-3 h+6\right) & \text{for }h\geq 3 \,,\ell\geq 2\, ,\\
\Gamma^{5}_{h, \ell} &= -8\lsp  \Xi_{h+\ell} \Xi_{h-1}2^\ell(1-(\shortminus 1)^{\ell })(h-2) (h-1) \left(h^2-3 h+4\right) & \text{for }h\geq 3 \,,\ell\geq 0\, ,\\
\Gamma^{6}_{h, \ell} &= -\frac{8}{3}\lsp  \Xi_{h+\ell} \Xi_{h-1}2^\ell(1+(\shortminus 1)^{\ell })(h-3) (h-2) (h-1) h & \text{for }h\geq 4 \, ,\ell\geq 0\, .\\
}[]
For all the other values of the twist the anomalous dimensions simply vanish, while for the remaining values of the spin our results do not agree with the ones of \cite{Dolan:2001tt}.\footnote{This is not an inconsistency since we recall that the inversion formula can fail for low spins.} Notice that the exchange of  either spin even or spin odd operators depends on the symmetry property of the corresponding tensor $\mathbb{T}_k$ under the exchange $1\leftrightarrow 2$.   To conclude, let us comment on the anomalous dimensions of the operators transforming in the $(0,4,0)$ representation of SU(4), also known as $\mathbf{105}$, for which we do expect to recover the well known results~---~see for example \cite{Alday:2017vkk,Caron-Huot:2018kta}. In this case we can write
\begin{align}
\frac{\langle a^{(0)}_{6, \Delta, \ell} \gamma^{(1)}_{6, \Delta, \ell} \rangle}{\langle a^{(0)}_{6, \Delta, \ell}\rangle}= -\frac{(h-3) (h-2) (h-1) h}{(\ell +1) (2 h+\ell -2)} \,  ,
\end{align}
which does agree with (2.13) of \cite{Alday:2017vkk} modulo shifting $h \to h+2$ and renaming $h=n+2$. At this same order the inversion formula also predicts corrections to the OPE coefficients. According to~\eqref{largeCInversion} these are encoded in the residue associated to the simple poles. Quite remarkably, they take a very simple form
\eqn{
\langle a_{k, \Delta, \ell}^{(1)}\rangle= \frac{1}{2} \partial_h \Gamma^k_{h, \ell}\,  ,
}[]
with the  exception of
\begin{align}
\begin{aligned}
\langle a_{4, \Delta, \ell}^{(1)}\rangle &=  -\frac{8}{3}\lsp  \Xi_{2+\ell} (1+(\shortminus 1)^{\ell }) && \text{ for } h=2\, , \ell \geq 2 \, ,\\
\langle a_{5, \Delta, \ell}^{(1)}\rangle &=  -2\lsp  \Xi_{2+\ell} (1-(\shortminus 1)^{\ell }) && \text{ for } h=2\, , \ell \geq 0 \, ,\\
\langle a_{6, \Delta, \ell}^{(1)}\rangle &=\begin{cases}
\frac{2}{3}\lsp  \Xi_{2+\ell} (1+(\shortminus 1)^{\ell }) \\
-  \Xi_{3+\ell} (1+(\shortminus 1)^{\ell })
\end{cases} && \begin{aligned}
\text{ for } h=2\, , \ell \geq 0\, , \\
\text{ for } h=3\, , \ell \geq 0 \, .\\ 
\end{aligned}
\end{aligned} 
\end{align}
\section{Results}\label{sec:Results}

\subsection[OPE data in the \texorpdfstring{$\hBPS\times\qBPS$}{2x4} OPE]{OPE data in the $\boldsymbol{\hBPS\times\qBPS}$ OPE}\label{sec:invMixed}

The OPE data in the $\hBPS \times \qBPS$ can be extracted by studying the correlator $\langle \hBPS \qBPS \qBPS \hBPS\rangle$.
However, the method of the inversion formula outlined above requires the knowledge of the protected parts of its $u$- and $t$-crossed version. Therefore also $\langle \qBPS \qBPS \hBPS \hBPS\rangle$ will be needed.  Explicitly, we have to consider
\eqna{
c^t_k(\Delta, \ell)&=\frac{\tilde{\kappa}^{(-2,2)}_{2h+2\ell} (-2)^\ell}{2}\int_0^1 \frac{dz}{z^2}\frac{d\zb}{\zb^2} \left[(1-z)(1-\zb)\right]^{2} \kappa_{4-2h}^{-2, 2}(z)\lsp \kappa_{2h+2\ell}^{-2, 2} (\zb)
\times \\ & \quad \; \times 
\text{dDisc}\mleft[\frac{\zb-z}{z\zb}  \frac{(z\zb)^3}{((1-z)(1-\zb))^4}  \left(\mathbb{M}_{\mathrm{III}\leftrightarrow \mathrm{II}}^\text{T}\right)_{k k^\prime} \mathcal{W}^t_{k^\prime}(1-z, 1-\zb) \mright] \, ,\\
\mathcal{W}^t_{k^\prime}(z, \zb)&=w_{k^\prime}^{\scriptscriptstyle  \langle \qBPS \qBPS \hBPS \hBPS\rangle}(z, \zb)+\mathcal{A}_{k^\prime}^{\scriptscriptstyle  \langle \qBPS \qBPS \hBPS \hBPS\rangle}(z, \zb) \, ,
}[corr4422]
where $\mathbb{M}_{\mathrm{III}\leftrightarrow \mathrm{II}}$ allows to pass from the \THQQH{k} to the   \TQQHH{k} basis.  Similarly
\eqna{
c^u_k(\Delta, \ell)&=\frac{\tilde{\kappa}^{(2,2)}_{2h+2\ell} (-2)^\ell}{2}\int_0^1 \frac{dz}{z^2}\frac{d\zb}{\zb^2} \left[(1-z)(1-\zb)\right]^{2} \kappa_{4-2h}^{2, 2}(z)\kappa_{2h+2\ell}^{2, 2} (\zb)
\times \\ & \quad \; \times 
\text{dDisc}\left[\frac{\zb-z}{z\zb}  \frac{(z\zb)^3}{((1-z)(1-\zb))^3}  \left(\mathbb{M}_{1\leftrightarrow 3}^\text{T}\right)_{k k^\prime} \mathcal{W}^u_{k^\prime}(1-z, 1-\zb) \right] \, ,\\
\mathcal{W}^u_{k^\prime}(z, \zb)&=(1-z)(1-\zb)\lsp w_{k^\prime}^{\scriptscriptstyle  \langle \hBPS \qBPS \qBPS \hBPS\rangle}\mleft(\frac{z}{z-1}, \frac{\zb}{\zb-1}\mright) \, ,
}[]
where in the third line we have used \eqref{uCrossing} with $\mathbb{M}_{1\leftrightarrow 2}=\mathds{1}$. Interestingly, by tracing back the contributions to $c^{t,u}_k(\Delta, \ell)$ from $\mathcal{W}^{t, u}_k(z, \zb)$, one discovers that the only operators that contribute are the twist-two terms and the identity in $\langle \qBPS \qBPS \hBPS \hBPS\rangle$, which we were able to completely fix.  This consequently implies that there is no dependence on the undetermined constant~$\kappa$.

Retracing the same steps of the all $\hBPS$ case, we have first of all checked that we were able to recover the disconnected OPE coefficients as given in the free theory.\footnote{As discussed in appendix \ref{app:InversionIntegrals}, the inversion integrals contributing to $c^u_k$ are not well defined for $h=1,2$, in these cases we have resorted to the standard conformal block expansion to fix the $a^{(0)}_{k, \Delta, \ell}$.}  Then we can pass to the study of anomalous dimensions.  These can arise only from \eqref{corr4422} since it is the only piece containing  $\log z$ terms once passed to the crossed channel.  Excluding the structures 3,5 and 9,  which simply vanish (see table~\ref{tab:freeExchangedHQQH}), we find
\eqnal{
\Gamma^{1}_{h, \ell} &= -\frac{4}{3}(\shortminus 2)^{ \ell} \lsp \Xi_{h+\ell}^{(-2,2)} \Xi_{h-1}^{(-2,-2)}(h-3)^2 h^2 && \text{for }h\geq 4 \, ,\\
\Gamma^{2}_{h, \ell} &= -\frac{8}{3}(\shortminus 2)^\ell\lsp \Xi_{h+\ell}^{(-2,2)} \Xi_{h-1}^{(-2,-2)} (h-3) h \left(h^2-3 h+5\right) && \text{for }h\geq 4 \, ,\\
\Gamma^{4}_{h, \ell} &= -\frac{4}{3}(\shortminus 2)^\ell\lsp \Xi_{h+\ell}^{(-2,2)} \Xi_{h-1}^{(-2,-2)}(h-3) h \left(h^2-3 h+8\right) && \text{for }h\geq 4 \, ,\\
\Gamma^{6}_{h, \ell} &= -\frac{4}{3}(\shortminus 2)^\ell\lsp \Xi_{h+\ell}^{(-2,2)} \Xi_{h-1}^{(-2,-2)} \left(h^2-4 h+6\right) \left(h^2-2 h+3\right) && \text{for }h\geq 3\,  ,\\
\Gamma^{7}_{h, \ell} &= -8(\shortminus 2)^\ell\lsp \Xi_{h+\ell}^{(-2,2)} \Xi_{h-1}^{(-2,-2)}(2 h^4-12 h^3+38 h^2-60 h+25) && \text{for }h\geq 3 \, ,\\
\Gamma^{8}_{h, \ell} &= -8(\shortminus 2)^\ell\lsp \Xi_{h+\ell}^{(-2,2)} \Xi_{h-1}^{(-2,-2)} (h^4-6 h^3+21 h^2-36 h+21)&& \text{for }h\geq 3\,  ,\\
\Gamma^{10}_{h, \ell} &= -\frac{40}{3}(\shortminus 2)^\ell\lsp \Xi_{h+\ell}^{(-2,2)} \Xi_{h-1}^{(-2,-2)} (2 h^4-12 h^3+46 h^2-84 h+63) && \text{for }h\geq \, 3\,  .
}[]

Among all the possible representations exchanged in the OPE,  the $(1,2,1)$ seems rather special. Indeed $c^u_6(\Delta, \ell)$ vanishes at any order in the $1/c$ expansion and as a consequence all the OPE data we can extract from the inversion formula will depend only on the $\langle \qBPS \qBPS \hBPS \hBPS \rangle$ correlator and thus take a very simple form
\eqn{ \frac{\langle a^{(0)}_{6, \Delta, \ell} \gamma^{(1)}_{6, \Delta, \ell}\rangle}{\langle a^{(0)}_{6, \Delta, \ell} \rangle}=-\frac{\left(h^2-4 h+6\right) \left(h^2-2 h+3\right)}{(\ell +1) (2 h+\ell -2)}\, , \qquad h\geq 3\,.
}[]
This observation suggests that this representation might be interpreted  as the analogous of the $(0,4,0)$ in the all $\hBPS$ case. Namely a representation where the superconformal block\footnote{By superconformal block in this context one usually means the familiar $u^{-2}g_{\Delta+4,\ell}$ that appears in the unprotected part of the $\hBPS$ correlator. Here we are talking about the same object but expanded in the full four-point function. In the all $\CO_2$ case, all representations but the $(0,4,0)$ have superblocks consisting of sums of conformal blocks, each associated to a different superdescedant. The $(0,4,0)$ is the only representation where there is only a single superdescendant contribution, namely $Q^4\Qb{}^4\CO$, which has the same spin as the superprimary and the dimension is increased by four.\label{foot:040note}} can be written as a \emph{single} conformal block, possibly with shifted quantum numbers.

Let us now analyze the correction to the OPE coefficients
\eqnal{
\langle a^{(1)}_{1, \Delta, \ell}\rangle &=(\shortminus 2)^\ell\Xi^{(-2,2)}_{h+\ell}\Xi^{(-2,2)}_{h-1}\begin{cases}
-\frac{4 (\shortminus 1)^{\ell } (\ell +1) (\ell +4)}{(\ell +2) (\ell +3)}  \\
-\frac{4 (-1)^{\ell } (\ell +1) (\ell +6)}{3 (\ell +3) (\ell +4)}-\frac{64}{3}+\frac{1}{2}\partial_h \Gamma^1_{h, \ell}\\
-\frac{2 (-1)^{\ell } (\ell +1) (\ell +8)}{3 (\ell +4) (\ell +5)}-\frac{136}{3} +\frac{1}{2}\partial_h \Gamma^1_{h, \ell} \\
8 (-1)^{\ell } \left(\frac{1}{(h+\ell -1) (h+\ell )}-\frac{1}{(h-2) (h-1)}\right)+\frac{1}{2}\partial_h \Gamma^1_{h, \ell}  
\end{cases} && \begin{array}{ll}
h =3\\  [0.35em]
h =4 \\[0.35em]
h =5\\[0.35em]
h  \geq 6
\end{array}
\\
\langle a^{(1)}_{2, \Delta, \ell}\rangle &=(\shortminus 2)^\ell\Xi^{(-2,2)}_{h+\ell}\Xi^{(-2,2)}_{h-1}\begin{cases}
\frac{12 (-1)^{\ell } (\ell +1) (\ell +4)}{(\ell +2) (\ell +3)}-20 \\
\frac{4 (-1)^{\ell } (\ell +1) (\ell +6)}{(\ell +3) (\ell +4)}-\frac{188}{3}+\frac{1}{2}\partial_h \Gamma^1_{h, \ell} \\
\frac{2 (-1)^{\ell } (\ell +1) (\ell +8)}{(\ell +4) (\ell +5)}-\frac{272}{3}+\frac{1}{2}\partial_h \Gamma^1_{h, \ell} \\
-24 (-1)^{\ell }\left( \frac{1}{(h +\ell -1) (h +\ell )} -\frac{1}{(h-2) (h-1)} \right)+\frac{1}{2}\partial_h \Gamma^1_{h, \ell} 
\end{cases}
&& \begin{array}{ll}
h =3\\  [0.35em]
h =4 \\[0.35em]
h =5\\[0.35em]
h  \geq 6
\end{array}\\
\langle a^{(1)}_{4, \Delta, \ell}\rangle &=(\shortminus 2)^\ell\Xi^{(-2,2)}_{h+\ell}\Xi^{(-2,2)}_{h-1}\begin{cases}
-\frac{12 (-1)^{\ell } (\ell +1) (\ell +4)}{(\ell +2) (\ell +3)}-16 \\
-\frac{4 (-1)^{\ell } (\ell +1) (\ell +6)}{(\ell +3) (\ell +4)}-\frac{112}{3}+\frac{1}{2}\partial_h \Gamma^1_{h, \ell} \\
-\frac{2 (-1)^{\ell } (\ell +1) (\ell +8)}{(\ell +4) (\ell +5)}-\frac{136}{3}+\frac{1}{2}\partial_h \Gamma^1_{h, \ell} \\
24 (-1)^{\ell }\left( \frac{1}{(h +\ell -1) (h +\ell )} -\frac{1}{(h-2) (h-1)} \right)+\frac{1}{2}\partial_h \Gamma^1_{h, \ell} 
\end{cases}
&& \begin{array}{ll}
h =3\\  [0.35em]
h =4 \\[0.35em]
h =5\\[0.35em]
h  \geq 6
\end{array}\\
\langle a^{(1)}_{6, \Delta, \ell}\rangle &=(\shortminus 2)^\ell\Xi^{(-2,2)}_{h+\ell}\Xi^{(-2,2)}_{h-1}\begin{cases}
-16 +\frac{1}{2}\partial_h \Gamma^1_{h, \ell} \\
-\frac{112}{3}+\frac{1}{2}\partial_h \Gamma^1_{h, \ell} \\
-\frac{136}{3}+\frac{1}{2}\partial_h \Gamma^1_{h, \ell} \\
+\frac{1}{2}\partial_h \Gamma^1_{h, \ell} 
\end{cases}
&& \begin{array}{ll}
h =3\\  [0.35em]
h =4 \\[0.35em]
h =5\\[0.35em]
h  \geq 6
\end{array}\\
\langle a^{(1)}_{7, \Delta, \ell}\rangle &=(\shortminus 2)^\ell\Xi^{(-2,2)}_{h+\ell}\Xi^{(-2,2)}_{h-1}\begin{cases}
-\frac{108 (-1)^{\ell } (\ell +1) (\ell +4)}{(\ell +2) (\ell +3)}-240 +\frac{1}{2}\partial_h \Gamma^1_{h, \ell} \\
-\frac{36 (-1)^{\ell } (\ell +1) (\ell +6)}{(\ell +3) (\ell +4)}-496+\frac{1}{2}\partial_h \Gamma^1_{h, \ell}\\
-\frac{18 (-1)^{\ell } (\ell +1) (\ell +8)}{(\ell +4) (\ell +5)}-544+\frac{1}{2}\partial_h \Gamma^1_{h, \ell} \\
216 (-1)^{\ell }\left( \frac{1}{(h +\ell -1) (h +\ell )} -\frac{1}{(h-2) (h-1)} \right)+\frac{1}{2}\partial_h \Gamma^1_{h, \ell} 
\end{cases}
&& \begin{array}{ll}
h =3\\  [0.35em]
h =4 \\[0.35em]
h =5\\[0.35em]
h  \geq 6
\end{array}\\
\langle a^{(1)}_{8, \Delta, \ell}\rangle &=(\shortminus 2)^\ell\Xi^{(-2,2)}_{h+\ell}\Xi^{(-2,2)}_{h-1}\begin{cases}
\frac{36 (-1)^{\ell } (\ell +1) (\ell +4)}{(\ell +2) (\ell +3)}-144+\frac{1}{2}\partial_h \Gamma^1_{h, \ell} \\
\frac{12 (-1)^{\ell } (\ell +1) (\ell +6)}{(\ell +3) (\ell +4)}-272+\frac{1}{2}\partial_h \Gamma^1_{h, \ell} \\
\frac{6 (-1)^{\ell } (\ell +1) (\ell +8)}{(\ell +4) (\ell +5)}-272+\frac{1}{2}\partial_h \Gamma^1_{h, \ell} \\
-72(-1)^{\ell }\left( \frac{1}{(h +\ell -1) (h +\ell )} -\frac{1}{(h-2) (h-1)} \right)+\frac{1}{2}\partial_h \Gamma^1_{h, \ell} 
\end{cases}
&& \begin{array}{ll}
h =3\\  [0.35em]
h =4 \\[0.35em]
h =5\\[0.35em]
h  \geq 6
\end{array}\\
\langle a^{(1)}_{10, \Delta, \ell}\rangle &=(\shortminus 2)^\ell\Xi^{(-2,2)}_{h+\ell}\Xi^{(-2,2)}_{h-1}\begin{cases}
-\frac{108 (-1)^{\ell } (\ell +1) (\ell +4)}{(\ell +2) (\ell +3)}-560+\frac{1}{2}\partial_h \Gamma^1_{h, \ell}\\
-\frac{36 (-1)^{\ell } (\ell +1) (\ell +6)}{(\ell +3) (\ell +4)}-\frac{2960}{3}+\frac{1}{2}\partial_h \Gamma^1_{h, \ell} \\
-\frac{18 (-1)^{\ell } (\ell +1) (\ell +8)}{(\ell +4) (\ell +5)}-\frac{2720}{3}+\frac{1}{2}\partial_h \Gamma^1_{h, \ell} \\
216(-1)^{\ell }\left( \frac{1}{(h +\ell -1) (h +\ell )} -\frac{1}{(h-2) (h-1)} \right)+\frac{1}{2}\partial_h \Gamma^1_{h, \ell} 
\end{cases}
&& \begin{array}{ll}
h =3\\  [0.35em]
h =4 \\[0.35em]
h =5\\[0.35em]
h  \geq 6
\end{array}\\
}[]
Notice again the simplicity of the OPE coefficient associated to the $(1,2,1)$ representation.
\subsection[OPE data in the \texorpdfstring{$\qBPS\times\qBPS$}{4x4} OPE]{OPE data in the $\boldsymbol{\qBPS\times\qBPS}$ OPE}\label{sec:invSingle}

The analysis for the OPE data contained in the $\qBPS\times\qBPS$ OPE proceeds analogously,  with the  difference that in this case we will consider only  $\langle \qBPS \qBPS \qBPS \qBPS \rangle$. The relevant ingredients are given in subsection~\ref{subsec:WI4444}.  Another important difference with respect to the previous case is that the twist-two operators and the identity are not the only ones contributing to the dDisc,  at order $c^{-1}$. This makes the answer not completely fixed and  it introduces a dependence on the unknown coefficients $\kappa_m$, $\lambda_m$ of equation~\eqref{eq:ambigAnsatzQQQQ}.  Remarkably, these constants do not appear in $c_1(\Delta, \ell)$: the function corresponding to the structure $(4,0,4)$.  For this reason and to avoid clutter, we will report explicitly  the anomalous dimension and $a^{(1)}_{k, \Delta, \ell}$ just for this interesting case.\footnote{The results for all the other structures can be found in an ancillary \textsl{Mathematica} file attached to the arXiv version of this paper.}  We find that, starting at twist $\tau= 10$ ($h\geq 5$) the operators exchanged in $\qBPS\times\qBPS$ transforming in the $(4,0,4)$ acquire an anomalous dimension
\eqna{
&\langle a^{(0)}_{1, \Delta, \ell} \gamma^{(1)}_{1, \Delta, \ell}\rangle=-\frac{1}{9} (-2)^{\ell } \left(1+(\shortminus 1)^{\ell }\right)  \Xi _{h+\ell }\Xi _{h-1} (h-4)_6 \Big( (h+\ell-2)_4 (h^2-3 h-2)\\ & \qquad \quad \;\; \;  -\frac{2}{3} (16 (h+\ell -1) (h+\ell )+(h-5) (h+2) (3 (h+\ell -1) (h+\ell )-2)-8)\Big)  \, .
}[]
The first term corresponds to the contributions coming from the identity and the twist-two operators and it is the leading one at large spin. The second term, on the other hand, may receive corrections from twist-four operators that we could not fix in the ambiguity --- for instance the scalar of dimension four in the $(0,4,0)$, or~\TQQQQ{13}. For this reason we quote the large spin asymptotic of the above result after having factorized away the free theory OPE coefficient
\eqn{
\frac{\langle a^{(0)}_{1, \Delta, \ell}\gamma^{(1)}_{1, \Delta, \ell}\rangle}{\langle a^{(0)}_{1, \Delta, \ell}\rangle} \xrightarrow{\ell \to \infty}-\frac{(h-4) (h+1) \left(h^2-3 h-2\right)}{\ell ^2} \qquad \text{for }h\geq 5 \, .
}[eq:a0gammaQQQQ]

\subsection{A note about the MRV limit}

The simplicity of the result~\eqref{eq:a0gammaQQQQ}, and the fact that it does not depend on the $\kappa_m$, $\lambda_m$ constants, suggests that the representation $(4,0,4)$ is the analog of the $(0,4,0)$ in the all $\hBPS$ case, in the sense explained around footnote~\ref{foot:040note}. Also note that the anomalous dimensions start from twist ten, contrary to the ones in the other structures that start from twist eight, which is the twist that one expects for a $[\qBPS\lsp\qBPS]_{0,\ell}$ operator. This is indicative of a shift in the dimension of the conformal block. Roughly speaking, this means that we expect the superconformal block in the structure $\TQQQQ1$ to be proportional to $g_{\Delta+2,\ell}$.

The analogy with the all half-BPS case can be made even more precise here. Indeed we can devise an analog of the Maximally R-symmetry Violating (MRV) limit introduced in~\cite{Alday:2020lbp}. In the $s$-channel this limit consists in sending $y_1\to y_2$. In our more general case we have
\eqn{
y_1\to y_2\,,\qquad S_1 \to S_2\,,\qquad \Sb_1 \to \Sb_2\,.
}[eq:MRV]
Taking this limit on the tensor structures $\TQQQQ{k}$ yields
\eqn{
\lim_{\mathrm{1\to2}}\TQQQQ{k} = \delta_{k1}\, (\strSSb{23})^2(\strSSb{32})^2(\strSSb{24})^2(\strSSb{42})^2\,,
}[]
namely it sends all structures to zero but the first, which is the one we are considering here. Even though we did not study these correlators in Mellin space, analogously to the findings of~\cite{Alday:2020lbp} we expect a zero in $u$ at twist eight when taking the $u$-channel limit (which is $1\to3$ instead of $1\to2$). Note that here we expect only one zero instead of two zeros.

We further note that the same reasoning will not work for the $\langle\hBPS\qBPS\qBPS\hBPS\rangle$ because no particular structure is isolated by taking the MRV limit. In that case however --- as we have seen in section~\ref{sec:invMixed} --- the anomalous dimensions start from twist six anyway, so we do not expect any additional zeros in Mellin space. Finally, taking the MRV limit~\eqref{eq:MRV} on $\langle\qBPS\qBPS\hBPS\hBPS\rangle$ precisely isolates the $(0,4,0)$ representation. This is consistent with one's expectations since the long multiplets in that OPE will be the same as those exchanged in the all half-BPS case.

\section{Double trace correlators from higher-point Witten diagrams}\label{sec:fivepoint}

\subsection{Five-point function and the OPE}

In this section we want to prove that the correlator $\langle \qBPS \hBPS\hBPS\hBPS\rangle$ is protected at lowest nontrivial order in the $1/N$ expansion.  This is not surprising since this correlator is ``next-to-extremal'' but there is no available proof of it yet because the usual arguments only apply to the half-BPS case~\cite{Eden:2000gg, Erdmenger:1999pz, Baggio:2012rr}. In order to achieve that, we will look at the five-point function $\langle\hBPS\hBPS\hBPS\hBPS\hBPS\rangle$. The procedure will involve two steps: first we need to decompose the five-point function into tensor structures that correspond to the exchange of a given representation between the first two operators and secondly we need to take the OPE limit and extract the most singular piece. Since, in the interacting theory, $\qBPS$ is the lightest operator in the $(2,0,2)$ sector of the $\hBPS\times\hBPS$ OPE, the leading singularity as $x_1\to x_2$ is guaranteed to give us the correlator that we need.

The five-point function was studied up to order $1/N^3$ in the supergravity limit in a recent paper~\cite{Goncalves:2019znr}. Let us report here their results for convenience. The free theory value at finite~$N$ reads
\eqna{
&\langle \hBPS(x_1,y_1)\ldots \hBPS(x_5,y_5)\rangle\big|_{\mathrm{free}} \equiv \CG^{(5)}_{\mathrm{free}}(x_1,\ldots,x_5;y_1,\ldots,y_5)=\\
&\qquad= 2 \sqrt{\frac{2}{N^2-1}}\Bigg[ \sum_{\mathrm{perm}} \frac{\CA_{ijk,lm}}{x_{ij}^2x_{jk}^2x_{ki}^2(x_{lm}^2)^2} + \frac{2}{N^2-1} \sum_{\mathrm{perm}} \frac{\CA_{ijklm}}{x_{ij}^2x_{jk}^2x_{kl}^2x_{lm}^2x_{mi}^2} \Bigg]\,,
}[]
where we defined
\eqn{
\CA_{ijk,lm} = \stry{ij}\lsp\stry{jk}\lsp\stry{ki}\lsp\stry{lm}^2\,,\qquad
\CA_{ijklm} = \stry{ij}\lsp\stry{jk}\lsp\stry{kl}\lsp\stry{lm}\lsp\stry{mi}\,,
}[]
and the sums $\sum_{\mathrm{perm}}$ are over all inequivalent permutations of five elements, each counted once. The large $N$ result in the supergravity approximation (namely the 't Hooft coupling $\lambda$ being sent to infinity) instead is given by
\eqn{
\langle \hBPS(x_1,y_1)\cdots \hBPS(x_5,y_5)\rangle\big|_{\mathrm{sugra}} =  \CG^{(5)}_{\mathrm{free}}\big|_{1\over N} + \frac{1}{N^3}\CG^{(5)}_{\mathrm{sugra}}\,,
}[]
where $\CG^{(5)}_{\mathrm{sugra}}$ is given in an auxiliary file of~\cite{Goncalves:2019znr} and is a linear combination of the following $D$-functions
\eqn{
\CG^{(5)}_{\mathrm{sugra}} \supset \big\{D_{11123}\,,\;D_{11233}\,,\;D_{11112}\,,\;D_{11222}\,,\;\mathrm{permutations}\big\}\,.
}[eq:Dset5pf]
The $D$-functions with $n$ labels represent an $n$-point contact Witten diagram or, in other words, the integral of $n$ bulk-to-boundary propagators
\eqn{
D_{\Delta_1\cdots\Delta_n}(x_1,\ldots,x_n) = \int_0^\infty\frac{\di z}{z^5}\int_{\BBR^4}\di^4x\,\prod^n_{i=1}\left(\frac{z}{z^2+(\vec{x}-\vec{x}_i)^2}\right)^{\Delta_i}\,.
}[eq:Ddef]

The four-point function we want to match is shown at zero coupling in equation~\eqref{eq:QHHH4pf}. Since we only need to extract the most singular piece in the OPE we need only the following term
\eqn{
\hBPS(x_1,y_1)\times\hBPS(x_2,y_2) \big|_{(2,0,2)} = \lambda_{\hBPS\hBPS\qBPS}\,\big(\strSYSb{\partial_3}{12}{\partial_3}\big)^2 \qBPS(x_2,S_3,\Sb_3) + O\big(x_{12}^2\big)\,.
}[]
Here we introduced a new notation: whenever a symbol $\partial_i$ replaces a label $i$ in a tensor structure --- defined in~\eqref{eq:buildingBlocks} --- we need to replace the polarization attached to that point with a covariant derivative --- defined in~\eqref{eq:diffop}
\eqn{
i \to \partial_i\quad\equiv\quad y_{i\llsp M} \to \CD_{i\lsp M}\,,\qquad S_i^m \to \bar\partial^m_i\,,\qquad S_{i\lsp m} \to \partial_{i\lsp m}\,.
}[]
As we remarked before, the differential operators defined in~\eqref{eq:diffop} can only be applied to $(q,0,\qb)$ or $(0,p,0)$ tensors. As we will explain in the next subsection, it is unnecessary to know their explicit expressions anyway because we can always adopt the Casimir method as in subsection~\ref{sec:casimir}.

\subsection{Projecting on R-symmetry representations}

In order to find our operators of interest we need to decompose the tensor structures $\CA_{ijklm}$ and $\CA_{ijk,lm}$ into a basis of projectors associated to the representations exchanged in the $\hBPS\times\hBPS$ OPE. Here it follows a systematic way to do it. In general the OPE can be written as
\eqn{
\hBPS(x_1,y_1)\times\hBPS(x_2,y_2)= \sum_{\CO_L}\lambda_{\hBPS\hBPS\CO_L}\,\frac{t^{\mathsf{OPE}}_{\CO_L}(y_1,y_2,\boldsymbol{\partial}_{\bfS_0})}{(x_{12}^2)^{\frac12(4-\Delta_L)}} \CO_L(x_2,\bfS_0) + O(x_{12}^2)\,,
}[]
with $t^{\mathsf{OPE}}_{\CO_L}$ being a differential operator in the polarizations $\bfS_0$ and the sum ranging over all lightest operators of dimension $\Delta_L$ within a given representation. If we use this equation inside a five-point function we obtain
\eqna{
&\lim_{x_1\to x_2}(x_{12}^2)^{\frac12(4-\Delta_L)}\langle\hBPS(x_1,y_1)\hBPS(x_2,y_2)\hBPS\hBPS\hBPS\rangle \\&\qquad=
\sum_{\CO_L}\lambda_{\hBPS\hBPS\CO_L}\,t^{\mathsf{OPE}}_{\CO_L}(y_1,y_2,\boldsymbol{\partial}_{\bfS_0}) \langle \CO_L(x_2,\bfS_0)\hBPS\hBPS\hBPS\rangle\,.
}[]
Since $t^{\mathsf{OPE}}_{\CO_L}$ are essentially three-point functions, they satisfy the Casimir equation in the first two points
\eqn{
\CC_2(\partial_{\bfS_1},\partial_{\bfS_2})\,t^{\mathsf{OPE}}_{\CO_L}(y_1,y_2;\partial_{\bfS_0})= C_2\,t^{\mathsf{OPE}}_{\CO_L}(y_1,y_2;\partial_{\bfS_0})\,.
}[]
Therefore we can expand the five-point function $\langle\hBPS\hBPS\hBPS\hBPS\hBPS\rangle$ into an arbitrary basis of monomials in $\stry{ij}$ and then separate the various contributions by rotating into a basis which diagonalizes $\CC_2$. In this particular case the quadratic Casimir is sufficient to fix all the structures. For the representations in which $pq=0$ we can alternatively apply the differential operator directly on the four-point tensor structures. For example, considering $(2,0,2)$ one has\footnote{The numerical factor in front is arbitrary. It was chosen to factor an overall constant multiplying~\eqref{eq:Ei202}.}
\eqn{
t_{(2,0,2)}^{\mathsf{OPE}}(y_1,y_2,\partial_{S_0},\partial_{\Sb_0}) = \frac1{2^93^2}\big(\strSYSb{\partial_0}{12}{\partial_0}\big)^2\,,
}[eq:tOPE202def]
which should act on the three different four-point structures of $\langle\qBPS\hBPS\hBPS\hBPS\rangle$. Here we pick a simpler linear combination of the $\TQHHH{k}$ of~\eqref{eq:TQHHHDef} since we do not care about the representations exchanged in the rest of the four-point function --- note that we shifted the point labels by one to the \hypertarget{eqref:TQHHHtildeDef}{right}
\eqn{
\TQHHHtilde1 = \strSYSb{0}{34}{0}\, \strSYSb{0}{35}{0}\,\stry{45}\,,\qquad
\TQHHHtilde2 = \strSYSb{0}{34}{0}\, \strSYSb{0}{45}{0}\,\stry{35}\,,\qquad
\TQHHHtilde3 = \strSYSb{0}{35}{0}\, \strSYSb{0}{45}{0}\,\stry{34}\,.
}[]
Calling $\BBE_i^{(2,0,2)}$ the resulting five-point structures, we have
\eqna{
\BBE_1^{(2,0,2)} \equiv t_{(2,0,2)}^{\mathsf{OPE}}\TQHHHtilde1 &= 
10 \CA_{123,45}+20 \CA_{145,23}+20 \CA_{245,13}+2 \CA_{345,12}-5 \CA_{12345}\\&\;\quad-5 \CA_{12354}-5 \CA_{12453}-5 \CA_{12543}-20 \CA_{13245}-20 \CA_{13254}
\,,\\
\BBE_2^{(2,0,2)} \equiv t_{(2,0,2)}^{\mathsf{OPE}}\TQHHHtilde2 &=
-10 \CA_{124,35}-20 \CA_{135,24}-20 \CA_{235,14}-2 \CA_{345,12}+5 \CA_{12354}\\&\;\quad+5 \CA_{12435}+5 \CA_{12453}+5 \CA_{12534}+20 \CA_{13524}+20 \CA_{14235}
\,,\\
\BBE_3^{(2,0,2)} \equiv t_{(2,0,2)}^{\mathsf{OPE}}\TQHHHtilde3 &=
10 \CA_{125,34}+20 \CA_{134,25}+20 \CA_{234,15}+2 \CA_{345,12}-5 \CA_{12345}\\&\;\quad-5 \CA_{12435}-5 \CA_{12534}-5 \CA_{12543}-20 \CA_{13425}-20 \CA_{14325}
\,.\\
}[eq:Ei202]
Similar computations can be done for the other representations. Once we have all of them, we can expand the free correlator as follows
\eqn{
\CG^{(5)}_{\mathrm{free}} = \sum_{\substack{2q+p\leq4,\\p\,\,\mathrm{even}}}\sum_k g_k^{(q,p,q)}(x_1,\ldots,x_5)\,\BBE_k^{(q,p,q)}(y_1,\ldots,y_5)\,.
}[]
\subsection[Remarks for finite \texorpdfstring{$N$}{N}]{Remarks for finite $\boldsymbol{N}$}\label{sec:check5point}

Before showing the computation for $\langle\qBPS\hBPS\hBPS\hBPS\rangle$ at large $N$, let us make a few remarks about the free theory at finite $N$. At zero coupling, the operator $\qBPS$ is not the only one appearing in the leading OPE singularity: there is also the $Q^2\Qb{}^2$ descendant of the free Konishi $K=\tr\lsp \varphi^2  $.\footnote{Note that this problem would also arise in perturbation theory at weak coupling, but not at large $N$ because the Konishi operator becomes heavy in the supergravity approximation.} This superdescendant, in terms of free fields, is precisely the pure single trace contraction that we ignored when constructing the operator~$\qBPS$ in subsection~\ref{sec:qBPSdef}. Among the operators exchanged in the $\hBPS\times\hBPS$ OPE there are no other supermultiplets that can have a dimension\nobreakdash-four scalar transforming in the $(2,0,2)$.

It follows that we should expect to find some (numerical) coefficients $\alpha_1$, $\alpha_2$ and $\alpha_3$ such that the following limit holds\footnote{Since $\Delta_L=4$ the power of $x_{12}^2$ vanishes.}
\eqna{
&\lim_{x_1\to x_2}\;\sum_{k=1}^3\alpha_k\lsp \TQHHHtilde{k}(y_1,\ldots,y_5) \lsp g_i^{(2,0,2)}(x_1,\ldots,x_5) =\\&\qquad= \lambda_{\hBPS\hBPS\qBPS}\, \langle\qBPS(x_2,\bfS_2)\hBPS(x_3,y_3)\hBPS(x_4,y_4)\hBPS(x_5,y_5)\rangle\\&\qquad\;\quad+
\lambda_{\hBPS\hBPS (Q^2\Qb{}^2\lnsp K)}\, \langle(Q^2\Qb{}^2\lnsp K)(x_2,\bfS_2)\hBPS(x_3,y_3)\hBPS(x_4,y_4)\hBPS(x_5,y_5)\rangle
\,,
}[]
where the four-point functions on the right hand side are computed by Wick contractions. The first one is given explicitly in~\eqref{eq:QHHH4pf}. The OPE coefficient of $\qBPS$ is known from~\cite{Dolan:2004iy} and the one of the Konishi can also be inferred from the $\hat{A}_{00,10}$ of~\cite{Dolan:2004iy}, although we recomputed it by using $\qBPS^{(\mathrm{st,1})}$ of subsection~\ref{sec:qBPSdef} for self-consistency. They read\footnote{The first differs from~\cite{Dolan:2004iy} by a factor of $3/4$ due to the normalization of the four-point tensor structures of $\langle\hBPS\hBPS\hBPS\hBPS\rangle$. The latter was instead recomputed by Wick contractions directly.}
\twoseqn{
\lambda_{\hBPS\hBPS\qBPS} &= \frac{1}{2\sqrt6}\left(1-\frac{3}{N^2-1}\right)^{1/2}\,,
}[]{
\lambda_{\hBPS\hBPS(Q^2\Qb{}^2\lnsp K)} &= \frac{1}{2\sqrt3}\frac{1}{\sqrt{N^2-1}}\,,
}[]
A simple computation shows that
\eqn{
\alpha_1 = \alpha_2 = \alpha_3  = \frac54\,.
}[]
The specific value of these coefficients depends on how we normalize the tensor structures $\BBE_i^{(q,p,q)}$, therefore it is not meaningful. The important thing is that they do not depend on $x_{ij}$ nor on $N$.

Another possible pitfall, that we report just as a side note, has to do with the $(0,4,0)$ scalar. It contributes to the OPE via the following differential operator
\eqn{
t_{(0,4,0)}^{\mathsf{OPE}}(y_1,y_2,\partial_{y_0}) = \frac1{2^73^2} \stry{1\partial_0}^2\,\stry{2\partial_0}^2\,,
}[eq:tOPE040def]
Unlike the previous case, there are no superdescendants that can contribute to this representation at dimension four. Thus the exchanged operator must be the superprimary of a half-BPS multiplet transforming in the $(0,4,0)$. However, there are two distinct such operators: one is an admixture of a single trace and a double trace operator and the other is a pure double trace
\twoseqn{
\hBPS[4] &= \frac{2\sqrt{N^2+1}}{\sqrt{(N^2-9)(N^2-4)(N^2-1)}}\,\left(\tr\lsp(\varphi\cdot y)^4 - \frac{2N^2-3}{N(N^2+1)}\big(\tr\lsp(\varphi\cdot y)^2\big)^2\right) \,,
}[eq:O4mixture]{
\hBPS[4]^{(\mathrm{dt})} &= \frac{\sqrt{2}}{\sqrt{N^4-1}}\,\big(\tr\lsp(\varphi\cdot y)^2\big)^2\,.
}[][]
The relative coefficient in~\eqref{eq:O4mixture} is obtained by requiring that the two combinations are orthogonal~\cite{Aprile:2018efk, Aprile:2020uxk} and the overall coefficients simply normalize the two-point functions to one. As argued in~\cite{Aprile:2018efk}, the three-point function $\langle\hBPS[4]\hBPS\hBPS\rangle$ should vanish. On the other hand, from the conformal block expansion in~\cite{Dolan:2004iy} we see a nonzero OPE coefficient, there called $C_2$. By doing Wick contractions explicitly we see that the operator with vanishing OPE coefficient is $\hBPS[4]$ and the one with coefficient $C_2$ is $\hBPS[4]^{(\mathrm{dt})}$. Therefore we conclude that the operator exchanged in the $\hBPS\times\hBPS$ OPE is the pure double trace one. If we match the OPE limit as explained earlier we again find simple numerical coefficients $\alpha_1=\alpha_2=\alpha_3=60$.

\subsection{OPE limit of the supergravity result}

Now we are ready to take the OPE $x_1\to x_2$ limit of the supergravity result of~\cite{Goncalves:2019znr}. From the representation in~\eqref{eq:Ddef} it is easy to take the limit at coincident points. Indeed, when the right hand side converges, we have
\eqn{
\lim_{x_1\to x_2}D_{\Delta_1\Delta_2\Delta_3\cdots\Delta_n}(x_1,x_2,\ldots,x_n) = D_{\Delta_1+\Delta_2\,\Delta_3\cdots\Delta_n}(x_2,\ldots,x_n)\,.
}[eq:sumLabel]
When the external dimensions are integers one can often compute the $D$-functions by using some differential recursion relations. For example, the $D$-functions appearing in the supergravity result of the five-point function can all be obtained by applying the following recursion relations
\eqn{
D_{\Delta_1\cdots\Delta_i+1\cdots\Delta_j+1\cdots\Delta_n} = \frac{2-\Sigma}{\Delta_i\Delta_j}\frac{\partial}{\partial x_{ij}^2}D_{\Delta_1\cdots\Delta_n}\,,
}[eq:recRelD]
on the seed function $D_{11112}$, where we defined ${\Sigma\equiv\frac12(\Delta_1+\Delta_2+\Delta_3+\Delta_4)}$. The integral $D_{11112}$ is a one-loop pentagon diagram, known from~\cite{Bern:1993kr, Bern:1992em}. The same strategy works also for four points. In that case, however, the seed function is taken to be $D_{1111}$, which is divergent. This obstacle is overcome by removing a divergent prefactor and applying the recursion relations on the reduced function $\Db_{1111}$, defined in appendix~\ref{app:Dfunctions}.

Now we project onto the representation $(2,0,2)$ using the tensors $\BBE_k^{(2,0,2)}$ defined in~\eqref{eq:Ei202} and we add up the first two labels of the $D$-functions appearing in~\eqref{eq:Dset5pf}. It is easy to see that most functions obtained this way can be written as derivatives acting on $\Db_{1111}$, using the relations in~\eqref{eq:recRelD}. In the $(2,0,2)$ sector there are only two exceptions up to permutations:
\eqn{
D_{3111}(z,\zb)\qquad \mathrm{and}\qquad D_{4112}(z,\zb)\,.
}[]
These functions are indeed divergent. We can see this using the representation found in~\cite{Dolan:2004iy}
\eqn{
\Db_{\Delta_1\Delta_2\Delta_3\Delta_4}(z,\zb) = \frac{\Gamma(\Delta_1-s)\Gamma(\Delta_2-s)}{(z\zb)^s} \overbar{E}_{\Delta_1\Delta_2\Delta_3\Delta_4}(z,\zb) + \Db_{\Delta_1\Delta_2\Delta_3\Delta_4}(z,\zb)_{\mathrm{reg.}}\,,
}[]
where we introduced
\eqna{
s &= \frac12 (\Delta_1+\Delta_2-\Delta_3-\Delta_4)\,,\qquad s\in\BBN\,,\\
\overbar{E}_{\Delta_1\Delta_2\Delta_3\Delta_4}(z,\zb) &=
 \frac{\Gamma(\Delta_3)\Gamma(\Delta_4)}{\Gamma(\Delta_3+\Delta_4)} \sum_{m=0}^{s-1} (-1)^m (s-m-1)! \frac{(\Delta_1-s)_m(\Delta_2-s)_m(\Delta_3)_m(\Delta_4)_m}{m!(\Delta_3+\Delta_4)_{2m}}\\&\;\quad \times\,u^m\,{}_2F_1(\Delta_2-s+m,\Delta_3+m;\Delta_3+\Delta_4+2m;z+\zb-z\zb)\,,
}[]
and $\Db_{\Delta_1\Delta_2\Delta_3\Delta_4}(z,\zb)_{\mathrm{reg.}}$ is finite. In order to regularize the divergent $D$-functions we can shift all dimensions by $\epsilon$. The $\Gamma(\Delta_2-s)$ function at the numerator is responsible for the divergence. If we denote for convenience $n_\epsilon \equiv n+\epsilon$, the regularized $\Db$ functions read
\eqna{\Db_{3_\epsilon\,1_\epsilon\,1_\epsilon\,1_\epsilon}(z,\zb) &= \frac{1}{u\lsp\epsilon} -\frac{2\gamma_E}{u}+\frac{v\log v}{u\lsp(1-v)}+ \Db_{3111}(z,\zb)_{\mathrm{reg.}}\,,\\
\Db_{4_\epsilon\,1_\epsilon\,1_\epsilon\,2_\epsilon}(z,\zb) &=\frac{1}{u\lsp\epsilon} -\frac{2 \gamma_E}{u}-\frac{v(1-v+v\log v)}{u\lsp(1-v)^2}+ \Db_{4112}(z,\zb)_{\mathrm{reg.}}\,,\\
\Db_{4_\epsilon\,1_\epsilon\,2_\epsilon\,1_\epsilon}(z,\zb) &=\frac{1}{u\lsp\epsilon} -\frac{2 \gamma_E}{u}+\frac{1}{u\lsp(1-v)}-\frac{v\left(v-2\right)\log v}{u (1-v)^2}+ \Db_{4121}(z,\zb)_{\mathrm{reg.}}\,,\\
\Db_{4_\epsilon\,2_\epsilon\,1_\epsilon\,1_\epsilon}(z,\zb) &=\frac{1}{u^2\lsp\epsilon} -\frac{2 \gamma_E}{u^2}-\frac{1+v}{u\lsp(1-v)^2}+\frac{v \left((1-v)^2-2 u\right) \log v}{u^2 (1-v)^3}+ \Db_{4211}(z,\zb)_{\mathrm{reg.}}\,.
}[]

Notice that, unlike the regular part, the singular  part of $\Db_{4121}$ does not satisfy the identities in~\eqref{eq:DpermIdentities} and to compute it we have to resort to the explicit formulas in \cite{Dolan:2004iy}. Quite nicely, the divergences cancel in the end result because the above $D$-functions appear only through the following combination
\eqn{
a\,\Db_{4112}+b\,\Db_{4121}+c\lsp u\,\Db_{4211} +d\,\Db_{3111} \sim \frac{a+b+c+d}{u\lsp\epsilon} + O(1)\,,
}[]
with $a+b+c+d=0$.\footnote{As a side remark, one could do the same computation for the $(0,4,0)$ and in that case the divergences cancel in the same way but $\Db_{3111}$ never appears, namely we have $d=a+b+c=0$.}${}^,$\footnote{Note that the cancellation of these divergences can be seen manifestly by using the representation of the five-point Witten diagram in terms of box integrals~\cite{Bern:1993kr}. We thank Xinan Zhou for point this out to us.} For the regular part of these $\Db$-functions we find
\twoseqn{
&\begin{aligned}
\Db_{3111}(z,\zb)_{\mathrm{reg.}}&= R_1(z, \zb)\left(2 \text{Li}_2(z)-2 \text{Li}_2(\zb)+\log \left(\frac{1-z}{1-\zb}\right) \log (z  \zb)\right)+\\
& \quad \; R_2(z, \zb)\log (z\zb)+ R_3(z, \zb)\log\left((1-z)(1-\zb)\right)\,,\\
R_1(z, \zb)&= \frac{2 (z-1) (\zb-1)}{(z-\zb)^3}\,,\\
R_2(z, \zb)&=-\frac{z+\zb-2}{(z-\zb)^2}\,,\\
R_3(z, \zb)&= \frac{(z-1)(\zb-1)}{z \zb}\left(\frac{z+\zb}{(z-\zb)^2}-\frac{1}{z+\zb-z\zb}\right)\,,
\end{aligned}
}[]{
&\begin{aligned}
\Db_{4112}(z,\zb)_{\mathrm{reg.}}&= \tilde{R}_1(z, \zb)\left(2 \text{Li}_2(z)-2 \text{Li}_2(\zb)+\log \left(\frac{1-z}{1-\zb}\right) \log (z  \zb)\right)+\\
& \quad \; \tilde{R}_2(z, \zb)\log (z\zb)+ \tilde{R}_3(z, \zb)\log\left((1-z)(1-\zb)\right)+\tilde{R}_4(z, \zb)\,,\\
\tilde{R}_1(z, \zb)&= -\frac{6 (z-1)^2 (\zb-1)^2 (z+\zb)}{(z-\zb)^5}\,,\\
\tilde{R}_2(z, \zb)&=\frac{1}{(z-\zb)^4}\left(z \left(z^2 (\zb-2)+z \left(5 \zb^2-16 \zb+9\right)+9 \zb-6\right)+z\leftrightarrow \zb \right)\,,\\
\tilde{R}_3(z, \zb)&= \frac{(z-1)^2 (\zb-1)^2}{z \zb}\left(\frac{1}{(z+\zb-z \zb)^2}-\frac{z^2+10 z \zb+\zb^2}{(z-\zb)^4}\right)\,,\\
\tilde{R}_4(z, \zb)&=\frac{(z-1) (\zb-1) (z+\zb-4)}{(z-\zb)^2 (z+\zb-z \zb)}\,.
\end{aligned}
}[]
The regular part of $\Db_{4121}(z,\zb)_{\mathrm{reg.}}$ and $\Db_{4211}(z,\zb)_{\mathrm{reg.}}$ can be obtained from the identities in~\eqref{eq:DpermIdentities}.

After having regularized all the $D$-function we can finally take the OPE limit and explicitly see that the free theory result is perfectly matched at order $1/N^3$ and the non-rational pieces in the $D$-functions all cancel each other. This observation supports conjectured claims that the three-point functions of all the exchanged short multiplets are protected~\cite{Goncalves:2019znr, Heslop:2003xu}. The very same situation occurs when we select $\hBPS[4]^{(\mathrm{dt})}$ in the OPE of $\hBPS \times \hBPS$, namely the supergravity result perfectly  reproduces  $\langle \hBPS[4]^{(\mathrm{dt})} \hBPS \hBPS \hBPS \rangle^{\mathrm{free}}$ up to order $1/N^3$.

\section{Outlook}\label{sec:outlook}
In this paper we have revived the study of quarter-BPS operators in the context of four dimensional $\mathcal{N}=4$ SYM. In particular we have found constraints imposed by the underlying chiral algebra on the structure of the protected part of the four-point correlators, involving one or more quarter-BPS operators. There are several directions that would be interesting to pursue. We list some below. 

\begin{description}[style=unboxed,leftmargin=0cm, labelsep=1pt] % \the\parindent]
\item[Superconformal blocks]: in order to fully exploit the power of superconformal symmetry, it would be very useful and interesting to find the form of the superconformal blocks. Differently from the half-BPS case~\cite{Dolan:2001tt}, they are not known. Being them eigenfunctions of the quartic and quadratic Casimir of the superconformal group, it could be possible to explore the superspace approach to solve the eigenvalue problem. 
\item[Numerical bootstrap]: it would be nice to  explore the mixed correlator system using numerical bootstrap techniques. Despite the fact that we do not use the full superconformal symmetry, it can still be possible to obtain information on the dimensions and OPE coefficients of the intermediate non protected operators. In particular, this method can be helpful to see if there are some OPE coefficients which are protected.
\item[Basis of function]: it would be interesting to understand if the basis of $\bar{D}$ functions is enough to compute the correlators involving double trace operators. It has been observed, in a holographic computation, that such a basis is not enough and needs to be supplemented with more general functions~\cite{Ceplak:2021wzz}. We hope to be able to clarify this point in the future. 
\item[{Higher point functions}]: in this paper we have made a comparison with a five-point correlator computation~\cite{Goncalves:2019znr}. It would be very interesting to be able to use our results together with supersymmetry constraints and more generally the results of~\cite{Antunes:2021kmm}, to find some structures of the six-point correlators of half-BPS operators. 
\item[{Triple trace operators}]: in the context of holographic computations to get the corrections in the large central charge limit of the four-point correlators of $\mathcal{O}_2$ as in~\cite{Bissi:2020woe,Bissi:2020wtv}, the anomalous dimensions and the OPE coefficients of triple trace operators are needed. It would be interesting to use the results of this paper for the leading twist, non protected, triple trace operator appearing in the OPE of $\mathcal{O}_2 \times \mathcal{O}_{02}$ and study the constraints imposed by crossing symmetry given by its presence, at the level of the four-point function.
\end{description}

\acknowledgments{We would like to thank Xinan Zhou for valuable comments on the draft.  AM would like to thank Leonardo Rastelli for discussions. This work is supported by Knut and Alice Wallenberg Foundation under grant KAW 2016.0129 and by VR grant 2018-04438.
Some computations in this work were enabled by resources in project SNIC 2020/15-320 provided by the Swedish National Infrastructure for Computing (SNIC) at UPPMAX, partially funded by the Swedish Research Council through grant agreement no. 2018-05973.
}

\Appendices

\section{Notation and conventions}\label{app:notation}

\subsection{Notation for R-symmetry structures}\label{app:conventions}

In this appendix we will show in detail the conventions used for tensor structures and polarizations.  In table~\ref{tab:indices} we summarize our naming convention for the various labels and indices that appear throughout the manuscript.

The tensor structures in four dimensions are a function of a six-dimensional complex vector $y$ and two four-dimensional complex twistors $S$ and $\Sb$. The tensor structures in the chiral algebra are instead functions of two spinors $\eta$ and $\etat$ associated to the R-symmetry and flavor $\SU(2)$ respectively. Since $y$ has to contract symmetric traceless tensors, it is subject to the constraints $y\cdot y = 0$. Similarly, if we want to describe tensors of $\SU(4)$ with $S^m$ and $\Sb_m$, we need to remove the traces by imposing $S\cdot \Sb = 0$. The polarizations $\eta,\etat$ do not need other constraints because they automatically square to zero.\footnote{See footnote~\ref{foot:etasq}.}

Let us first discuss the R-symmetry structures in four dimensions. Here follows our convention for the six dimensional Clifford algebra
\twoseqn{
\setlength{\arrayrulewidth}{.25pt}
\setstretch{1.2}
\Sigma^{A}_{mn} = \left(
\begin{array}{c:c}
0 & - \sigma^A \lsp \hat\epsilon \\
\hdashline
\sigmab^A \lsp \check\epsilon & 0 \\
\end{array}
\right)\,,\quad
\Sigma^{5}_{mn} = \left(
\begin{array}{c:c}
-i \hat\epsilon & 0 \\
\hdashline
0 & i\hat\epsilon \\
\end{array}
\right)\,,\quad
\Sigma^{6}_{mn} = \left(
\begin{array}{c:c}
\hat\epsilon & 0 \\
\hdashline
0 & \hat\epsilon \\
\end{array}
\right)\,.
}[]{
\overbarUp\Sigma_{A}^{mn} = \left(
\begin{array}{c:c}
0 & - \hat\epsilon \lsp \sigma^A \\
\hdashline
\check\epsilon \lsp\sigmab^A & 0 \\
\end{array}
\right)\,,\quad
\overbarUp\Sigma_{5}^{mn} = \left(
\begin{array}{c:c}
-i \hat\epsilon & 0 \\
\hdashline
0 & i\hat\epsilon \\
\end{array}
\right)\,,\quad
\overbarUp\Sigma_{6}^{mn} = \left(
\begin{array}{c:c}
-\hat\epsilon & 0 \\
\hdashline
0 & -\hat\epsilon \\
\end{array}
\right)\,,
}[][]
where $\sigma^{A=1,2,3}_{aa'}$ are Pauli matrices, $\sigma^4_{aa'} = i\lsp \delta_{aa'}$, $\sigmab^A =  -(\hat\epsilon \sigma^A \hat\epsilon)^T$, $\hat\epsilon = \epsilon^{ab} = -\epsilon_{ab} = \check\epsilon$ with $\epsilon^{12}=1$.
With this definition we can build the generators of $\SO(6)$
\eqna{
\bigl(\Sigma^{MN}\bigr)_m^{\phantom{m}n} &= \frac14\left(\Sigma^M_{mp}\,\overbarUp\Sigma^{N\lsp pn} - \Sigma^N_{mp}\,\overbarUp\Sigma^{M\lsp pn}\right)\,,\\
\bigl(\overbarUp\Sigma^{MN}\bigr)^m_{\phantom{m}n} &= \frac14\left(\overbarUp\Sigma^{M\lsp mp}\,\Sigma^N_{pn} - \overbarUp\Sigma^{N\lsp mp}\,\Sigma^M_{pn}\right)\,.
}[SigmaMNdef]
These matrices can also be used to relate tensors in the adjoint of $\SU(4)$ with tensors in the rank-two antisymmetric of $\SO(6)$. We can also map $\SO(6)$ fundamental indices to antisymmetric $\SU(4)$ indices as follows
\eqn{
\rmy_{mn} \equiv y_M \Sigma^M_{mn}\,,\qquad
\bar\rmy^{mn} \equiv y_M \overbarUp\Sigma^{M\lsp mn} \,.
}[]
Some useful identities of the $\Sigma$ matrices are
\twoseqn{
\Sigma^M_{mn}\Sigmab^{N\,np}+\Sigma^N_{mn}\Sigmab{}^{M\,np}&=\delta^p_m\delta^{MN}\,,&\qquad \epsilon^{mnpq}\,\Sigma^M_{pq} &= -2\Sigmab{}^{M\,mn}
}[]{
\Sigma^P_{mn}\Sigma_{P\,pq} &= 2\epsilon_{mnpq}\,,&\qquad \Sigma^P_{mn}\Sigmab_P^{pq} &= 2\bigl(\delta^q_m\delta^p_n-\delta^p_m\delta^q_n\bigr)\,.
}[eq:so6identities][]

If we have operators in the $(q,p,q)$ where all Dynkin labels are nonzero the conditions on the polarizations stated at the beginning are not enough and we need to impose further constraints. The complete list is the following
\eqn{
y\cdot y = 0\,,\qquad S\cdot \Sb = 0\,,\qquad \rmy S = 0\,,\qquad \bar\rmy \Sb = 0\,.
}[constraints]
As explained around~\eqref{eq:opScaling}, operators in the $(q,p,q)$ will be fields with homogeneity $p$ in $y$ and $q$ in $S,\Sb$
\eqn{
\CO_{pq}(\lambda S,\bar \lambda\Sb,\kappa y) = (\lambda \bar\lambda)^q \kappa^p \,\CO_{pq}(S,\Sb,y)\,.
}[]
In order to recover the tensor form of this operator we must differentiate with respect to the polarizations. However, we need to be careful because the polarizations are constrained and so their derivatives are not free. The following differential operators can be used to avoid this problem~\cite{Bargmann:1977gy, Nirschl:2004pa, Costa:2011mg, Cuomo:2017wme}
\eqna{
\CD_M &= \left(2+ y\cdot \frac{\partial}{\partial y}\right)\frac{\partial}{\partial y^M} - \frac12 y_M \frac{\partial^2}{\partial y \cdot \partial y}\,,\\
\partial_m &= \left(3+ S\cdot \frac{\partial}{\partial S} + \Sb\cdot\frac{\partial}{\partial\Sb}\right) \frac{\partial}{\partial S^m} - \Sb_m\,\frac{\partial^2}{\partial S \cdot \partial \Sb}\,,\\
\bar\partial^m &= \left(3+ S\cdot \frac{\partial}{\partial S} + \Sb\cdot\frac{\partial}{\partial\Sb}\right) \frac{\partial}{\partial \Sb_m} - S^m\,\frac{\partial^2}{\partial S \cdot \partial \Sb}\,.
}[eq:diffop]
With the aid of the above operators we would like to write
\eqn{
(\CO_{pq})^{m_1\cdots m_q}_{n_1\cdots n_q\,,\,M_1\cdots M_p} = \partial_{n_1}\cdots \partial_{n_q}\,\bar\partial^{m_1}\cdots \bar\partial^{m_q}\, \CD_{M_1}\cdots \CD_{M_p}\CO_{pq}(S,\Sb,y)\,.
}[]
However, unfortunately, the operators do not implement the constraints $\rmy S = \Sb\bar\rmy = 0$ and so the derivatives are not free and the above expression does not hold unless either $p$ or $q$ is zero.

The conventions for $\SU(2)_R\times\SU(2)_F$ are analogous to the usual spinor notation in four dimensions~\cite{WessnBagger} with the substitutions $a \to \alpha,\,a'\to\alphad,\,A\to\mu$ and the Euclidean signature. 
An operator in the $(R\,;F)$ will be a field with homogeneity $R$ in $\eta$ and $F$ in $\etat$
\eqn{
O_{R,F}(\lambda \eta,\tilde \lambda\etat) = \lambda^R \tilde\lambda^F \,O_{R,F}(\eta,\etat)\,.
}[]
Since there are no constraints on $\eta$ and $\etat$, in order to recover the indices one simply needs to take derivatives
\eqn{
(O_{R,F})_{a_1\cdots a_R\,,\, a'_1\cdots a'_F} =  \frac{\partial}{\partial \eta^{a_1}}\cdots \frac{\partial}{\partial \eta^{a_R}}\, \frac{\partial}{\partial \etat^{a'_1}}\cdots \frac{\partial}{\partial \etat^{a'_F}}\, O_{R,F}(\eta,\etat)\,.
}

\begin{table}[h]
\setstretch{1.2}
\centering
\begin{tabular}{lllll}
\Hhline
indices & range & polarization& group & representation\\
\hline
$M,N$ & $1,\ldots,6$ & $y^M$ &$\SO(6)_R$ & $\tiny\yng(1)$\\
$m,n$ & $1,\ldots,4$ & $S^m,\,\Sb_m,\,\rmy_{mn},\,\bar\rmy^{mn}$ &$\SU(4)_R$ & $\tiny\yng(1)\,$, $\overline{\tiny\yng(1)}\,$, $\tiny\Yvcentermath1\yng(1,1)$\\
$\mu,\nu$ & $1,\ldots,4$ &  &$\SO(4)$ &  $\tiny\yng(1)$\\
$A,B$ & $1,\ldots,4$ &  &$\SU(2)_R\times \SU(2)_F$ &  $(\tiny\yng(1)\lsp;\tiny\yng(1))$\\
$a,b$ & $1,2$        & $\eta^a,\,\eta_a$ & $\SU(2)_R$ & $\tiny\yng(1)$ $\cong$ $\overline{\tiny\yng(1)}\,$\\
$a',b'$ & $1,2$        & $\etat^{a'},\,\etat_{a'}$ & $\SU(2)_F$ & $\tiny\yng(1)$ $\cong$ $\overline{\tiny\yng(1)}\,$\\
$I,J$ & $1,\ldots, N^2-1$ & & $\SU(N)_{\mathrm{gauge}}$ & $\mathbf{adj}$\\
$i,j$ & $1,\ldots,4$ & & operator label\\
$k$ & $1,\ldots,N_\mathrm{str}$ & & tensor structure label\\
\Hhline
\end{tabular}

\caption{Conventions for the polarizations, indices of the various symmetry groups and other labels.\label{tab:indices}}
\end{table}

\subsection{Convention for the conformal blocks}\label{app:notationBlocks}

The OPE coefficients that we show in the text are meaningful only if we specify the normalization of the conformal blocks. The definition that we use is the following
\eqn{
g_{\Delta,\ell}(z,\zb) = g_{\Delta,\ell}^{0,0}(z,\zb)\,,
}[]
\eqn{
g^{\Delta_{12},\Delta_{34}}_{\Delta,\ell}(z,\zb) = \frac{z \zb}{z-\zb}\frac{1}{(-2)^\ell}\left(\kappa_{\Delta+\ell}^{\Delta_{12},\Delta_{34}}(z)\kappa_{\Delta-\ell-2}^{\Delta_{12},\Delta_{34}}(\zb) - (z\leftrightarrow\zb)\right)\,,
}[]
with
\eqn{
\kappa_{\beta}^{a,b}(z) = z^{\beta/2} \,{}_2F_1\mleft(\frac{\beta-a}{2},\frac{\beta+b}{2};\beta;z\mright)\,.
}[]

\section{Ward identities for the \texorpdfstring{$\boldsymbol{(2,0,2)}$}{(2,0,2)} four-point function}\label{app:supplement}

Here we show the nonzero entries of the vector $\hat{v}_k^{(m)}(z,\zb)$ introduced in~\eqref{eq:vhatdef4444} in terms of seven polynomials $P_1$, $\ldots$, $P_7$ which we will define below
\eqnal{
\hat{v}_{41}^{(1)} &= \frac{112 P_6 P_1\lnsp[1]}{15 z^3 \zb^3}	\,,\qquad & \hat{v}_{42}^{(1)} &= \frac{P_5}{7560 z^3 \zb^3}
\,,\\\hat{v}_{41}^{(2)} &= -\frac{112 P_2}{9 z^2 \zb^2}	\,,\qquad & \hat{v}_{42}^{(2)} &= -\frac{P_7 P_1\lnsp[1]}{6 z^2 \zb^2}
\,,\\\hat{v}_{41}^{(3)} &= -\frac{56 P_2}{9 z^2 \zb^2}	\,,\qquad & \hat{v}_{42}^{(3)} &= -\frac{P_7 P_1\lnsp[1]}{12 z^2 \zb^2}
\,,\\\hat{v}_{41}^{(4)} &= \frac{16 P_1\lnsp[1]}{z \zb}	\,,\qquad & \hat{v}_{42}^{(4)} &= \frac{P_1\lnsp\big[\tfrac{2}{3}\big]+2}{42 z \zb}
\,,\\\hat{v}_{41}^{(5)} &= \frac{8 P_1\lnsp[1]}{z \zb}	\,,\qquad & \hat{v}_{42}^{(5)} &= \frac{P_1\lnsp\big[\tfrac{2}{3}\big]+2}{84 z \zb}
\,,\\\hat{v}_{41}^{(6)} &= \frac{1008 P_1\lnsp[1]}{25 z \zb}	\,,\qquad & \hat{v}_{42}^{(6)} &= \frac{3 \left(P_1\lnsp\big[\tfrac{17}{28}\big]+2\right)}{50 z \zb}
\,,\\\hat{v}_{41}^{(7)} &= -\frac{112}{15}
\,,\\\hat{v}_{41}^{(8)} &= \frac{56}{15}	
\,,\\\hat{v}_{41}^{(9)} &= \frac{2 P_3}{z^2 \zb^2}	\,,\qquad & \hat{v}_{42}^{(9)} &= \frac{P_4}{336 z^2 \zb^2}
\,,\\\hat{v}_{41}^{(11)} &= \frac{28 P_1\lnsp[1]}{15 z \zb}	\,,\qquad & \hat{v}_{42}^{(11)} &= \frac{P_1\lnsp\big[\tfrac{4}{7}\big]+2}{360 z \zb}
\,,\\\hat{v}_{41}^{(12)} &= \frac{4 \left(7 P_1\lnsp[1] z \zb-P_3\right)}{15 z^2 \zb^2}	\,,\qquad & \hat{v}_{42}^{(12)} &= -\frac{P_7 P_1\lnsp[1]}{30 z^2 \zb^2}\,,
\\& & \hat{v}_{42}^{(13)} &= -\frac{1}{720}	
\,,\\\hat{v}_{41}^{(14)} &= -\frac{448 P_1\lnsp\big[\tfrac{8}{7}\big]}{5 z \zb}	\,,\qquad & \hat{v}_{42}^{(14)} &= -\frac{2 \left(P_1\lnsp\big[\tfrac{2}{3}\big]+2\right)}{15 z \zb}
\,,\\\hat{v}_{41}^{(15)} &= \frac{224 P_1\lnsp\big[\tfrac{8}{7}\big]}{5 z \zb}	\,,\qquad & \hat{v}_{42}^{(15)} &= \frac{P_1\lnsp\big[\tfrac{2}{3}\big]+2}{15 z \zb}
\,,\\\hat{v}_{41}^{(17)} &= \frac{56 P_1\lnsp\big[\tfrac{6}{7}\big]}{5 z \zb}	\,,\qquad & \hat{v}_{42}^{(17)} &= \frac{P_1\lnsp\big[\tfrac{2}{3}\big]+2}{60 z \zb}
\,,\\\hat{v}_{41}^{(18)} &= \frac{128 P_1\lnsp[1]}{225 z \zb}	\,,\qquad & \hat{v}_{42}^{(18)} &= \frac{4 \left(P_1\lnsp\big[\tfrac{2}{3}\big]+2\right)}{4725 z \zb}
\,,\\\hat{v}_{41}^{(19)} &= \frac{56 P_1\lnsp\big[\tfrac{6}{7}\big]}{5 z \zb}	\,,\qquad & \hat{v}_{42}^{(19)} &= \frac{P_1\lnsp\big[\tfrac{2}{3}\big]+2}{60 z \zb}
\,,\\\hat{v}_{41}^{(20)} &= -\frac{128}{1575}	
\,,\\\hat{v}_{41}^{(21)} &= \frac{16 P_1\lnsp\big[\tfrac{10}{9}\big]}{5 z \zb}	\,,\qquad & \hat{v}_{42}^{(21)} &= \frac{P_1\lnsp\big[\tfrac{2}{3}\big]+2}{210 z \zb}
\,,\\\hat{v}_{41}^{(22)} &= 7	\,,\qquad & \hat{v}_{42}^{(22)} &= -\frac{1}{144}\,,
\\& & \hat{v}_{42}^{(24)} &= -\frac{1}{432}
\,,\\\hat{v}_{41}^{(25)} &= -\frac{7}{3}	
\,,\\\hat{v}_{41}^{(26)} &= \frac{42 P_1\lnsp\big[\tfrac{4}{3}\big]}{5 z \zb}	\,,\qquad & \hat{v}_{42}^{(26)} &= \frac{P_1\lnsp\big[\tfrac{8}{15}\big]+2}{80 z \zb}
\,,\\\hat{v}_{41}^{(27)} &= -\frac{168 P_1\lnsp[1]}{5 z \zb}	\,,\qquad & \hat{v}_{42}^{(27)} &= -\frac{P_1\lnsp\big[\tfrac{11}{15}\big]+2}{20 z \zb}
\,,\\\hat{v}_{41}^{(29)} &= -\frac{168 P_1\lnsp[1]}{5 z \zb}	\,,\qquad & \hat{v}_{42}^{(29)} &= -\frac{P_1\lnsp\big[\tfrac{11}{15}\big]+2}{20 z \zb}
\,,\\\hat{v}_{41}^{(30)} &= \frac{6356 P_1\lnsp[1]}{225 z \zb}	\,,\qquad & \hat{v}_{42}^{(30)} &= \frac{227 \left(P_1\lnsp\big[\tfrac{497}{681}\big]+2\right)}{5400 z \zb}
\,,\\\hat{v}_{41}^{(32)} &= \frac{1148 P_1\lnsp[1]}{825 z \zb}	\,,\qquad & \hat{v}_{42}^{(32)} &= \frac{41 \left(P_1\lnsp\big[\tfrac{68}{123}\big]+2\right)}{19800 z \zb}
\,,\\\hat{v}_{41}^{(33)} &= -\frac{7756 P_1\lnsp[1]}{10725 z \zb}	\,,\qquad & \hat{v}_{42}^{(33)} &= -\frac{277 \left(P_1\lnsp\big[\tfrac{574}{831}\big]+2\right)}{257400 z \zb}
\,,\\\hat{v}_{41}^{(34)} &= -\frac{14}{75}	
\,,\\\hat{v}_{41}^{(35)} &= -\frac{6}{5}	
\,,\qquad & \hat{v}_{42}^{(37)} &= -\frac{1}{2520}	
\,,\\\hat{v}_{41}^{(38)} &= 2	\,,\qquad & \hat{v}_{42}^{(38)} &= -\frac{1}{672}
\,,\\&\qquad & \hat{v}_{42}^{(40)} &= -\frac{1}{1344}\,.
}[]
The polynomials appearing in the above expressions are given by
\eqna{
P_1\lnsp[a](z,\zb) &= a\lsp z\zb -z -\zb\,,\\
P_2(z,\zb) &= 11 z^2 \zb^2-27 z^2 \zb+18 z^2-27 z \zb^2+18 z \zb+18 \zb^2\,,\\
P_3(z,\zb) &= 108 z^2 \zb^2-259 z^2 \zb+168 z^2-259 z \zb^2+168 z \zb+168 \zb^2\,,\\
P_4(z,\zb) &= 88 z^2 \zb^2-259 z^2 \zb+168 z^2-259 z \zb^2+686 z \zb-336 z+168 \zb^2-336 \zb\,,\\
P_5(z,\zb) &= 979 z^3 \zb^3-4872 z^3 \zb^2+7560 z^3 \zb-3780 z^3-4872 z^2 \zb^3+17304 z^2 \zb^2-18900 z^2 \zb\\&\;\quad+7560 z^2+7560 z \zb^3-18900 z \zb^2+7560 z \zb-3780 \zb^3+7560 \zb^2\,,\\
P_6(z,\zb) &= 13 z^2 \zb^2-45 z^2 \zb+45 z^2-45 z \zb^2+45 \zb^2\,,\\
P_7(z,\zb) &= (z-2) (\zb-2)\,.
}[]

\section{Proof of vanishing of low-lying operators}\label{app:low_ops}

In section~\ref{sec:generalities} we argued that the simplest quarter-BPS operator of $B\overbar{B}$ type is $\qBPS$. In this appendix we show why it is impossible to construct $\qBPS[11]$ and $\qBPS[21]$ and  why $\CO^{(\mathrm{st},2)}_{02}$ in~\eqref{Osingtrace2} vanishes as well.

Let us start from the first two. Both of these operators must be written solely out of the six scalars in $\varphi^M$. The latter case has potentially two options: it could be of single trace type
\eqn{
\qBPS[21]^{(\mathrm{st})}(S,\Sb,y)=\tr\lsp\mleft(\varphi^{M_1}\varphi^{M_2}\varphi^{M_3}\varphi^{M_4}\mright)\,S\cdot\Sigma_{M_1M_4}\lnsp\cdot\Sb \,\lsp y_{M_2}y_{M_3}\,,
}[]
or the analogous one with $(M_1M_3)$\lsp$(M_2M_4)$ pairing. Or it could be of double trace type
\eqn{
\qBPS[21]^{(\mathrm{dt})}(S,\Sb,y)=\tr\lsp\mleft(\varphi^{M_1}\varphi^{M_2})\,\tr\lsp(\varphi^{M_3}\varphi^{M_4}\mright)\,S\cdot\Sigma_{M_1M_4}\lnsp\cdot\Sb \,\lsp y_{M_2}y_{M_3}\,.
}[]
The second trivially vanishes as can be shown by a $(14)(23)$ permutation, recalling that the $\Sigma_{M_1M_4}$ are antisymmetric. The first also vanishes in $\SU(N)$ due to the fact that the generators are Hermitian and the trace of four of them is real, which implies
\eqn{
\tr\lsp(T^{I_1}T^{I_2}T^{I_3} T^{I_4}) = \tr\lsp(T^{I_4}T^{I_3} T^{I_2} T^{I_1})\,.
}[]
The former case instead can only be single trace since it has three fields
\eqn{
\qBPS[11] = \tr\lsp\mleft(\varphi^{M_1}\varphi^{M_2}\varphi^{M_3}\mright)\,S\cdot\Sigma_{M_1M_3}\lnsp\cdot\Sb \,\lsp y_{M_2}\,.
}[]
This operator would seem allowed, however a quick computation shows that its two-point function identically vanishes and thus the operator must be absent in a unitary theory.

We will now show why the single trace operator in \eqref{Osingtrace2} actually vanishes.  For simplicity  let us rewrite here is explicit form
\begin{align}
\mathcal{O}^{(\mathrm{st}, 2)}_{02} = \varphi^{M_1}_{a} \varphi^{M_2}_{b}\varphi^{M_3}_{c}\varphi^{M_4}_{d}\,\tr\lsp(T^a T^b T^cT^d)\,S\cdot\Sigma_{M_1M_3}\cdot\bar{S}S \cdot\Sigma_{M_2M_4}\cdot\bar{S}\, .
\end{align}
The SU($N$) generators satisfy two important identities 
\begin{align}
\label{Identity1}
&f^{abe}f^{cde}=\frac{2}{N}(\delta^{ac}\delta^{bd}-\delta^{ad}\delta^{bc})+d^{ace}d^{bde}-d^{ade}d^{bce}\, , \\
\label{Identity2}
&f^{abe}d^{cde}+f^{cbe}d^{dae}+f^{dbe}d^{ace}=0\,,
\end{align}
where $d^{abc}$ is a totally symmetric tensor and $f^{abc}$ is the completely antisymmetric structure constant. By using these relations one can write
\begin{align} \nonumber
&\text{tr}(T^aT^bT^cT^d)=\frac{\delta^{ab}\delta^{cd}}{4N}+\frac{1}{8}(d^{abe}d^{cde}+i\, d^{cde}f^{abe}+i\,d^{abe}f^{cde}-\underbrace{f^{abe}f^{cde}}_{\text{using \eqref{Identity1}}})\\ \nonumber
&=\frac{\delta^{ab}\delta^{cd}-\delta^{ac}\delta^{bd}+\delta^{ad}\delta^{bc}}{4N}+\frac{1}{8}(d^{abe}d^{cde}-d^{ace}d^{dbe}+d^{ade}d^{bce})+\frac{i}{8}(d^{abe}f^{cde}+\underbrace{d^{cde}f^{abe}}_{\text{using \eqref{Identity2}}})\\  \nonumber
&=\frac{\delta^{ab}\delta^{cd}-\delta^{ac}\delta^{bd}+\delta^{ad}\delta^{bc}}{4N}+\frac{1}{8}(d^{abe}d^{cde}-d^{ace}d^{dbe}+d^{ade}d^{bce})\\& \label{trace4}\;\quad+\frac{i}{8}(d^{abe}f^{cde}-d^{ace}f^{dbe}+d^{ade}f^{bce})
\end{align}
Now that we have rewritten the trace appearing in $\mathcal{O}^{(\mathrm{st}, 2)}_{02} $ in this way,  it is straightforward to argue that given the antisymmetry under  $M_2\leftrightarrow M_4$  (or equivalently for $M_1\leftrightarrow M_3$) the first two terms in~\eqref{trace4}, being symmetric, they vanish.  The last piece, instead, is antisymmetric, so the same argument can not be applied.  By using \eqref{Identity2}  it is possible to rewrite this term of the trace as 
\begin{align} \begin{aligned}
d^{bae}f^{dce}-d^{bde}f^{cae}+d^{bce}f^{ade}&=-d^{abe}f^{cde}-d^{bde}f^{cae}+d^{bce}f^{ade}\\
&=-d^{abe}f^{cde}-(f^{bce}d^{dae}+f^{dce}d^{abe})-(f^{cde}d^{bae}+f^{bde}d^{ace})\\
&=-d^{abe}f^{cde}+d^{ace}f^{dbe}-d^{ade}f^{bce}\, , 
\end{aligned}
\end{align}
which is clearly antisymmetric under  $a\leftrightarrow b$ together with $c\leftrightarrow d$. However under the corresponding simultaneous exchange of $(1,2)(3,4)$ the $\mathrm{SO}(6)$ part is  symmetric, so finally their product vanishes.

\section{Useful integrals for computing the inversion formula}\label{app:InversionIntegrals}
In this appendix we collect all the relevant computations necessary to obtain the OPE coefficients and anomalous dimensions in section~\ref{sec:Results}.

Let us start from the $\zb$ integral in \eqref{cT}. Since in the studied examples we have at most $\log (1-\zb)$, which has vanishing double discontinuity,  we will only focus on
\eqn{
\overline{\mathcal{I}}^{\Do, \Dt}(\lambda)=\int_0^1 \frac{d\zb}{\zb^2} (1-\zb)^{\frac{\Dt-\Do}{2}}\kappa_{2h+2\ell}^{\Do, \Dt} (\zb) \lsp \zb^{-\frac{\Dt}{2}} \text{dDisc}\left[\left(\frac{1-\zb}{\zb}\right)^\lambda\right] \, ,
}[]
where we recall that we have defined $h$ as half the twist $2h=\tau=\Delta-\ell$.
This integral can be performed by replacing the hypergeometric function in $\kappa_{2h+2\ell}^{\Do, \Dt}(\zb)$ with its integral transform by introducing a fictitious integration variable $V$ and then by performing the change of variables $\zb \to\frac{T}{(T-1) V+1}$. The result is
\begin{align} \nonumber
\overline{\mathcal{I}}_\lambda^{\Do, \Dt}&=\frac{2 \Gamma (2 (h+\ell )) \Gamma \mleft(h+\ell -\frac{\Dt}{2}-\lambda -1\mright)}{\Gamma
   \mleft(h+\ell -\frac{\Do}{2}\mright) \Gamma \mleft(h+\ell -\frac{\Dt}{2}\mright) \Gamma \mleft(h+\ell +\frac{\Dt}{2}+\lambda +1\mright)} \times
   \\  \label{eq:intDiscRational}& \;\quad\times \sin (\pi  \lambda ) \Gamma (\lambda +1) \sin \mleft(\pi  \frac{\Dt-\Do}{2}+\pi\lambda \mright)\, \Gamma \mleft(-\frac{\Do}{2}+\frac{\Dt}{2}+\lambda +1\mright)\\ \nonumber
   &=\frac{2 \pi ^2 \Gamma (2 h+2 \ell ) \Gamma \mleft(h+\ell -\frac{\Dt}{2}-\lambda
   -1\mright)}{\Gamma (-\lambda ) \Gamma \mleft(\frac{\Do}{2}-\frac{\Dt}{2}-\lambda \mright) \Gamma \mleft(h+\ell -\frac{\Do}{2}\mright) \Gamma \mleft(h+\ell
   -\frac{\Dt}{2}\mright) \Gamma \mleft(h+\ell +\frac{\Dt}{2}+\lambda
   +1\mright)} \, ,
\end{align}
where in the second line we have assumed $\lambda<0$ and we have used Euler's reflection formula $\Gamma(1-x)\Gamma(x)=\pi/\sin(\pi x)$. Notice that the $\Gamma$ functions that appear are regular for all the relevant values of $\lambda$ and $\Delta_i$. Translated to $c_k(\Delta, \ell)$ this means that the $\zb$ integral just provides the spin dependence of the OPE data that we want to extract and does not give us any information about the twist of the possible exchange operators.

The information we are after is instead encoded in the $z$ integrals.  These can be distinguished in three different types
\threeseqn{
\mathcal{I}_1^{\Do, \Dt}(\lambda)&=\int_0^1 \frac{d z}{z^2}(1-z)^{\frac{\Dt-\Do}{2}}\kappa_{4-2h}^{\Do, \Dt}(z)\left[\left(\frac{z}{1-z}\right)^\lambda z^{-\frac{\Dt}{2}}\right] \, ,
}[]{
\mathcal{I}_2^{\Do, \Dt}(\lambda)&=\int_0^1 \frac{d z}{z^2}(1-z)^{\frac{\Dt-\Do}{2}}\kappa_{4-2h}^{\Do, \Dt}(z)\lsp \left[ z^\lambda z^{-\frac{\Dt}{2}} \right]\, ,
}[]{
\mathcal{I}_3^{\Do, \Dt}(\lambda)&=\int_0^1 \frac{d z}{z^2}(1-z)^{\frac{\Dt-\Do}{2}}\kappa_{4-2h}^{\Do, \Dt}(z)\left[\log z \left(\frac{z}{1-z}\right)^\lambda z^{-\frac{\Dt}{2}}\right] \, .
}[][]
In all these cases $\lambda$ has to be considered positive and we are assuming $h \in \mathbb{N},\, h \geq 1$.\footnote{When solving these integrals spurious poles can appear at half integer values of $h$. We will ignore them having in mind that they can be cancelled by adding a reflected block with $h \to 1-h$ \cite{Caron-Huot:2017vep,Alday:2017vkk}.} The first integral can be computed similarly to the $\zb$ one and gives
\eqn{
\mathcal{I}_1^{\Do, \Dt}(\lambda)=-\frac{\pi \lsp r^{\Do, \Dt}_{h, \lambda}  \sin \mleft(\pi  \left(\frac{\Do}{2}+h\right)\mright) \sin \mleft(\pi 
   \left(\frac{\Dt}{2}+h\right)\mright) \sin \mleft(\pi  \left(-\frac{\Dt}{2}+h+\lambda \right)\mright)}{\sin (2 \pi  h) \sin (\pi  \lambda ) \sin \mleft( \pi  \left(\frac{\Do-\Dt}{2}+ \lambda \right)\mright) \sin \mleft(\pi  \left(\frac{\Dt}{2}+h-\lambda \right)\mright)}\, , 
}[]
where we have collected ratios of $\Gamma$ functions  appearing throughout  these computations in a single function 
\eqn{
r^{\Do, \Dt}_{h, \lambda} =\frac{\Gamma \mleft(h+\frac{\Do}{2}-1\mright) \Gamma \mleft(h+\frac{\Dt}{2}-1\mright) \Gamma \mleft(h-\frac{\Dt}{2}+\lambda -2\mright)}{\Gamma (2 h-3) \Gamma
   (\lambda ) \Gamma \mleft(\frac{\Do}{2}-\frac{\Dt}{2}+\lambda \mright)
   \Gamma \mleft(h+\frac{\Dt}{2}-\lambda \mright)} \, .
}[]
We are now going to analyze in detail the poles that $\mathcal{I}_1^{\Do, \Dt}(\lambda)$ develops and the corresponding residues for all the values of the external dimensions appearing in the main text.
\begin{itemize}
\item $\mathbf{\Do=\Dt=0}$: we have simple poles for $h=\lambda+1+n$, $n \in \mathbb{N}$ with residues
\eqn{
\text{Res}_{h=\lambda+1+n}\lsp \mathcal{I}_1^{0,0}(\lambda)=-r^{0,0}_{\lambda+n+1, \lambda}\,.
}[]
\item $\mathbf{\Do=-\Dt=-2}$: we have simple poles for $h=\lambda+n$, $n \in \mathbb{N}$ with residues
\eqn{
\text{Res}_{h=\lambda+n} \lsp \mathcal{I}_1^{-2,2}(\lambda)=-r^{-2,2}_{\lambda+n, \lambda}\,.
}[]
\item $\mathbf{\Do=\Dt=2}$: we have simple poles for $h=\lambda+1+n$, $n \in \mathbb{N}$ with residues
\eqn{
\text{Res}_{h=\lambda+n}\lsp \mathcal{I}_1^{2,2}(\lambda)=-r^{2,2}_{\lambda+n, \lambda}\,.
}[]
There is a small caveat in this case: the function $r_{h, \lambda}^{2,2}$ is not well defined for $\lambda=1$ at $h=1,\,2$.  More generally it is not possible to properly define the residue of the corresponding integral for these specific values. To overcome this problem,  in the main text, when computing the contributions from  $\langle\qBPS \hBPS \qBPS \hBPS\rangle$ to $\langle a^{(0)}_{k, \Delta, \ell}\rangle$, we have used the usual expansion in conformal blocks. These results seem to suggest that  $\mathrm{Res}_{h=1,2} \mathcal{I}_1^{2,2}(1)=0$, as well as $\mathrm{Res}_{h=2} \mathcal{I}_1^{2,2}(2)=0$.
\end{itemize}
Moving to the second kind of integral,  the integration can now  be performed by replacing the hypergeoemetric function contained in $\kappa_{4-2h}^{\Do, \Dt}$ with its series representation. By resumming the results after integration we get 
\eqna{
\mathcal{I}_2^{\Do, \Dt}(\lambda)&=\frac{\Gamma \mleft(\frac{1}{2} (-\Do+\Dt+2)\mright) \Gamma \mleft(-\frac{\Dt}{2}-h+\lambda +1\mright)}{\Gamma
   \mleft(-\frac{\Do}{2}-h+\lambda +2\mright)} \times \\& \quad \;  \times 
  {} _3F_2\left( \begin{array}{cc}
   -\frac{\Do}{2}-h+2\,, \quad \frac{\Dt}{2}-h+2\, , \quad -\frac{\Dt}{2}-h+\lambda +1 \\
   4-2 h\, , \quad -\frac{\Do}{2}-h+\lambda +2
   \end{array} ; 1\right) \, .
   }[]
   By using some identities for the generalized hypergeometric function it is easy to study the residues associated to this integral for the interesting values of the external dimensions.
   \begin{itemize}
 \item $\boldsymbol{\Do=\Dt=0}$: we have simple poles for $h=\lambda+n+1, \, n \in \mathbb{N}$  with residues
 \eqn{
 \mathrm{Res}_{h=\lambda+n+1} \mathcal{I}_2^{0,0}(\lambda)=(\shortminus 1)^{n+1}r_{\lambda+n+1, \lambda}^{0,0}\,.
 }[]
  \item $\boldsymbol{\Do=\Dt=2}$: we have simple poles for $h=\lambda+n \, n \in \mathbb{N}$  with residues
 \eqn{
 \mathrm{Res}_{h=\lambda+n} \mathcal{I}_2^{2,2}(\lambda) = \frac{(\lambda -2) (\lambda -1) (-1)^{n+1}}{(\lambda +n-2) (\lambda +n-1)}r_{\lambda+n, \lambda}^{2,2}\,.
 }[]
 Notice that the residue vanishes for $\lambda=1, \, 2$ for any twist.
   \end{itemize}
The last integral, the one containing $\log z$, is the hardest one and it was not possible to find a closed formula valid for any value of $\Do$ and $\Dt$. Thus we will report only the results necessary to reproduce the computations of the main discussion.
\begin{itemize}
\item $\boldsymbol{\Do=\Dt=0}$: for $\lambda>0$ and $h\geq 1$
\eqna{
\mathcal{I}^{0,0}_3(\lambda)&= -\frac{\pi ^2 r_{h,\lambda }^{0,0} }{\sin ^2(\pi  (\lambda -h))}+\frac{ \pi  \tan (\pi  h) \sin (\pi  (h+\lambda )) ) r_{h,\lambda }}{2  \sin ^2(\pi  \lambda )\sin (\pi  (\lambda -h)}\Big(H_{h-\lambda }  +H_{\lambda +h}\\& \quad \; -2 H_{h }-\frac{1}{\lambda +h -2}-\frac{1}{\lambda +h -1}-\frac{1}{\lambda +h}+\frac{1}{\lambda -h
   }+\frac{2}{h-1}+\frac{2}{h } \Big)\,,
}[]
where $H_n = \sum_{k=1}^n 1/k$ is the $n$-th harmonic number, or rather its analytic continuation to the complex plane.
\item $\boldsymbol{-\Do=\Dt \in \mathbb{N}>0}$: for $\lambda>0$ and $h\geq 1$
\eqna{
\mathcal{I}^{-\Dt,\Dt}_3(\lambda)&=\frac{-\pi ^2 r_{h, \lambda}^{-\Dt, \Dt}}{\sin ^2\left(\pi  \left(\frac{\Dt }{2}-\lambda +h \right)\right)}+\Pi^{(\Dt)}_{h, \lambda} r_{h, \lambda}^{-\Dt, \Dt} \Big( -\frac{2}{\Dt -2 \lambda +2 h }  \\ &+
 \sum_{\alpha=\pm 1}\left( \frac{2}{2 h+ \alpha  \Dt  -2}+\frac{2}{2 h -\alpha  \Dt }\right)+ \sum_{k=0}^2 \frac{2}{\Dt -2 (-k+\lambda +h )} \\& 
+H_{\frac{\Dt }{2}-\lambda +h }+H_{-\frac{\Dt }{2}+\lambda +h }-H_{h -\frac{\Dt
   }{2}}-H_{\frac{\Dt}{2}+h } \Big)\, ,
}[]
where we have defined
\eqn{
\Pi^{(\Dt)}_{h, \lambda}= \frac{\pi \sin \left(\frac{1}{2} \pi  (\Dt -2 h )\right) \sin \left(\frac{1}{2} \pi  (\Dt +2 h )\right)
   \sin \left(\frac{1}{2} \pi  (\Dt -2 (\lambda +h ))\right)}{\sin (\pi  \lambda ) \sin (2 \pi  h ) \sin (\pi  (\Dt -\lambda )) \sin \left(\pi  \left(\frac{\Dt
   }{2}-\lambda +h \right)\right)} \, .
}[]
\end{itemize}
In both cases notice the appearance of double poles: these are the signs of anomalous dimensions.

\section{Useful tools for resolving the ambiguity}
\subsection{Recursion relations for conformal blocks}\label{app:blocksRecRel}

In subsection~\ref{sec:ambiguity} we explained how to fix part of the ambiguity by imposing that some twist-two operators vanish. The way in which this happens in practice is that a twist-four addition to the ambiguity functions $\CA_m(z,\zb)$ can result into twist-two and higher contributions to the correlator. This is a consequence of the entries of $v_k^{(m)}(z,\zb)$ which can multiply the blocks and shift their dimension. An explicit example that shows how this comes about was given in equation~\eqref{eq:recrelExample}.

All these contributions with shifted twist can be obtained thanks to the recursion relations satisfied by the conformal blocks. The special case with equal external dimensions was given in~\cite{Dolan:2001tt}. For a general scalar conformal block in four dimensions one has the following recursion relations
\twoseqn{
\frac{1+v}{u} g_{\Delta,\ell}^{(a,b)} &= \begin{aligned}[t]
&-\frac12 \lsp g_{\Delta -1,\ell -1}^{(a,b)}-2\lsp g_{\Delta -1,\ell +1}^{(a,b)} -
\frac{\big(J^2-a^2\big)\big(J^2-b^2\big)}{8 (J -1) J^2 (J +1)}\,g_{\Delta +1,\ell +1}^{(a,b)}
\\&-
\frac{\big((\tau -2)^2-a^2\big) \big((\tau -2)^2-b^2\big)}{32 (\tau -1) (\tau -2)^2 (\tau -3)}\,g_{\Delta +1,\ell -1}^{(a,b)}
+\\&+
\frac{a b \left((\Delta -2)^2+\ell  (\ell +2)\right) }{(\tau -2) (\tau -4) (J -2) J}\,g_{\Delta ,\ell }^{(a,b)}
\end{aligned}
}[]{
\frac{1-v}{u} g_{\Delta,\ell}^{(a,b)} &=  \begin{aligned}[t]
&2 \lsp g_{\Delta -2,\ell }^{(a,b)}-\frac{a b \left(a^2-J^2\right) \left(J^2-b^2\right)}{8 (J-1) J^2 (J+1) (2-\tau ) (4-\tau )}\, g_{\Delta +1,\ell +1}^{(a,b)}
+\\&+
\frac{\left(J^2-a^2\right) \left(J^2-b^2\right)}{2 (J-1) J^2 (J+1)}\,g_{\Delta ,\ell +2}^{(a,b)}
+\\&+
\frac{1}{2} \left(\frac{a^2 b^2}{(J-2) J (\tau -2) (\tau -4)}+1\right) g_{\Delta ,\ell }^{(a,b)}
+\\&-
\frac{a b \left((\tau -2)^2-a^2\right) \left((\tau -2)^2-b^2\right)}{32 (J-2) J (\tau -1) (\tau -2)^2 (\tau -3)}\,g_{\Delta +1,\ell -1}^{(a,b)}
+\\&+
\frac{\left((\tau -2)^2-a^2\right) \left((\tau -2)^2-b^2\right) }{32 (\tau -1) (\tau -2)^2 (\tau -3)}\,g_{\Delta ,\ell -2}^{(a,b)}
+\\&+
\frac{\left((\tau -2)^2-a^2\right) \left((\tau -2)^2-b^2\right)  (J^2-b^2)(J^2-a^2) }{128 (J-1) J^2 (J+1) (\tau -1) (\tau -2)^2 (\tau -3)}\,g_{\Delta +2,\ell }^{(a,b)}
+\\&-
\frac{2 a b }{(J-2) J}\,g_{\Delta -1,\ell +1}^{(a,b)}-\frac{a b }{2 (\tau -2) (\tau -4)}\,g_{\Delta -1,\ell -1}^{(a,b)}
 \end{aligned}
}[][]
where $a=\Delta_{12}$, $b=\Delta_{34}$, $\tau=\Delta-\ell$, $J=\Delta+\ell$ and $u=z\zb$, $v=(1-z)(1-\zb)$. These two relations are sufficient for all cases considered here.

\subsection{Resumming blocks with coefficients}\label{app:resumming}

The ambiguity resolution for the correlators $\langle\qBPS\qBPS\hBPS\hBPS\rangle$ and $\langle\qBPS\qBPS\qBPS\qBPS\rangle$ involves an infinite family of conformal blocks of fixed twist
\eqn{
\CA_m(z,\zb) = \sum_{\ell=\ell_0}^\infty\,a_{\tau+\ell,\ell}\,g_{\tau+\ell,\ell}(z,\zb)\,,
}[]
with some given coefficients $a_{\tau+\ell,\ell}$ and fixed integer $\tau$. The goal of this appendix is to show how to perform this type of sums. First we separate the $z$ and the $\zb$ dependent parts by multiplying the sum by $z-\zb$
\eqn{
(z-\zb)\CA_m(z,\zb) = z\zb\,\kappa_{\tau-2}(\zb)\sum_{\ell=\ell_0}^\infty \frac{a_{\tau+\ell,\ell}}{(-2)^\ell}\lsp \kappa_{2\ell+\tau}(z) - (z\leftrightarrow\zb)\,,
}[]
Next we use the integral representation for the hypergeometric function inside $\kappa_{2\ell+\tau}(z)$. Under the assumption that the coefficients are suppressed enough to make the series convergent, we can swap the sum and the integral signs to obtain
\eqn{
S(z,t) = \frac{t^{\frac{\tau+b-1}2}(1-t)^{\frac{\tau-b-1}2}}{(1-tz)^{\frac{\tau-a}2}}\,\sum_{\ell=\ell_0}^\infty  \frac{a_{\tau+\ell,\ell}}{(-2)^\ell}\lsp\frac{\Gamma(2\ell+\tau)}{\Gamma\mleft(\ell+\frac{\tau+b}2\mright)\Gamma\mleft(\ell+\frac{\tau-b}2\mright)}\,\left(\frac{t(1-t)}{1-tz}\right)^\ell\,,
}[]
with $a=\Delta_{12}$ and $b=\Delta_{34}$.  If the coefficients $a_{\tau+\ell,\ell}$ come from a free theory OPE, we expect their expressions to involve ratios of $\Gamma$ functions and polynomials in $\ell$, possibly with some $(-1)^\ell$. If that is the case, the sum $S(z,t)$ can be performed in terms of hypergeometric functions. When $\tau$ is an integer, such hypergeometric functions will reduce to rational functions. Then we only need to perform the final integration in $\di t$ over the interval $[0,1]$. 
\eqn{
\CA_m(z,\zb) = \frac{z\zb}{z-\zb}\,\kappa_{\tau-2}(\zb) \int_0^1\di t\, S(z,t) + (z\leftrightarrow\zb)\,.
}[]
For this purpose, it is convenient to make the change of variables $t = \frac{s-z}{(s-1)z}$ which makes the integral somewhat easier to perform.

\section{\texorpdfstring{$\boldsymbol{D}$}{D}-functions}\label{app:Dfunctions}

It is possible to strip a kinematic prefactor from the $D$-functions defined in~\eqref{eq:Ddef} so as to obtain a function of the cross ratios $z,\zb$ only. Furthermore we can also remove the pole in $\Sigma=2$ that renders $D_{1111}$ ill defined. This leads to a ``reduced'' $D$-function denoted as $\Db$. The standard definition is the following
\eqn{
\Db_{\Delta_1\Delta_2\Delta_3\Delta_4}(z,\zb) = \frac{(x_{13}^2)^{\Sigma-\Delta_4}(x_{24}^2)^{\Delta_2}}{(x_{14}^2)^{\Sigma-\Delta_1-\Delta_4}(x_{34}^2)^{\Sigma-\Delta_3-\Delta_4}}\frac{2\lsp\prod_{i=1}^4\Gamma(\Delta_i)}{\pi^2\lsp\Gamma(\Sigma-2)}\,D_{\Delta_1\Delta_2\Delta_3\Delta_4}(x_1,\ldots,x_4)\,.
}[]
The recursion relations~\eqref{eq:recRelD} can be written as derivatives with respect to $z$ and $\zb$ of this reduced function as follows~\cite{Arutyunov:2002fh, Rastelli:2017udc}
\eqnal{
\Db_{\Delta_1+1\,\Delta_2+1\,\Delta_3\,\Delta_4} &= \frac{(1-z)\partial_z-(1-\zb)\partial_\zb}{z-\zb}\Db_{\Delta_1\Delta_2\Delta_3\Delta_4}\,,\\
\Db_{\Delta_1\,\Delta_2\,\Delta_3+1\,\Delta_4+1} &= \left(z\zb\,\frac{(1-z)\partial_z-(1-\zb)\partial_\zb}{z-\zb}+\Sigma-\Delta_1-\Delta_2\right)\Db_{\Delta_1\Delta_2\Delta_3\Delta_4}\,,\\
\Db_{\Delta_1\,\Delta_2+1\,\Delta_3+1\,\Delta_4} &= \frac{z\partial_z-\zb\partial_\zb}{z-\zb}\Db_{\Delta_1\Delta_2\Delta_3\Delta_4}\,,\\
\Db_{\Delta_1+1\,\Delta_2\,\Delta_3\,\Delta_4+1} &= \left((1-z)(1-\zb)\,\frac{z\partial_z-\zb\partial_\zb}{z-\zb}+\Sigma-\Delta_2-\Delta_3\right)\Db_{\Delta_1\Delta_2\Delta_3\Delta_4}\,,\\
\Db_{\Delta_1\,\Delta_2+1\,\Delta_3\,\Delta_4+1} &= \left(\frac{z(z-1)\partial_z-\zb(\zb-1)\partial_\zb}{z-\zb}+\Delta_2\right)\Db_{\Delta_1\Delta_2\Delta_3\Delta_4}\,,\\
\Db_{\Delta_1+1\,\Delta_2\,\Delta_3+1\,\Delta_4} &= \left(\frac{z(z-1)\partial_z-\zb(\zb-1)\partial_\zb}{z-\zb}+\Sigma-\Delta_4\right)\Db_{\Delta_1\Delta_2\Delta_3\Delta_4}\,.
}[]
with $\Sigma = \frac12(\Delta_1+\Delta_2+\Delta_3+\Delta_4)$.
The seed $\Db_{1111}$ is known in closed form and it reads
\eqn{
\Db_{1111}(z,\zb) = \frac{1}{z-\zb}\left(2\lsp\mathrm{Li}_2(z) -2\lsp\mathrm{Li}_2(\zb) + \log(z\zb) \,\log\!\Biggish(\frac{1-z}{1-\zb}\Biggish)\right)\,.
}[]
The $\Db$-functions satisfy the same permutation identities as conformal four-point functions. Namely
\eqna{
\Db_{\Delta_1\Delta_2\Delta_3\Delta_4}(z,\zb) &= 
\big((1-z)(1-\zb)\big)^{-\Delta_2}\Db_{\Delta_1\Delta_2\Delta_4\Delta_3}\Biggish(\frac{z}{z-1},\frac\zb{\zb-1}\Biggish)\,,\\&=
\big((1-z)(1-\zb)\big)^{\Delta_4-\Sigma}\Db_{\Delta_2\Delta_1\Delta_3\Delta_4}\Biggish(\frac{z}{z-1},\frac\zb{\zb-1}\Biggish)\,,\\&=
\big((1-z)(1-\zb)\big)^{\Delta_1+\Delta_4-\Sigma}\Db_{\Delta_2\Delta_1\Delta_4\Delta_3}(z,\zb)\,,\\&=
\Db_{\Delta_3\Delta_2\Delta_1\Delta_4}(1-z,1-\zb)\,, \\&=
(z\zb)^{\Delta_3+\Delta_4-\Sigma}\Db_{\Delta_4\Delta_3\Delta_2\Delta_1}(z,\zb)
\,.
}[eq:DpermIdentities]
They also satisfy the following identity~\cite{Dolan:2004iy}
\eqn{
\Db_{\Delta_1\Delta_2\Delta_3\Delta_4}(z\zb) = \Db_{\Sigma-\Delta_3\,\Sigma-\Delta_4\,\Sigma-\Delta_1\,\Sigma-\Delta_2}(z\zb)\,.
}[]

\section[Some comments on  \texorpdfstring{$\boldsymbol{\langle\qBPS[0q]\qBPS[0q]\hBPS\hBPS\rangle}$}{<qq 22>}]{Some comments on $\boldsymbol{\langle\qBPS[0q]\qBPS[0q]\hBPS\hBPS\rangle}$}\label{app:QqQqHH}
In Section~\ref{sec:QQHH} we have discussed the correlator $\langle \qBPS \qBPS \hBPS \hBPS \rangle $ and we have enumerated all the possible representations exchanged, as reported in table~\ref{tab:irrepsQQHH},  and the corresponding tensor structures  $\protect\TQQHH{k}$.  If now one considers a similar four-point function but with a generic quarter-BPS operator $\qBPS[0q]$,  as expected there is a proliferation of exchanged representations\footnote{In order to find this expression we did explicitly the cases for $q=1,\ldots7$ with \texttt{LieART} and tried to extrapolate a reasonable pattern.  We also checked that the dimensions agree for many values of $q$.}
\eqna{
(q,0,q) \otimes (q,0,q) &= \bigoplus_{\delta = -q}^q \bigoplus_{n=0}^{2q - 2|\delta|}\bigoplus_{\substack{m=|\delta| \\ m\equiv \delta\;\mathrm{mod}\;2}}^{2q-n-|\delta|}\,\mu_{n,m}^{(\delta)}\,(n+\delta+|\delta|,m,n-\delta+|\delta|)\,, \\
\mu_{n,m}^{(\delta)} & = \min(n+1,2q-|\delta|-n-m+1)\,.
}[reprsO0q]
However,  when considering the representations appearing in the intersection with $(0,2,0) \otimes (0,2,0)$,  \eqref{reprsO0q} reduces  to the ones in  table~\ref{tab:irrepsQQHH} with the exact same multiplicities. This observation tells us that the tensor structures in  $\langle\qBPS[0q]\qBPS[0q]\hBPS\hBPS\rangle$ should be related to the ones found for the $q=2$ case and indeed we obtain\footnote{Notice that this is not true for $\langle \hBPS \qBPS[0q] \qBPS[0q] \hBPS \rangle$.}
\eqn{
\mathbb{T}^{\langle\qBPS[0q]\qBPS[0q]\hBPS\hBPS\rangle}_k=(\strSSb{12} \strSSb{21})^{q-2}\lsp \protect\TQQHH{k} \, .
}[]
Given the simplicity of the tensor structures a similar analysis as the one for the $\qBPS$ case can be performed in this channel and we leave this to future works.  Extending our results to $\langle \hBPS \qBPS[0q] \qBPS[0q] \hBPS \rangle$ and eventually to $\langle \qBPS[0q] \qBPS[0q] \qBPS[0q]  \qBPS[0q] \rangle$ seems instead way harder to achieve and would require a more efficient way of dealing with the increasing number of representations exchanged.

\Bibliography[refs.bib]


\providecommand{\href}[2]{#2}\begingroup\raggedright\begin{thebibliography}{10}

\bibitem{Beem:2013qxa}
C.~Beem, L.~Rastelli and B.C.~van Rees, \emph{{The $\mathcal N=4$
  Superconformal Bootstrap}},
  \href{https://doi.org/10.1103/PhysRevLett.111.071601}{\emph{Phys. Rev. Lett.}
  {\bfseries 111} (2013) 071601}
  [\href{https://arxiv.org/abs/1304.1803}{{\ttfamily 1304.1803}}].

\bibitem{Beem:2016wfs}
C.~Beem, L.~Rastelli and B.C.~{van Rees}, \emph{{More ${\mathcal N}=4$
  superconformal bootstrap}},
  \href{https://doi.org/10.1103/PhysRevD.96.046014}{\emph{Phys. Rev. D}
  {\bfseries 96} (2017) 046014}
  [\href{https://arxiv.org/abs/1612.02363}{{\ttfamily 1612.02363}}].

\bibitem{Beem:2014zpa}
C.~Beem, M.~Lemos, P.~Liendo, L.~Rastelli and B.C.~van Rees, \emph{{The $
  \mathcal{N}=2 $ superconformal bootstrap}},
  \href{https://doi.org/10.1007/JHEP03(2016)183}{\emph{JHEP} {\bfseries 03}
  (2016) 183} [\href{https://arxiv.org/abs/1412.7541}{{\ttfamily 1412.7541}}].

\bibitem{Poland:2018epd}
D.~Poland, S.~Rychkov and A.~Vichi, \emph{{The Conformal Bootstrap: Theory,
  Numerical Techniques, and Applications}},
  \href{https://doi.org/10.1103/RevModPhys.91.015002}{\emph{Rev. Mod. Phys.}
  {\bfseries 91} (2019) 015002}
  [\href{https://arxiv.org/abs/1805.04405}{{\ttfamily 1805.04405}}].

\bibitem{Bissi:2021spj}
A.~Bissi, P.~Dey and G.~Fardelli, \emph{{Two Applications of the Analytic
  Conformal Bootstrap: A Quick Tour Guide}},
  \href{https://doi.org/10.3390/universe7070247}{\emph{Universe} {\bfseries 7}
  (2021) 247} [\href{https://arxiv.org/abs/2107.10097}{{\ttfamily
  2107.10097}}].

\bibitem{Aharony:2016dwx}
O.~Aharony, L.F.~Alday, A.~Bissi and E.~Perlmutter, \emph{{Loops in AdS from
  Conformal Field Theory}},
  \href{https://doi.org/10.1007/JHEP07(2017)036}{\emph{JHEP} {\bfseries 07}
  (2017) 036} [\href{https://arxiv.org/abs/1612.03891}{{\ttfamily
  1612.03891}}].

\bibitem{Alday:2017xua}
L.F.~Alday and A.~Bissi, \emph{{Loop Corrections to Supergravity on $AdS_5
  \times S^5$}},
  \href{https://doi.org/10.1103/PhysRevLett.119.171601}{\emph{Phys. Rev. Lett.}
  {\bfseries 119} (2017) 171601}
  [\href{https://arxiv.org/abs/1706.02388}{{\ttfamily 1706.02388}}].

\bibitem{Aprile:2017bgs}
F.~Aprile, J.M.~Drummond, P.~Heslop and H.~Paul, \emph{{Quantum Gravity from
  Conformal Field Theory}},
  \href{https://doi.org/10.1007/JHEP01(2018)035}{\emph{JHEP} {\bfseries 01}
  (2018) 035} [\href{https://arxiv.org/abs/1706.02822}{{\ttfamily
  1706.02822}}].

\bibitem{Goncalves:2019znr}
V.~Gon\c{c}alves, R.~Pereira and X.~Zhou, \emph{{$20'$ Five-Point Function from
  $AdS_5\times S^5$ Supergravity}},
  \href{https://doi.org/10.1007/JHEP10(2019)247}{\emph{JHEP} {\bfseries 10}
  (2019) 247} [\href{https://arxiv.org/abs/1906.05305}{{\ttfamily
  1906.05305}}].

\bibitem{Beem:2013sza}
C.~Beem, M.~Lemos, P.~Liendo, W.~Peelaers, L.~Rastelli and B.C.~van Rees,
  \emph{{Infinite Chiral Symmetry in Four Dimensions}},
  \href{https://doi.org/10.1007/s00220-014-2272-x}{\emph{Commun. Math. Phys.}
  {\bfseries 336} (2015) 1359}
  [\href{https://arxiv.org/abs/1312.5344}{{\ttfamily 1312.5344}}].

\bibitem{Dolan:2001tt}
F.A.~Dolan and H.~Osborn, \emph{{Superconformal symmetry, correlation functions
  and the operator product expansion}},
  \href{https://doi.org/10.1016/S0550-3213(02)00096-2}{\emph{Nucl. Phys. B}
  {\bfseries 629} (2002) 3}
  [\href{https://arxiv.org/abs/hep-th/0112251}{{\ttfamily hep-th/0112251}}].

\bibitem{Cordova:2016emh}
C.~Cordova, T.T.~Dumitrescu and K.~Intriligator, \emph{{Multiplets of
  Superconformal Symmetry in Diverse Dimensions}},
  \href{https://doi.org/10.1007/JHEP03(2019)163}{\emph{JHEP} {\bfseries 03}
  (2019) 163} [\href{https://arxiv.org/abs/1612.00809}{{\ttfamily
  1612.00809}}].

\bibitem{Dolan:2002zh}
F.A.~Dolan and H.~Osborn, \emph{{On short and semi-short representations for
  four-dimensional superconformal symmetry}},
  \href{https://doi.org/10.1016/S0003-4916(03)00074-5}{\emph{Annals Phys.}
  {\bfseries 307} (2003) 41}
  [\href{https://arxiv.org/abs/hep-th/0209056}{{\ttfamily hep-th/0209056}}].

\bibitem{DHoker:2003csh}
E.~D'Hoker, P.~Heslop, P.~Howe and A.V.~Ryzhov, \emph{{Systematics of quarter
  BPS operators in N=4 SYM}},
  \href{https://doi.org/10.1088/1126-6708/2003/04/038}{\emph{JHEP} {\bfseries
  04} (2003) 038} [\href{https://arxiv.org/abs/hep-th/0301104}{{\ttfamily
  hep-th/0301104}}].

\bibitem{Ryzhov:2001bp}
A.V.~Ryzhov, \emph{{Quarter BPS operators in N=4 SYM}},
  \href{https://doi.org/10.1088/1126-6708/2001/11/046}{\emph{JHEP} {\bfseries
  11} (2001) 046} [\href{https://arxiv.org/abs/hep-th/0109064}{{\ttfamily
  hep-th/0109064}}].

\bibitem{DHoker:2001jzy}
E.~D'Hoker and A.V.~Ryzhov, \emph{{Three point functions of quarter BPS
  operators in N=4 SYM}},
  \href{https://doi.org/10.1088/1126-6708/2002/02/047}{\emph{JHEP} {\bfseries
  02} (2002) 047} [\href{https://arxiv.org/abs/hep-th/0109065}{{\ttfamily
  hep-th/0109065}}].

\bibitem{Heslop:2003xu}
P.~Heslop and P.~Howe, \emph{{Aspects of N=4 SYM}},
  \href{https://doi.org/10.1088/1126-6708/2004/01/058}{\emph{JHEP} {\bfseries
  01} (2004) 058} [\href{https://arxiv.org/abs/hep-th/0307210}{{\ttfamily
  hep-th/0307210}}].

\bibitem{Heslop:2001gp}
P.J.~Heslop and P.S.~Howe, \emph{{OPEs and three-point correlators of protected
  operators in N=4 SYM}},
  \href{https://doi.org/10.1016/S0550-3213(02)00023-8}{\emph{Nucl. Phys. B}
  {\bfseries 626} (2002) 265}
  [\href{https://arxiv.org/abs/hep-th/0107212}{{\ttfamily hep-th/0107212}}].

\bibitem{Nirschl:2004pa}
M.~Nirschl and H.~Osborn, \emph{{Superconformal Ward identities and their
  solution}},
  \href{https://doi.org/10.1016/j.nuclphysb.2005.01.013}{\emph{Nucl. Phys. B}
  {\bfseries 711} (2005) 409}
  [\href{https://arxiv.org/abs/hep-th/0407060}{{\ttfamily hep-th/0407060}}].

\bibitem{Costa:2011mg}
M.S.~Costa, J.~Penedones, D.~Poland and S.~Rychkov, \emph{{Spinning Conformal
  Correlators}}, \href{https://doi.org/10.1007/JHEP11(2011)071}{\emph{JHEP}
  {\bfseries 11} (2011) 071} [\href{https://arxiv.org/abs/1107.3554}{{\ttfamily
  1107.3554}}].

\bibitem{Elkhidir:2014woa}
E.~Elkhidir, D.~Karateev and M.~Serone, \emph{{General Three-Point Functions in
  4D CFT}}, \href{https://doi.org/10.1007/JHEP01(2015)133}{\emph{JHEP}
  {\bfseries 01} (2015) 133} [\href{https://arxiv.org/abs/1412.1796}{{\ttfamily
  1412.1796}}].

\bibitem{Cuomo:2017wme}
G.F.~Cuomo, D.~Karateev and P.~Kravchuk, \emph{{General Bootstrap Equations in
  4D CFTs}}, \href{https://doi.org/10.1007/JHEP01(2018)130}{\emph{JHEP}
  {\bfseries 01} (2018) 130}
  [\href{https://arxiv.org/abs/1705.05401}{{\ttfamily 1705.05401}}].

\bibitem{Bargmann:1977gy}
V.~Bargmann and I.T.~Todorov, \emph{{Spaces of Analytic Functions on a Complex
  Cone as Carries for the Symmetric Tensor Representations of SO(N)}},
  \href{https://doi.org/10.1063/1.523383}{\emph{J. Math. Phys.} {\bfseries 18}
  (1977) 1141}.

\bibitem{Simmons-Duffin:2012juh}
D.~Simmons-Duffin, \emph{{Projectors, Shadows, and Conformal Blocks}},
  \href{https://doi.org/10.1007/JHEP04(2014)146}{\emph{JHEP} {\bfseries 04}
  (2014) 146} [\href{https://arxiv.org/abs/1204.3894}{{\ttfamily 1204.3894}}].

\bibitem{Karateev:2017yoq}
D.~Karateev, \emph{{Kinematics of 4D Conformal Field Theories}}, Ph.D. thesis,
  SISSA, Trieste,
  \href{https://www.sissa.it/tpp/phdsection/AlumniThesis/Denis%20Karateev.pdf}{\ttfamily
  www.sissa.it/tpp/phdsection/AlumniThesis/Denis Karateev.pdf}, 2017.

\bibitem{Perelomov2:1966}
A.~Perelomov and V.~Popov, \emph{{Casimir Operators for $u(n)$ and $su(n)$}},
  {\emph{Sov. J. of Nucl. Phys.} {\bfseries 3} (1966) 676}.

\bibitem{Hartwell:1994rp}
G.~Hartwell and P.S.~Howe, \emph{{(N, p, q) harmonic superspace}},
  \href{https://doi.org/10.1142/S0217751X95001820}{\emph{Int. J. Mod. Phys. A}
  {\bfseries 10} (1995) 3901}
  [\href{https://arxiv.org/abs/hep-th/9412147}{{\ttfamily hep-th/9412147}}].

\bibitem{Andrianopoli:1999vr}
L.~Andrianopoli, S.~Ferrara, E.~Sokatchev and B.~Zupnik, \emph{{Shortening of
  primary operators in N extended SCFT(4) and harmonic superspace
  analyticity}}, \href{https://doi.org/10.4310/ATMP.1999.v3.n4.a8}{\emph{Adv.
  Theor. Math. Phys.} {\bfseries 4} (2000) 1149}
  [\href{https://arxiv.org/abs/hep-th/9912007}{{\ttfamily hep-th/9912007}}].

\bibitem{Beem:2014kka}
C.~Beem, L.~Rastelli and B.C.~van Rees, \emph{{$ \mathcal{W} $ symmetry in six
  dimensions}}, \href{https://doi.org/10.1007/JHEP05(2015)017}{\emph{JHEP}
  {\bfseries 05} (2015) 017} [\href{https://arxiv.org/abs/1404.1079}{{\ttfamily
  1404.1079}}].

\bibitem{Chester:2014mea}
S.M.~Chester, J.~Lee, S.S.~Pufu and R.~Yacoby, \emph{{Exact Correlators of BPS
  Operators from the 3d Superconformal Bootstrap}},
  \href{https://doi.org/10.1007/JHEP03(2015)130}{\emph{JHEP} {\bfseries 03}
  (2015) 130} [\href{https://arxiv.org/abs/1412.0334}{{\ttfamily 1412.0334}}].

\bibitem{Beem:2016cbd}
C.~Beem, W.~Peelaers and L.~Rastelli, \emph{{Deformation quantization and
  superconformal symmetry in three dimensions}},
  \href{https://doi.org/10.1007/s00220-017-2845-6}{\emph{Commun. Math. Phys.}
  {\bfseries 354} (2017) 345}
  [\href{https://arxiv.org/abs/1601.05378}{{\ttfamily 1601.05378}}].

\bibitem{Beem:2014rza}
C.~Beem, W.~Peelaers, L.~Rastelli and B.C.~van Rees, \emph{{Chiral algebras of
  class S}}, \href{https://doi.org/10.1007/JHEP05(2015)020}{\emph{JHEP}
  {\bfseries 05} (2015) 020} [\href{https://arxiv.org/abs/1408.6522}{{\ttfamily
  1408.6522}}].

\bibitem{Beem:2017ooy}
C.~Beem and L.~Rastelli, \emph{{Vertex operator algebras, Higgs branches, and
  modular differential equations}},
  \href{https://doi.org/10.1007/JHEP08(2018)114}{\emph{JHEP} {\bfseries 08}
  (2018) 114} [\href{https://arxiv.org/abs/1707.07679}{{\ttfamily
  1707.07679}}].

\bibitem{Rastelli:2017ymc}
L.~Rastelli and X.~Zhou, \emph{{Holographic Four-Point Functions in the (2, 0)
  Theory}}, \href{https://doi.org/10.1007/JHEP06(2018)087}{\emph{JHEP}
  {\bfseries 06} (2018) 087}
  [\href{https://arxiv.org/abs/1712.02788}{{\ttfamily 1712.02788}}].

\bibitem{Behan:2021pzk}
C.~Behan, P.~Ferrero and X.~Zhou, \emph{{More on holographic correlators:
  Twisted and dimensionally reduced structures}},
  \href{https://doi.org/10.1007/JHEP04(2021)008}{\emph{JHEP} {\bfseries 04}
  (2021) 008} [\href{https://arxiv.org/abs/2101.04114}{{\ttfamily
  2101.04114}}].

\bibitem{Aprile:2017xsp}
F.~Aprile, J.~Drummond, P.~Heslop and H.~Paul, \emph{{Unmixing Supergravity}},
  \href{https://doi.org/10.1007/JHEP02(2018)133}{\emph{JHEP} {\bfseries 02}
  (2018) 133} [\href{https://arxiv.org/abs/1706.08456}{{\ttfamily
  1706.08456}}].

\bibitem{Drukker:2009sf}
N.~Drukker and J.~Plefka, \emph{{Superprotected n-point correlation functions
  of local operators in N=4 super Yang-Mills}},
  \href{https://doi.org/10.1088/1126-6708/2009/04/052}{\emph{JHEP} {\bfseries
  04} (2009) 052} [\href{https://arxiv.org/abs/0901.3653}{{\ttfamily
  0901.3653}}].

\bibitem{Maldacena:2011jn}
J.~Maldacena and A.~Zhiboedov, \emph{{Constraining Conformal Field Theories
  with A Higher Spin Symmetry}},
  \href{https://doi.org/10.1088/1751-8113/46/21/214011}{\emph{J. Phys. A}
  {\bfseries 46} (2013) 214011}
  [\href{https://arxiv.org/abs/1112.1016}{{\ttfamily 1112.1016}}].

\bibitem{Alba:2013yda}
V.~Alba and K.~Diab, \emph{{Constraining conformal field theories with a higher
  spin symmetry in d=4}},  \href{https://arxiv.org/abs/1307.8092}{{\ttfamily
  1307.8092}}.

\bibitem{Dolan:2004iy}
F.A.~Dolan and H.~Osborn, \emph{{Conformal partial wave expansions for N=4
  chiral four point functions}},
  \href{https://doi.org/10.1016/j.aop.2005.07.005}{\emph{Annals Phys.}
  {\bfseries 321} (2006) 581}
  [\href{https://arxiv.org/abs/hep-th/0412335}{{\ttfamily hep-th/0412335}}].

\bibitem{Caron-Huot:2017vep}
S.~Caron-Huot, \emph{{Analyticity in Spin in Conformal Theories}},
  \href{https://doi.org/10.1007/JHEP09(2017)078}{\emph{JHEP} {\bfseries 09}
  (2017) 078} [\href{https://arxiv.org/abs/1703.00278}{{\ttfamily
  1703.00278}}].

\bibitem{Alday:2017vkk}
L.F.~Alday and S.~Caron-Huot, \emph{{Gravitational S-matrix from CFT dispersion
  relations}}, \href{https://doi.org/10.1007/JHEP12(2018)017}{\emph{JHEP}
  {\bfseries 12} (2018) 017}
  [\href{https://arxiv.org/abs/1711.02031}{{\ttfamily 1711.02031}}].

\bibitem{Caron-Huot:2018kta}
S.~Caron-Huot and A.-K.~Trinh, \emph{{All tree-level correlators in
  AdS$_{5}$\texttimes{}S$_{5}$ supergravity: hidden ten-dimensional conformal
  symmetry}}, \href{https://doi.org/10.1007/JHEP01(2019)196}{\emph{JHEP}
  {\bfseries 01} (2019) 196}
  [\href{https://arxiv.org/abs/1809.09173}{{\ttfamily 1809.09173}}].

\bibitem{Alday:2020lbp}
L.F.~Alday and X.~Zhou, \emph{{All Tree-Level Correlators for M-theory on
  $AdS_7 \times S^4$}},
  \href{https://doi.org/10.1103/PhysRevLett.125.131604}{\emph{Phys. Rev. Lett.}
  {\bfseries 125} (2020) 131604}
  [\href{https://arxiv.org/abs/2006.06653}{{\ttfamily 2006.06653}}].

\bibitem{Eden:2000gg}
B.U.~Eden, P.S.~Howe, E.~Sokatchev and P.C.~West, \emph{{Extremal and
  next-to-extremal n point correlators in four-dimensional SCFT}},
  \href{https://doi.org/10.1016/S0370-2693(00)01181-3}{\emph{Phys. Lett. B}
  {\bfseries 494} (2000) 141}
  [\href{https://arxiv.org/abs/hep-th/0004102}{{\ttfamily hep-th/0004102}}].

\bibitem{Erdmenger:1999pz}
J.~Erdmenger and M.~Perez-Victoria, \emph{{Nonrenormalization of
  next-to-extremal correlators in N=4 SYM and the AdS / CFT correspondence}},
  \href{https://doi.org/10.1103/PhysRevD.62.045008}{\emph{Phys. Rev. D}
  {\bfseries 62} (2000) 045008}
  [\href{https://arxiv.org/abs/hep-th/9912250}{{\ttfamily hep-th/9912250}}].

\bibitem{Baggio:2012rr}
M.~Baggio, J.~de~Boer and K.~Papadodimas, \emph{{A non-renormalization theorem
  for chiral primary 3-point functions}},
  \href{https://doi.org/10.1007/JHEP07(2012)137}{\emph{JHEP} {\bfseries 07}
  (2012) 137} [\href{https://arxiv.org/abs/1203.1036}{{\ttfamily 1203.1036}}].

\bibitem{Aprile:2018efk}
F.~Aprile, J.~Drummond, P.~Heslop and H.~Paul, \emph{{Double-trace spectrum of
  $N=4$ supersymmetric Yang-Mills theory at strong coupling}},
  \href{https://doi.org/10.1103/PhysRevD.98.126008}{\emph{Phys. Rev. D}
  {\bfseries 98} (2018) 126008}
  [\href{https://arxiv.org/abs/1802.06889}{{\ttfamily 1802.06889}}].

\bibitem{Aprile:2020uxk}
F.~Aprile, J.M.~Drummond, P.~Heslop, H.~Paul, F.~Sanfilippo, M.~Santagata
  et~al., \emph{{Single particle operators and their correlators in free $
  \mathcal{N} $ = 4 SYM}},
  \href{https://doi.org/10.1007/JHEP11(2020)072}{\emph{JHEP} {\bfseries 11}
  (2020) 072} [\href{https://arxiv.org/abs/2007.09395}{{\ttfamily
  2007.09395}}].

\bibitem{Bern:1993kr}
Z.~Bern, L.J.~Dixon and D.A.~Kosower, \emph{{Dimensionally regulated pentagon
  integrals}}, \href{https://doi.org/10.1016/0550-3213(94)90398-0}{\emph{Nucl.
  Phys. B} {\bfseries 412} (1994) 751}
  [\href{https://arxiv.org/abs/hep-ph/9306240}{{\ttfamily hep-ph/9306240}}].

\bibitem{Bern:1992em}
Z.~Bern, L.J.~Dixon and D.A.~Kosower, \emph{{Dimensionally regulated one loop
  integrals}}, \href{https://doi.org/10.1016/0370-2693(93)90400-C}{\emph{Phys.
  Lett. B} {\bfseries 302} (1993) 299}
  [\href{https://arxiv.org/abs/hep-ph/9212308}{{\ttfamily hep-ph/9212308}}].

\bibitem{Ceplak:2021wzz}
N.~Ceplak, S.~Giusto, M.R.R.~Hughes and R.~Russo, \emph{{Holographic
  correlators with multi-particle states}},
  \href{https://arxiv.org/abs/2105.04670}{{\ttfamily 2105.04670}}.

\bibitem{Antunes:2021kmm}
A.~Antunes, M.S.~Costa, V.~Goncalves and J.V.~Boas, \emph{{Lightcone Bootstrap
  at higher points}},  \href{https://arxiv.org/abs/2111.05453}{{\ttfamily
  2111.05453}}.

\bibitem{Bissi:2020woe}
A.~Bissi, G.~Fardelli and A.~Georgoudis, \emph{{All loop structures in
  supergravity amplitudes on AdS5 \texttimes{} S5 from CFT}},
  \href{https://doi.org/10.1088/1751-8121/ac0ebf}{\emph{J. Phys. A} {\bfseries
  54} (2021) 324002} [\href{https://arxiv.org/abs/2010.12557}{{\ttfamily
  2010.12557}}].

\bibitem{Bissi:2020wtv}
A.~Bissi, G.~Fardelli and A.~Georgoudis, \emph{{Towards all loop supergravity
  amplitudes on AdS5\texttimes{}S5}},
  \href{https://doi.org/10.1103/PhysRevD.104.L041901}{\emph{Phys. Rev. D}
  {\bfseries 104} (2021) L041901}
  [\href{https://arxiv.org/abs/2002.04604}{{\ttfamily 2002.04604}}].

\bibitem{WessnBagger}
J.~Wess and J.~Bagger, \emph{Supersymmetry and Supergravity}, Princeton
  University Press (1992).

\bibitem{Arutyunov:2002fh}
G.~Arutyunov, F.~Dolan, H.~Osborn and E.~Sokatchev, \emph{{Correlation
  functions and massive Kaluza-Klein modes in the AdS / CFT correspondence}},
  \href{https://doi.org/10.1016/S0550-3213(03)00448-6}{\emph{Nucl. Phys. B}
  {\bfseries 665} (2003) 273}
  [\href{https://arxiv.org/abs/hep-th/0212116}{{\ttfamily hep-th/0212116}}].

\bibitem{Rastelli:2017udc}
L.~Rastelli and X.~Zhou, \emph{{How to Succeed at Holographic Correlators
  Without Really Trying}},
  \href{https://doi.org/10.1007/JHEP04(2018)014}{\emph{JHEP} {\bfseries 04}
  (2018) 014} [\href{https://arxiv.org/abs/1710.05923}{{\ttfamily
  1710.05923}}].

\end{thebibliography}\endgroup
\end{document}